\documentclass[12pt]{article}
\usepackage{color,rotate,graphics,epsf}
\usepackage{hyperref,amssymb}

\newcommand{\insertfig}[2]{\mbox{\epsfxsize=#1cm \epsfbox{#2.eps}}}
\newcommand{\ft}[2]{{\textstyle\frac{#1}{#2}}}

\newcommand{\Sym}{\mathop{\mbox{\Large\bf S}}}
\newcommand{\sym}{\mathop{\mbox{\bf S}}}

\newcommand{\bit}[1]{\mbox{\boldmath$#1$}}
\font\cmss=cmss12 
\def\1{\hbox{{1}\kern-.25em\hbox{l}}}
\def\bfZ{\relax{\hbox{\cmss Z\kern-.4em Z}}}

\begin{document}

\begin{titlepage}

\begin{flushright}
\begin{tabular}{l}
DOE/ER/40762-269 \\
UMD-PP\#04-005 \\
LPT-Orsay-03-83
\end{tabular}
\end{flushright}

\vspace{15mm}

\centerline{\large \bf Superconformal operators
                       in ${\cal N} = 4$ super-Yang-Mills theory}

\vspace{10mm}

\centerline{\sc A.V. Belitsky$^a$, S.\'E. Derkachov$^{b,c}$,
                G.P. Korchemsky$^b$, A.N. Manashov$^d$}

\vspace{10mm}

\centerline{\it $^a$Department of Physics, University of Maryland at College Park}
\centerline{\it College Park, MD 20742-4111, USA}

\vspace{3mm}

\centerline{\it $^b$Laboratoire de Physique Th\'eorique\footnote{Unit\'e
                    Mixte de Recherche du CNRS (UMR 8627).},
                    Universit\'e de Paris XI}
\centerline{\it 91405 Orsay C\'edex, France}

\vspace{3mm}

\centerline{\it $^c$Department of Mathematics, St.-Petersburg
                    Institute of Technology}
\centerline{\it 198103, St.-Petersburg, Russia}

\vspace{3mm}

\centerline{\it $^d$Institut f\"ur Theoretische Physik II,
                    Ruhr-Universit\"at Bochum}
\centerline{\it 44780 Bochum, Germany}

\vspace{5mm}

\centerline{\bf Abstract}

\vspace{0.5cm}

We construct, in the framework of the $\mathcal{N}=4$ SYM theory, a
supermultiplet of twist-two conformal operators and study their
renormalization properties. The components of the supermultiplet have
the same anomalous dimension and enter as building blocks into
multi-particle quasipartonic operators. The latter are determined by
the condition that their twist equals the number of elementary
constituent fields from which they are built. A unique feature of the
$\mathcal{N}=4$ SYM is that all quasipartonic operators with different
$SU(4)$ quantum numbers fall into a single supermultiplet. Among them
there is a subsector of the operators of maximal helicity, which has
been known to be integrable  in the multi-color limit in QCD,
independent of the presence of supersymmetry. In the $\mathcal{N}=4$
SYM theory, this symmetry is extended to the whole supermultiplet of
quasipartonic operators and the one-loop dilatation operator coincides
with a Hamiltonian of integrable $SL(2|4)$ Heisenberg spin chain.
\end{titlepage}

\section{Introduction}

Quantum chromodynamics---the theory of strong interactions---enjoys a
number of space-time symmetries at the classical level: it is invariant
under the $SO(4,2)$ group of conformal transformations in four dimensions
\cite{MacSal69}. When the theory is quantized, five of the charges, the
dilatation and four conformal boosts cease to conserve and the conformal
group is reduced down to its Poincar\'e subgroup. Still, the conformal
symmetry has important consequences in QCD and provides powerful tools
in various applications (see the review \cite{BraMulKor03} and references
therein).

It turns out that the conformal symmetry is not the only symmetry of QCD.
A few years ago it has been found that evolution equations for scattering
amplitudes in high-energy QCD possess a hidden symmetry in the multi-color
limit \cite{Lip93,FadKor94}. Namely, the partial waves of the scattering
amplitudes satisfy a Schrodinger-like equation which has a large number
of conserved charges and is completely integrable. The underlying integrable
structure has been identified as corresponding to the celebrated Heisenberg
spin magnet. Later, similar integrable structures have been discovered in
studies of renormalization group equations (or the dilatation operator,
phrased differently \cite{CalSym70,Ohr82}) for multi-particle Wilson
operators in QCD \cite{BraDerMan98,BraDerKorMan99,Bel99a,Bel99b,DerKorMan00}.
The Wilson operators ${\cal W}_N^{\mu_1 \dots \mu_j}(0)$ are gauge-invariant
local composite operators built from $N$ fundamental fields $X_i$, carrying
in general Lorentz indices, and covariant derivatives ${\cal D}^\mu =
\partial^\mu - i g A^\mu$ acting on them
\begin{equation}
\label{WilsonOperators}
{\cal W}_N^{\mu_1 \dots \mu_j}
=
{\rm tr} \,
\Big\{
\left({\cal D}^{\mu_1} \dots
{\cal D}^{\mu_{k}} X_1 (0)\right) \left({\cal D}^{\mu_{k + 1}} \dots {\cal
D}^{\mu_{n}} X_2(0)\right)\cdots \left({\cal D}^{\mu_{m+1}} \dots {\cal
D}^{\mu_{j}} X_N(0)\right)
\Big\}
\, .
\end{equation}
According to their tensor structure these operators can be decomposed into
irreducible components with respect to the Lorentz group. The component with the
maximal Lorentz spin has a symmetric traceless Lorentz structure. It is projected
from ${\cal W}_N^{\mu_1 \dots \mu_j}(0)$ via
\begin{equation}
\label{MaxSpinWilson}
{\cal O}_N^{\mu_1 \dots \mu_j}
=
\Sym_{\mu_1 \dots \mu_j}
{\cal W}_N^{\mu_1 \dots \mu_j}
\, ,
\end{equation}
where the operation $\sym$ symmetrizes corresponding indices and subtracts traces,
e.g., $\sym_{\mu \nu} T^{\mu \nu} = \ft1{2} (T^{\mu \nu} + T^{\nu \mu}) - \ft14
g^{\mu \nu} T_\sigma{}^\sigma$. A distinguished property of the maximal-spin
operators is that their twist, i.e., dimension minus spin, equals to the sum of
twists of the individual fields $X_i$. The operators (\ref{MaxSpinWilson}) mix
under renormalization and the properties of their mixing matrix depend on the
choice of fundamental constituent $X_i$. As was shown in the above-mentioned
papers, for specific helicities of $N$ elementary fields entering the composite
operator, the corresponding one-loop dilatation operator admits, in multi-color
limit, $N - 2$ extra nontrivial conserved charges in addition to the conformal
Casimir and its third projection. Thus, the evolution equations for such operators
are completely integrable in QCD.

It has been found that the one-loop dilatation operator in QCD gives rise
to two essentially different integrable structures: For operators with
aligned helicities, like three-quark baryons of helicity-$3/2$
\cite{BraDerMan98,BraDerKorMan99} or spin-three ``glueballs'' \cite{Bel99b},
the dilatation operator is equivalent to a Hamiltonian of a closed non-compact
$SL(2)$ Heisenberg spin chain \cite{BraDerKorMan99,Bel99b}. The second
structure has emerged in the multi-color limit from the renormalization of
operators built from the quark-antiquark pair and gluon strength tensors
\cite{BraDerMan98,BraDerKorMan99,Bel99a,Bel99b,DerKorMan00}. The corresponding
dilatation operator is integrable as well and is also related to a Heisenberg
spin magnet. Since the QCD quarks belong to the fundamental representation
of the gauge group, the interaction between the quark and antiquark fields
is suppressed for $N_c \to \infty$. As a consequence, the arising Hamiltonian
turns out to be of an inhomogeneous open Heisenberg spin chain
\cite{BraDerMan98,Bel99a,Bel99b,DerKorMan00}. In both cases, the
spin operators acting on the sites of the chain are the generators of the
$SL(2)$ group. The latter corresponds, on the QCD side, to the so-called
collinear subgroup of the full conformal group.

We would like to stress that in one-loop approximation the non-conformal nature
of QCD is irrelevant and the one-loop dilatation operator ought to be conformally
invariant. At the same time, integrability emerges as yet another symmetry of the
dilatation operator in four-dimensional Yang-Mills theory which has immediate
consequences for renormalization of composite operators in QCD and its
supersymmetric extensions. To understand its origin one can consider the
dilatation operator in maximally supersymmetric extension of QCD --- the
$\mathcal{N}=4$ SYM theory \cite{GliSchOli77,BriSchSch77}. As compared to QCD,
this model exhibits a number of exceptional properties which simplify its
structure enormously. The ${\cal N} = 4$ SYM is an example of a full-fledged
interacting conformal field theory in four dimensions invariant under
superconformal $PSU(2, 2 | 4)$ transformations \cite{SohWes81,Man83}. More
recently it was proposed that in the multi-color limit the ${\cal N} = 4$ SYM
theory admits a dual description in terms of a superstring theory on $AdS_5
\times S^5$ background \cite{Mal97,GunKlePol98,Wit98}. This suggests that
integrability of Yang-Mills theory is in one-to-one correspondence with
symmetries of the string theory \cite{Gorsky,MinZar02,BeiSta03}.

The Wilson operators in the ${\cal N} = 4$ SYM theory have the same form
(\ref{WilsonOperators}) with the only difference being that they carry the
additional $SO(6) \sim SU(4)$ charge.  Among them there is a subclass of
the Berenstein-Maldacena-Nastase (BMN) operators \cite{BerMalNas02} which,
in the simplest form, do not involve covariant derivatives and have a
near-to-extremal charge with respect to an $SO(2)$ subgroup of the
$R$-symmetry group $SO(6)$. These operators do not have their counter-partners
in QCD and reflect the supersymmetric nature of the ${\cal N} = 4$ model.

Recently it has been found that the one-loop dilatation operator of the ${\cal N}
= 4$ model possesses integrable structures analogous to the those observed in QCD
in two different sectors of Wilson operators. Namely, renormalization of the BMN
operators is driven by an evolution operator which can be identified as a
Hamiltonian of a Heisenberg spin magnet with spins belonging to the $SO(6)$ group
\cite{MinZar02}. As we outlined above, the one-loop dilatation operator in QCD is
integrable in the $SL(2)$ sector and gives rise to the $SL(2)$ Heisenberg spin
magnet. The same $SL(2)$ integrable sector arises in the ${\cal N} = 4$
super-Yang-Mills theory \cite{BelGorKor03}. The subsequent studies of
renormalization properties of ``mixed'' operators involving covariant derivatives
and possessing a large $R$-charge have allowed to the authors of
Refs.~\cite{Bei03a,BeiSta03} to conclude that for an arbitrary Wilson operator
(\ref{WilsonOperators}) the one-loop dilatation operator in the ${\cal N} = 4$
super-Yang-Mills theory is equivalent to the Hamiltonian of the $PSU(2,2|4)$
Heisenberg spin chain. These intriguing results indicate the potential, if
extended to all orders, as was pointed out in Refs.\ \cite{BeiKriSta03,Bei03b},
of complete integrability of the ${\cal N} = 4$ super-Yang-Mills theory.
Additional support for this conjecture comes from the gauge/string duality:
recently is has been found that the classical world-sheet sigma model on the
$AdS_5 \times S^5$ admits an infinite set of conserved charges
\cite{ManSurWad02,BenPolRoi03,Val03,Ald03}. In Ref.\ \cite{DolNapWit03} it was
found that the Yangian structure of the superstring sigma model maps into
symmetries of the dilatation operator in the Yang-Mills theory. Several
successful matchings, starting from Refs.\ \cite{GubKlePol02,TseFro02}, of the
gauge-theory results to the rotating string solutions serve as a very nontrivial
verification of the AdS/CFT correspondence \cite{Sta03}.

Let us elaborate more on the Wilson operators (\ref{WilsonOperators}). As we
already mentioned, their maximal-spin component corresponds to a symmetric,
traceless tensor and has the lowest twist in each $N$-particle sector. Below
we discuss only this specific subclass of operators. To simplify the analysis
even further, one projects out the Lorentz tensor onto the light-cone. This is
accomplished by contracting the indices with a light-like vector $n_\mu$
($n_\mu^2= 0$), which has the virtue of automatic symmetrization and trace
subtraction. Thus the resulting tensor transforms in an irreducible
representation of the Lorentz group. This makes the covariance manifest
without the need to deal with open indices. Thus, the dynamics is effectively
projected on the light cone and the full superconformal group is reduced to
its collinear subgroup. This results into the following single-trace
multi-particle operators:
\begin{equation}
\label{MultiParticleOperators}
{\cal O}_{i_1, \dots , i_N} (\xi_1, \dots , \xi_N) = {\rm tr} \, X_{i_1} (n\xi_1)
\dots X_{i_k} (n\xi_k) X_{i_{k + 1}} (n\xi_{k + 1}) \dots X_{i_N} (n\xi_N) \, ,
\end{equation}
which is built from the ``good" field components of the ${\cal N} = 4$ model
$X = \{ F^+{}^\perp_\mu , \lambda_{+\alpha}^A , \bar\lambda^{\dot\alpha}_{+A},
\phi^{AB} \}$, defined in the next section. By making such a choice we
automatically restrict ourselves to the so-called quasipartonic operators. The
Taylor expansion of the non-local operators (\ref{MultiParticleOperators})
with respect to the field separations on the light-cone produces sets of local
gauge-invariant operators (\ref{MaxSpinWilson}). The gauge invariance of
(\ref{MultiParticleOperators}) is restored by inserting the gauge links
$[\xi_{k + 1}, \xi_k] = P \exp \left( i g \int^{\xi_{k+1}}_{\xi_k} d \xi'\,
n \cdot A (n\xi') \right)$ between the fields. We do not display them in
(\ref{MultiParticleOperators}) since later on we will be using the
light-like gauge $A^+ \equiv (n \cdot A) = 0$. In this gauge, the above
projections of fields represent on-shell partons; hence, the name
quasipartonic operators. These operators form a closed set under the
action of the dilatation operator and have the following unique properties
\cite{BukFroKurLip85}: They carry a definite twist which is equal to the number
of constituents. Since the twist is conserved under renormalization, the
dilatation operator does not change the number of partons.

In the present paper we will be dealing with two-particle quasipartonic
operators. The latter serve as building blocks for the multi-particle operators
(\ref{MultiParticleOperators}) discussed above and, correspondingly, to the
one-loop dilatation operator acting on this space: only pair-wise interactions
are relevant to this order. Moreover, we restrict ourselves to the discussion of
the color-singlet two-particle blocks because the gauge invariance of the Wilson
operators (\ref{MaxSpinWilson}) allows one to reduce the action of the pair-wise
dilatation generator in the color space in the multi-color limit to the quadratic
Casimirs, $ t^a_n \otimes t^a_{n+1} \to - \ft12 N_c$ for $N_c\to \infty$. Here an
$SU(N_c)$ generator $t^a_n$ is assumed to act on a constituent placed on the
$n$-th site of an $N$-particle colorless operator. Two-particle quasipartonic
operators have obviously twist-two, and, according to their quantum numbers, they
can be separated into different sectors. The operators belonging to each sector
mix only among themselves and their renormalization is driven by an evolution
kernel, or equivalently a dilatation operator. As was already mentioned, the
one-loop dilatation operator is completely integrable in QCD in the sector of
operators with maximal helicity. However, this sector is autonomous in QCD and
there exists no relation between it and the dilatation operators acting in other
sectors. To obtain such a relation one has to employ supersymmetric extensions
of QCD.

Supersymmetry connects Wilson operators belonging to different sectors and,
in addition, imposes severe constraints on their mixing matrix. These constraints
tighten up as we go from non-supersymmetric theories all the way up to the
maximal $\mathcal{N}=4$ supersymmetry. Ultimately this allows one to resolve
efficiently the operator mixing since the supersymmetry regroups all operators
into supermultiplets whose components evolve autonomously under scale
transformations and possess, at the same time, identical anomalous dimensions.
The number of supermultiplets and their structure depend on how much
supersymmetry a theory possesses. In the present paper we construct a
supermultiplet of twist-two operators for the $\mathcal{N}=4$ model by a
direct calculation and identify its anomalous dimension. We demonstrate that
the supermultiplet is unique; that is, its components span all twist-two
operators transforming in different irreducible representations of the
Lorentz and internal symmetry groups.%
\footnote{A subset of $SU(4)$ flavor-singlet twist-two operators was identified
in Ref.\ \cite{KotLip02} by a diagonalization of the anomalous dimension matrix
deduced from calculations of Feynman diagrams.}

The crucial difference between ${\cal N} = 4$ and other Yang-Mills theories is
that the former allows one to relate the sector of the maximal helicity operators
to all other sectors. This is a mere consequence of the span in helicities of
one-particle massless states which enter a given multiplet in supersymmetric
models. In ${\cal N} = 1$ supersymmetry, the gauge multiplet is formed by two
CPT-conjugated sets of states with the helicities $\lambda = 1, \ft12$ and
$\lambda = -1, -\ft12$, respectively. Supersymmetric transformations change the
helicity by a half, so that for a composite operator ${\cal O}^{(\lambda,
\lambda')}(\xi_1,\xi_2) \sim X^{(\lambda)}(\xi_1) X^{(\lambda')}(\xi_2)$ built
from fields of helicities $\lambda$ and $\lambda'$ one has
\begin{equation}
[{\cal Q}, {\cal O}^{(\lambda, \lambda')}]
\sim
{\cal O}^{(\lambda + 1/2, \lambda')}
+
{\cal O}^{(\lambda, \lambda' + 1/2)}
\, .
\end{equation}
This is not sufficient to link the aligned-helicity two-fermion operators to
the counter-aligned ones since for that one needs to flip the helicity ($\lambda$
or $\lambda'$) by one unit. The ${\cal N} = 4$ super-Yang-Mills, having the
maximal possible number of global supersymmetries, is CPT self-conjugate and
spans all possible helicities of particle states $\lambda = \pm 1, \pm \ft12, 0$.
Thus, to connect the above two sectors it is enough to apply the supersymmetric
transformations twice
\begin{equation}
\label{AlignedNonaligned}
[ {\cal Q}^A, [{\cal Q}^B, {\cal O}^{(\lambda, \lambda')}] ]
\sim
{\cal O}^{(\lambda + 1, \lambda')}
+
{\cal O}^{(\lambda + 1/2, \lambda' + 1/2)}
+
{\cal O}^{(\lambda, \lambda' + 1)}
\, .
\end{equation}
This explains why in the minimal supersymmetric extension of QCD  ---
${\cal N} = 1$ super-Yang-Mills --- there are two independent
supermultiplets of conformal operators: one of aligned helicities
\cite{BukFroKurLip85,BelMul99} and another one of opposite helicities
\cite{BukFroKurLip84,BukFroKurLip85,BelMulSch98,BelMul00}. The former
supermultiplet inherits integrability of the one-loop dilatation
operator in QCD. The exceptional place of the maximally supersymmetric
gauge theory in a row of Yang-Mills theories is that both types of the
operators enter the same supermultiplet. Therefore the one-loop
dilatation operator is integrable in the $\mathcal{N}=4$ model
\cite{BeiSta03} and anomalous dimension of the supermultiplet is
identical at one-loop order to the anomalous dimension of the
maximal-helicity quark \cite{BraDerMan98,BraDerKorMan99} and gluon
\cite{Bel99b} QCD conformal operators.

Our presentation is organized as follows. In the next section,
after recalling the definition of conformal operators in
a generic case, we turn to the classification of all two-particle conformal
blocks in maximally supersymmetric gauge theory. Subsequently in section
\ref{SuperMultiplet}, we discuss the projection of the ${\cal N} = 4$
supersymmetry algebra on the light-cone: one needs only this subalgebra for
discussion of twist-two conformal blocks. Then we construct a supermultiplet
which embeds all of them. The anomalous dimension of the supermultiplet is fixed
in section \ref{ADsupermultiplet} to two-loop order by computing one of its
components and using the result of Ref.\ \cite{KotLipVel03}. Section
\ref{Discussion} is devoted to discussions and conclusions. A few appendices
contain some details of the calculations.

\section{Conformal operators in ${\cal N} = 4$ super-Yang-Mills}
\label{CountingOperators}

As we outlined in the introduction, the nonlocal light-cone operators
(\ref{MultiParticleOperators}) serve as generating functions for the Wilson
operators of the lowest twist. Indeed, expanding Eq.\ (\ref{MultiParticleOperators})
in powers of the ``light-cone distances'' $\xi_j - \xi_k$, one can obtain an
infinite set of local composite operators (\ref{MaxSpinWilson}) with covariant
derivatives projected with the light-like vector%
\footnote{In addition to $n^\mu$, one introduces also an orthogonal light-cone
vector $n^{\ast \mu}$, $n^\ast_\mu n^{\ast \mu} = 0$ normalized such that
$n^\ast_\mu n^\mu = 1$. Then, an arbitrary vector can be decomposed as
$v^\mu \equiv v^+ n^{\ast \mu} + v^- n^\mu + v^\mu_\perp$, where the ``plus''
and ``minus'' superscripts stand for contraction with the vectors $n^\mu$
and $n^{\ast \mu}$, respectively.} $n^\mu$. The main advantage of dealing
with the nonlocal light-cone operators (\ref{MultiParticleOperators}) is
that they have simple transformation properties under a (super)conformal
group.

Since the fields entering Eq.\ (\ref{MultiParticleOperators}) ``live'' on the
light-cone, defined by the four-vector $n^\mu$, we can significantly simplify our
considerations by restricting the full $SO (4,2)$ conformal symmetry group down to
its collinear conformal subgroup $SO(2,1) \sim SL(2)$, for a review (see Ref.\
\cite{BraMulKor03}). The latter is a group of projective transformations on a
line;
\begin{equation}
\label{sl2}
\xi \to \frac{a \xi + b}{c \xi + d}
\, , \qquad
ad - bc = 1
\, ,
\end{equation}
with $\xi$ defining the position of a field operator on the light-cone, $X(\xi
n^\mu) \equiv X(\xi)$. Its algebra is formed by projections of the generators
of translations ${\cal L}^+ = i {\cal P}^+$, conformal boosts ${\cal L}^- =
\ft{i}2 {\cal K}^-$, and a combination of Lorentz transformations and dilatations
${\cal L}^0 = \ft{i}2 \left( {\cal D} + {\cal M}^{-+} \right)$. Yet another
combination of the latter two determines the operator of the twist, ${\cal E} =
i \left( {\cal D} - {\cal M}^{-+} \right)$, which commutes with ${\cal L}^0$,
${\cal L}^+$ and ${\cal L}^-$. Primary fields are transformed under the $SL(2)$
transformations
(\ref{sl2}) as
\begin{equation}
\label{X-trans}
X (\xi) \to X' (\xi) = (c \xi + d)^{- 2j} X \left( \frac{a\xi + b}{c\xi + d} \right)
\, , \qquad
j = \ft12 (d + s)
\end{equation}
where $j$ is the conformal spin of the field. It is given by half the sum of the
canonical dimension of the field, $d$, and its spin projection on the light-cone
${\mit\Sigma}_{+-} X(\xi) = s X(\xi)$. The operator ${\cal E}$ counts the twist
$t = (d - s)$ of the field $X$, i.e.,
\begin{equation}
{} [{\cal E} , X(\xi)] = (d-s) X(\xi)\,.
\end{equation}
Obviously, the twist of the multi-particle operator (\ref{MultiParticleOperators})
is equal to the sum of the twists of elementary fields. In what follows, as we
pointed out above, we shall consider the operators of the minimal twist.

Let us examine the transformation properties of the elementary fields in the
$\mathcal{N}=4$ theory. For scalars one has $d_{\rm sc}=1$ and $s_{\rm sc}=0$
leading to $j_{\rm sc}=1/2$ and $t_{\rm sc}=1$. For vectors and fermions, one has
to separate the corresponding fields into components with different values of the
spin projection $s_q=\pm 1/2$ and $s_g=\pm 1, 0$, respectively. As before, the
minimal twist corresponds to the component with the maximal spin. Let us discuss
the fermion and gauge fields separately.

The four-component fermion field is decomposed into two components with $s=\pm 1/2$
by means of the projection operators
\begin{equation}
\label{FermionLCdec}
\psi = \psi_+ + \psi_- \, , \qquad \psi_\pm \equiv {\mit\Pi}^{\pm} \psi\,,\qquad
{\mit\Pi}^{\pm} \equiv \ft12 \gamma^{\mp} \gamma^{\pm} \, ,
\end{equation}
so that $\psi_\pm$ have the following spins:
\begin{equation}
{\mit\Sigma}^{-+} \psi_{\pm} = \pm \ft12\, \psi_{\pm} \, ,
\end{equation}
where we have used the light-cone projected spin tensor ${\mit\Sigma}^{\mu\nu}
\equiv \ft{1}4 [\gamma^\mu , \gamma^\nu]$. Since the canonical dimension of the
fermion field is $d_q = 3/2$, one finds that the $\psi_+$-component has the
conformal spin $j_q=1$ and the twist $t_q=1$. Similarly, for the $\psi_-$-component
one finds $j_q=1/2$ and $t_q=2$. Therefore, only the $\psi_+$ component enters a
nonlocal operator of the minimal twist. Notice that Eq.\ (\ref{FermionLCdec}) is
identical to the decomposition used in the light-cone quantization \cite{KogSop70}
and we will use the latter formalism momentarily to point out important
modifications when the $\psi_-$ enters the composite operator. So far we did not
refer to the specifics of the maximally supersymmetric Yang-Mills theory. Its
Lagrangian is defined in appendix \ref{N=4Lagrangian} in terms of the Weyl spinors
in order to preserve the $SU(4)$ covariance. Going from the four-component
notations to the Weyl gaugino fields, the projection of the ``good'' components
converts into%
\footnote{The conventions for four-dimensional Clifford algebra is fixed in
appendix \ref{DiracAlgebra}.}
\begin{equation}
\lambda_{+ \alpha} \equiv \ft12 \bar\sigma^-{}_{\alpha\dot\beta} \, \sigma^{+ \;
\dot\beta\gamma} \lambda_\gamma \, , \qquad
\bar\lambda_+^{\dot\alpha}
\equiv \ft12 \sigma^{- \; \dot\alpha\beta} \, \bar\sigma^+{}_{\beta\dot\gamma}
\bar\lambda^{\dot\gamma}
\, .
\end{equation}
Each Weyl spinor $\lambda_{+ \alpha}$ and $\bar\lambda_+^{\dot\alpha}$ has only
one nonvanishing component which describes a state with a definite helicity,
$+1$ and $-1$, respectively. For brevity, we suppress in what follows the
subscript ``plus" on the fermion field designating its ``good'' light-cone
projections, i.e., $\lambda_+ \to \lambda$.

For the gauge field, one has to project the Lorentz indices of the strength tensor
$F^{\mu\nu}$ onto the longitudinal light-cone directions and the transverse space
with the two-dimensional metric $g^\perp_{\mu\nu} = g_{\mu\nu} - n_\mu n^\ast_\nu
- n_\nu n^\ast_\mu$. The spin assignments for different projections of $F^{\mu\nu}$
are as follows
\begin{equation}
{\mit\Sigma}^{-+} F^{\pm \mu}_{\phantom{+} \, \perp} = \pm F^{\pm
\mu}_{\phantom{+} \, \perp} \, , \qquad {\mit\Sigma}^{-+}  F^{+ -} =
{\mit\Sigma}^{-+}  F^{\mu\nu}_{\perp\perp} = 0 \, ,
\end{equation}
where ${\mit\Sigma}^{\mu\nu} F^{\rho\sigma} = g^{\mu\rho} F^{\nu\sigma} -
g^{\nu\rho} F^{\mu\sigma} - (\rho \leftrightarrow \sigma)$. Since the canonical
dimension of the strength tensor is $d_g=2$, the $F^{+ \mu}_{\phantom{+} \,
\perp}$-component has the conformal spin $j_g=3/2$ and twist $t_g=1$. For
$F^{+ -}$ and $F^{\mu\nu}_{\perp\perp}$-components one gets $j_g=1$ and $t_g=2$,
while for the $F^{- \mu}_{\phantom{+} \, \perp}$-component one has $j_g = 1/2$
and $t_g = 3$. Thus, the minimal twist is associated with the
$F^{+ \mu}_{\phantom{+} \, \perp}$-component only.

To summarize, the leading twist multi-particle operators are constructed solely
from the ``good'' fields $\psi_+$, $F^{+\mu}_{\phantom{+} \, \perp}$ and scalars
$\phi$ living on the light cone. Let us reiterate that the twist of such nonlocal
operators equals the number of primary fields involved and they are known in QCD
as quasipartonic operators \cite{BukFroKurLip85}. The ``good'' fields are
transformed as conformal primaries with respect to the collinear conformal
subgroup and carry the following conformal spins:
\begin{equation}
j_q = 1 \, , \qquad j_g = \ft32 \, , \qquad j_s = \ft12 \, ,
\end{equation}
for gauginos, gluons and scalars, respectively, as we established earlier. A
unique feature of quasipartonic operators is that they form a closed set with
respect to the action of the dilatation operator. Since the twist is preserved
under dilatations, the quasipartonic operators can only be transformed into
operators of the same twist. Therefore, the total number of constituents is
conserved. In other words, the dilatation operator acts ``elastically'' on the
space of quasipartonic operators --- there is no annihilation/production of the
``good'' fields --- and, therefore, it can be represented as a quantum-mechanical
Hamiltonian.

As was already mentioned above, the nonlocal light-cone quasipartonic operators
are generating functions of the Wilson operators (\ref{MaxSpinWilson}) of the
lowest twist or, equivalently, the maximal Lorentz spin. For obvious reasons, the
nonlocal light-cone operators corresponding to non-maximal spin components of the
Wilson operators (\ref{WilsonOperators}) and/or containing ``bad'' components of
fields belong to the class of non-quasipartonic operators. Their twist is higher
compared to the twist of quasipartonic operators built from the same number of
constituent fields. The transformation of non-quasipartonic operators under
dilatations takes a more complicated form. One of the reasons for this is that
the operators containing ``bad'' components are not dynamically independent and
can be re-expressed nonlocally in terms of ``good'' fields integrating the former
out in the functional integral in the light-like gauge $A^+ = 0$. This
effectively results into the use of equations of motion, e.g.,
\begin{equation}
\psi_-(\xi)
=
\ft12
( \partial^+ )^{-1} {\not\!\!{\cal D}}_\perp \gamma^+ \psi_+(\xi)
+
\dots
=
\ft{i}2
\int \frac{d\xi'}{2\pi}
\int\frac{d\nu}{\nu}
\,
{\rm e}^{-i\nu(\xi-\xi')}
{\not\!\!{\cal D}}_\perp \gamma^+ \psi_+(\xi')
+
\dots
 \, ,
\end{equation}
where the ellipses stand for contributions from scalars. This relation implies
that the ``bad'' component $\psi_-(\xi)$ can be treated as a (multi-particle)
composite state built from the ``good'' components $\psi_+(\xi')$ and
$A_\perp^\mu(\xi')=\int d\xi''\,  F^{+ \mu}_{\phantom{+} \, \perp}(\xi'+\xi'')$
smeared along the light-cone. This implies that, in distinction with
quasipartonic operators, the total set of nonquasipartonic operators of a given
twist is overcomplete. Construction of the basis of operators in this sector
remains an open problem.

\subsection{Two-particle conformal primaries}

In general, the Wilson operators (\ref{MaxSpinWilson}) do not transform
covariantly under the action of the $SO(4,2)$ conformal group and, as a
consequence, they do not have an autonomous renormalization under dilatations.
To identify the operators which are eigenfunctions of the dilatation operator,
one makes use of the collinear $SL(2)$ subgroup of the conformal group to
define the so-called conformal operators. The conformal operator
$\mathcal{O}_J$ is determined by requiring that it is transformed under the
$SL(2)$ transformations as a primary field of the conformal spin $J$,
Eqs.~(\ref{sl2}) and (\ref{X-trans}). The same condition can be expressed as
\begin{equation}
\label{PrimaryField}
[{\cal L}^- , {\cal O}_J(0)] = 0 \, , \qquad [ \bit{\cal L}^2 , {\cal O}_J(0)] =
J (J - 1) {\cal O}_J(0) \, ,
\end{equation}
where $\bit{\cal L}^2$ is the quadratic Casimir of the $SL(2)$ subgroup.

To construct a two-particle conformal operator, one considers the product of
two fields on the light-cone $X_1(0)X_2(\xi)$. It is transformed in accordance
with the direct product of two $SL(2)$ representations labelled by the
conformal spins of the fields, $j_1$ and $j_2$. Decomposing this product into
a sum of irreducible components, $[j_1] \otimes [j_2] = \sum_{n \ge 0} [n +
j_1 + j_2]$, one identifies the conformal operator $\mathcal{O}_{j}(0)$ with
$j = n + j_1 + j_2 - 2$ as the highest weight of the spin-$(j + 2)$ component.
In this way, one obtains the explicit expression for $\mathcal{O}_j (0)$ in
terms of the Jacobi polynomials
\begin{equation}
\label{TwoParticleConformalOp}
{\cal O}_{j} (0) = X_1 (0) (i \partial_+ )^n P_n^{(2j_1-1 , 2j_2-1)} \left(
\stackrel{\leftrightarrow}{\cal D}{\!}^+ \! / \partial^+ \right) X_2 (0) \, ,
\end{equation}
with $J = j + 2 = n + j_1 + j_2$ being its conformal spin defined by Eq.\
(\ref{PrimaryField}). Here we have used the notations
\begin{eqnarray*}
\partial
\equiv \ \stackrel{\rightarrow}{\partial} + \stackrel{\leftarrow}{\partial} \, ,
\qquad \stackrel{\leftrightarrow}{\cal D} \ \equiv \ \stackrel{\rightarrow}{\cal
D} - \stackrel{\leftarrow}{\cal D} \, .
\end{eqnarray*}
for the total and left-right derivatives, respectively. The spin-$(j + 2)$
representation space is spanned by the conformal operator ${\cal O}_{j} (0)$
and its descendants, generated with the step-up operator ${\cal L}^+$,
\begin{equation}
{\cal O}_{jl}(0)
=
i^{l-n}
[ {\cal L}^+ , \ldots , [{\cal L}^+ , [{\cal L}^+ , \mathcal{O}_j(0)]] \dots ]
=
( i \partial_+ )^{l - n} {\cal O}_{j}(0)
\, , \qquad
(l \ge n)
\, ,
\end{equation}
from the vacuum state ${\cal O}_{jj}(0) \equiv {\cal O}_{j}(0)$. Then, the product
$X_1(0) X_2(\xi)$ can be expanded over the conformal operators as follows (we recall
that we are using the $A^+ = 0$ gauge):
\begin{equation}
\label{NonlocalExpansionConformal}
X_1 (0) X_2 (\xi)
=
\sum_{n = 0}^\infty C_n (j_1, j_2)
(- i \xi)^n \int_0^1 d u \, u^{n + 2 j_1 - 1} (1 - u)^{n + 2 j_2 - 1}
{\cal O}_{j = n + j_1 + j_2 - 2} (u \xi)
\, ,
\end{equation}
where we introduced the ``reduced'' Clebsch-Gordan coefficients
$$
C_n (j_1, j_2)
=
\frac{
(2 n +  2 j_1 + 2 j_2 - 1)
{\mit\Gamma} (n + 2 j_1 + 2 j_2 - 1)
}{
{\mit\Gamma} (n + 2 j_1)
{\mit\Gamma} (n + 2 j_2)
}
\, .
$$
As we can see, the definition of conformal operators does not rely on supersymmetry
and makes use of the invariance of the underlying Yang-Mills theory under the
conformal $SO(4,2)$ transformations. In addition, the conformal symmetry protects
mixing between operators of different conformal spin. In QCD this property is
valid to one-loop order only due to the conformal anomaly, while in ${\cal N} =
4$ super-Yang-Mills the conformal symmetry holds to all orders provided that the
theory is regularized in the manner that preserves the symmetry of the classical
Lagrangian. We will return to this issue later in section \ref{ADsupermultiplet}
where we discuss subtleties of the dimension regularization of loop corrections.

In general, there exists a set of conformal operators possessing the same conformal
spin $j$ and other quantum numbers with respect to the Lorentz and internal flavor
groups. Combining them together, one can define a vector $\bit{{\cal O}}_{jl}$
which obeys a matrix renormalization group equation%
\footnote{The superscript $R$ stands for a subtracted operator $\bit{{\cal O}}^R
= \bit{\cal Z} \bit{{\cal O}}$. The latter generates finite Green functions with
elementary field operators.}
\begin{equation}
\frac{d}{d \ln\mu} \bit{{\cal O}}^R_{jl} = - \bit{\gamma}_j (g^2) \,
\bit{{\cal O}}^R_{jl} \, ,
\end{equation}
where the anomalous dimension matrix $\bit{\gamma}_j (g^2)$ has an infinite
series expansion in perturbation theory;
\begin{equation}
\label{ADinPT}
\bit{\gamma}_j (g^2) = \sum_{n = 0}^\infty \left( \frac{g^2}{8 \pi^2} \right)^{n
+ 1} \bit{\gamma}^{(n)}_j \, .
\end{equation}
Conformal symmetry implies that $\bit{\gamma}_j (g^2)$ depends only on the
conformal spin $j$, but it does not fix its functional form. Additional
constraints on the mixing matrix are imposed by supersymmetry. As we will show
below, it allows one to determine the eigenstates of the matrix $\bit{\gamma}_j
(g^2)$ in ${\cal N} = 4$ theory to all orders in the coupling constant without
actual calculations of Feynman diagrams.

The rest of this section is devoted to the counting and construction of all
twist-two operators in ${\cal N} = 4$ super-Yang-Mills by classifying them with
respect to irreducible representations of the Lorentz and flavor group. The
generic expression is given by Eq.\ (\ref{TwoParticleConformalOp}) with $X_i$
being one of the ``good'' fields $X = \{ F^+{}^\perp_\mu , \lambda_{+\alpha}^A ,
\bar\lambda^{\dot\alpha}_{+A}, \phi^{AB} \}$. For the sake of convenience, we
split all bilinears into bosonic and fermionic operators. The first set involves
gaugino-gaugino, gluon-gluon, scalar-scalar and gluon-scalar ones, while the
second contains gaugino-gluon and gaugino-scalar bilinears. Later using the
supersymmetry algebra we construct a supermultiplet which embraces all of them.

\subsection{Two-gaugino operators}

Let us start with the two-gaugino operators. The gaugino fields
$\lambda_{+\alpha}^A$ and $\bar\lambda^{\dot\alpha}_{+A}$ carry the conformal
spin $j_q=1$ and the corresponding conformal operator is given by
(\ref{TwoParticleConformalOp}) with the Jacobi polynomial being reduced to
the Gegenbauer polynomial $C_j^{3/2}$. One can assign a definite spatial parity
to the two-gaugino operators. In the Majorana notations, the difference between
the even and odd parity bilinears is encoded in the chirality matrix $\gamma_5$,
so that $\bar\psi \gamma^+ \psi$ and $\bar\psi \gamma^+ \gamma_5 \psi$ are the
vector and the axial-vector, respectively. When expressed in terms of two-component
Weyl spinors, $\lambda_{+\alpha}^A$ and $\bar\lambda^{\dot\alpha}_{+A}$, the
bilinears will be accompanied by the signature factors
\begin{equation}
\label{SignatureFactor}
\sigma_j \equiv 1 - (- 1)^j \, ,
\end{equation}
which trace back to the vanishing of the corresponding operators with even/odd
spatial parity. Next, since the gaugino transforms in the $\bit{4}$ of
$SU(4)$, the bilinear built from it and its complex conjugate is decomposed
into two irreducible representations $\bit{4} \otimes \bit{\bar 4} = \bit{1}
\oplus \bit{15}$. The latter is extracted with the help of the projector
\begin{equation}
\label{Proj15}
{}[ P_{\bit{\scriptstyle 15}} ]^{BC}_{AD} = \delta^C_A \delta^B_D - \frac{1}{4}
\delta^B_A \delta^C_D \, .
\end{equation}
The maximal-helicity gaugino operators, with fields having aligned helicities,
have two flavor components $\bit{4} \otimes \bit{4} = \bit{6} \oplus \bit{10}$,
projected with
\begin{equation}
[P_{\bit{\scriptstyle 10}}]^{CD}_{AB} = [P_{\bit{\scriptstyle \overline{10 \!} \,
}}]^{CD}_{AB} = \delta_A^C \delta_B^D + \delta_A^D \delta_B^C \, , \qquad
[P_{\bit{\scriptstyle 6}}]^{CD}_{AB} = [P_{\bit{\scriptstyle \bar 6}}]^{CD}_{AB}
= \delta_A^C \delta_B^D - \delta_A^D \delta_B^C \, .
\end{equation}
Therefore, in total we have the following two-fermion conformal operators:
\begin{itemize}
\item the singlet even- and odd-parity operators, respectively,
\begin{eqnarray}
\label{BiLambda}
{\cal O}^{qq}_{jl}
\!\!\!&=&\!\!\!
\sigma_j \,
{\rm tr} \,
\bar\lambda_{\dot\alpha A}
\sigma^{+ \; \dot\alpha\beta}
\big( i \partial^+ \big)^l
C_j^{3/2}
\left(
\stackrel{\leftrightarrow}{\cal D}{\!}^+ \! / \partial^+
\right)
\lambda^A_{\beta}
\, , \nonumber\\
\widetilde{\cal O}^{qq}_{jl}
\!\!\!&=&\!\!\!
\sigma_{j + 1} \,
{\rm tr} \,
\bar\lambda_{\dot\alpha A}
\sigma^{+ \; \dot\alpha\beta}
\big( i \partial^+ \big)^l
C_j^{3/2}
\left( \stackrel{\leftrightarrow}{\cal D}{\!}^+ \! / \partial^+ \right)
\lambda^A_{\beta} \, ,
\end{eqnarray}
\item the even- and odd-parity operators in $\bit{15}$
\begin{eqnarray}
{}[ {\cal O}_{jl}^{qq,\bit{\scriptstyle 15}} ]_A{}^B
\!\!\!&=&\!\!\!
\sigma_{j + 1} \,
{\rm tr} \,
{}[ P_{\bit{\scriptstyle 15}} ]^{BC}_{AD} \,
\bar\lambda_{\dot\alpha C} \sigma^{+ \; \dot\alpha\beta}
\big( i \partial^+ \big)^l
C_j^{3/2}
\left( \stackrel{\leftrightarrow}{\cal D}{\!}^+ \! / \partial^+ \right)
\lambda^D_{\beta}
\, , \nonumber\\
{}[ \widetilde{\cal O}_{jl}^{qq, \bit{\scriptstyle 15}} ]_A{}^B
\!\!\!&=&\!\!\!
\sigma_j \,
{\rm tr} \,
{}[ P_{\bit{\scriptstyle 15}} ]^{BC}_{AD} \,
\bar\lambda_{\dot\alpha C} \sigma^{+ \; \dot\alpha\beta}
\big( i \partial^+ \big)^l C_j^{3/2} \left( \stackrel{\leftrightarrow}{\cal
D}{\!}^+ \! / \partial^+ \right) \lambda^D_{\beta} \, ,
\end{eqnarray}
\item the maximal-helicity operators in antisymmetric $\bit{n} = \bit{6}$,
symmetric $\bit{n} = \bit{10}$ and their complex-conjugated $\bit{\bar n}
= \bit{\bar 6}, \bit{ \overline{10 \!} \,}$ representations
\begin{eqnarray}
{}[ {\cal T}^{qq, \bit{\scriptstyle n}}_{jl} ]^{\mu AB}
\!\!\!&=&\!\!\!
{\rm tr} \, [P_{\bit{\scriptstyle n}}]^{AB}_{CD} \,
\lambda^{\alpha C}
\bar\sigma^+{}_{\alpha\dot\beta}
\sigma^\mu_{\,\perp}{}^{\dot\beta \gamma} \big( i \partial^+ \big)^l C_j^{3/2}
\left( \stackrel{\leftrightarrow}{\cal D}{\!}^+ \! / \partial^+ \right)
\lambda_\gamma^D
\, , \nonumber\\
{}[ \bar {\cal T}^{qq, \bit{\scriptstyle \bar n}}_{jl}
]^\mu_{AB} \!\!\!&=&\!\!\! {\rm tr} \,
{}[P_{\bit{\scriptstyle \bar n}}]^{CD}_{AB} \,
\bar\lambda_{\dot\alpha C}
\sigma^{+ \, \dot\alpha\beta}
\bar\sigma^\mu_{\,\perp \beta \dot\gamma}
\big( i \partial^+ \big)^l C_j^{3/2} \left( \stackrel{\leftrightarrow}{\cal
D}{\!}^+ \! / \partial^+ \right)
\bar\lambda^{\dot\gamma}_D
\, .
\label{q-maxhel}
\end{eqnarray}
\end{itemize}
We also introduce $SU(4)$-conjugated operators
\begin{equation}
[ {\cal T}^{qq, \bit{\scriptstyle \bar 6}}_{jl} ]^\mu_{AB}
=
{}\frac{1}{2}
\varepsilon_{ABCD} {}[ {\cal T}^{qq, \bit{\scriptstyle 6}}_{jl} ]^{\mu CD}
\, , \qquad
{}[ \bar {\cal T}^{qq, \bit{\scriptstyle 6}}_{jl} ]^{\mu AB}
=
{}\frac{1}{2} \varepsilon^{ABCD}
{}[ \bar {\cal T}^{qq, \bit{\scriptstyle \bar 6}}_{jl} ]^\mu_{CD}
\, .
\end{equation}
As we will see in the next section, they naturally arise in supersymmertic
variations of other operators which form the supermultiplet. Notice that
in order to simplify our notations we have introduced a superscript of
the flavor representation only for non-singlet cases. Analogously, below
we will omit this label for the lowest-dimensional representation in a
given set of operators with the same field content or when the $SU(4)$
representation is obvious from the particle content.

In the above expressions, the signature factors $\sigma_j$ ($\sigma_{j+1}$) imply
that the corresponding operators are identically zero for even (odd) conformal
spins. The Fermi statistics of gaugino fields has a consequence that ${}[
{\cal T}^{qq, \bit{\scriptstyle 6}}_{jl} ]^{\mu AB}$ and ${}[ {\cal T}^{qq,
\bit{\scriptstyle 10}}_{jl} ]^{\mu AB}$ also vanish for odd and even $j$'s,
respectively.

\subsection{Two-gluon operators}

Let us address now gluonic operators. The vector field $F^{+ \mu}_{\phantom{+}
\, \perp}$ has the conformal spin $j_g = 3/2$ and carries no charge with respect
to the internal $R$-symmetry group. The two-gluon conformal operators are given
by (\ref{TwoParticleConformalOp}) for $j_1=j_2=3/2$, so that $P_j^{(2,2)}\sim
C_j^{5/2}$. The product of two vectors, $g^{\mu\alpha}_\perp g^{\nu\beta}_\perp
F^{+ }_{\phantom{+} \, \alpha}F^{+ }_{\phantom{+} \, \beta}$, can be decomposed
into irreducible representations of the Lorentz group as $\left( \ft12, \ft12
\right) \otimes \left( \ft12, \ft12 \right) = \left( 0, 0 \right) \oplus \left(
\left( 1, 0 \right) \oplus \left( 0, 1 \right) \right) \oplus \left( 1, 1
\right)$, or, equivalently,
\begin{eqnarray*}
g^{\mu\alpha}_\perp g^{\nu\beta}_\perp = \ft12 g^{\mu\nu}_\perp
g^{\alpha\beta}_\perp + \ft12 \varepsilon^{\mu\nu}_\perp
\varepsilon^{\alpha\beta}_\perp + \tau^{\mu\nu;\rho\sigma}_\perp
\tau^{\alpha\beta;}_\perp{}_{\rho\sigma} \, ,
\end{eqnarray*}
with $\varepsilon^\perp_{\mu\nu} \equiv \varepsilon^{\alpha\beta\rho\sigma}
g^\perp_{\alpha\mu} g^\perp_{\beta\nu} n^\ast_\rho n_\sigma$ normalized as
$\varepsilon^{0123} = 1$ and
$ \tau^{\mu\nu;\rho\sigma}_\perp = \ft12 \Big( g^{\mu\rho}_\perp
g^{\nu\sigma}_\perp + g^{\mu\sigma}_\perp g^{\nu\rho}_\perp - g^{\mu\nu}_\perp
g^{\rho\sigma}_\perp \Big)
$
being a totally symmetric and traceless tensor in each pair of its indices.
Thus, the two-gluon twist-two operators may have only three independent Lorentz
structures.

For further convenience we introduce the gluon fields of positive- and
negative-helicity $F^{+ \mu}_{\phantom{+} \, \perp} + i \widetilde F^{+
\mu}_{\phantom{+} \, \perp}$ and $F^{+ \mu}_{\phantom{+} \, \perp} - i \widetilde
F^{+ \mu}_{\phantom{+} \, \perp}$, respectively. Here the dual gluon field
strength is defined as $\widetilde F^{\mu\nu} \equiv \ft12
\varepsilon^{\mu\nu\rho\sigma} F_{\rho\sigma}$ and $\widetilde F^+{}^\mu_\perp
= \varepsilon^{\mu\nu}_\perp F^+{}^\perp_\nu$. From the positive- and
negative-helicity operators we build the even- and odd-parity combinations
$$
\ft14 g^\perp_{\mu\nu} \left\{ \left( F^{+ \mu}_{\phantom{+} \, \perp} - i
\widetilde F^{+ \mu}_{\phantom{+} \, \perp} \right) \left( F^{+ \nu}_{\phantom{+}
\, \perp} + i \widetilde F^{+ \nu}_{\phantom{+} \, \perp} \right) \pm \left( F^{+
\mu}_{\phantom{+} \, \perp} + i \widetilde F^{+ \mu}_{\phantom{+} \, \perp}
\right) \left( F^{+ \nu}_{\phantom{+} \, \perp} - i \widetilde F^{+
\nu}_{\phantom{+} \, \perp} \right) \right\} = - \left\{ \!
\begin{array}{r}
g^\perp_{\mu\nu} \\
i \varepsilon^\perp_{\mu\nu}
\end{array}
\! \right\} F^{+ \mu}_{\phantom{+} \, \perp} F^{\nu +}_{\, \perp} \, ,
$$
as well as the maximal-helicity operator
\begin{equation}
\label{max-hel}
\ft14 \left\{ \left( F^{+ \mu}_{\phantom{+} \, \perp} + i \widetilde F^{+
\mu}_{\phantom{+} \, \perp} \right) \left( F^{+ \nu}_{\phantom{+} \, \perp} + i
\widetilde F^{+ \nu}_{\phantom{+} \, \perp} \right) \pm \left( F^{+
\mu}_{\phantom{+} \, \perp} - i \widetilde F^{+ \mu}_{\phantom{+} \, \perp}
\right) \left( F^{+ \nu}_{\phantom{+} \, \perp} - i \widetilde F^{+
\nu}_{\phantom{+} \, \perp} \right) \right\} = - \tau^{\mu\nu; \rho\sigma}
\left\{ \!
\begin{array}{r}
F^{+ \perp}_{\phantom{+} \, \rho} F^{\perp +}_{\, \sigma}
\\
i \widetilde F^{+ \perp}_{\phantom{+} \, \rho} F^{\perp +}_{\, \sigma}
\end{array}
\! \right\} \, ,
\end{equation}
where the sign assignment corresponds to the upper and lower component of the
tensor structure on the right-hand side.

According to the above nomenclature, we introduce the following two-gluon
conformal operators:
\begin{itemize}
\item the even-parity operator%
\footnote{Here, compared to Eq.\ (\ref{TwoParticleConformalOp}), we shifted
the argument of the Gegenbauer polynomial to ensure that the total conformal
spin of the operator $J = (j - 1) + 3$ coincides with the conformal spin of
the gaugino operators.}
\begin{equation}
\label{BiGauge}
{\cal O}^{gg}_{jl}
=
\ft12 \sigma_j\,{\rm tr} \, F^{+ \mu}_{\phantom{+} \, \perp} \,
g^\perp_{\mu\nu} \, \big( i \partial^+ \big)^{l - 1} C_{j - 1}^{5/2} \left(
\stackrel{\leftrightarrow}{\cal D}{\!}^+ \! / \partial^+ \right) F_{\,
\perp}^{\nu +} \, ,
\end{equation}
\item the odd-parity operator
\begin{equation}
\widetilde{\cal O}^{gg}_{jl}
=
\ft12 \sigma_{j+1}\, {\rm tr} \, F^{+
\mu}_{\phantom{+} \, \perp} \, i \, \varepsilon^\perp_{\mu\nu} \big( i \partial^+
\big)^{l - 1} C_{j - 1}^{5/2} \left( \stackrel{\leftrightarrow}{\cal D}{\!}^+ \!
/ \partial^+ \right) F_{\, \perp}^{\nu +} \, ,
\end{equation}
\item the maximal-helicity operators
\begin{eqnarray}
{}[ {\cal T}^{gg}_{jl} ]^{\mu\nu}
\!\!\!&=&\!\!\!
{\rm tr} \,
\left( F^{+ \mu}_{\phantom{+} \, \perp} + i \widetilde F^{+ \mu}_{\phantom{+} \, \perp}
\right)
\big( i \partial^+ \big)^{l - 1} C_{j - 1}^{5/2}
\left(
\stackrel{\leftrightarrow}{\cal D}{\!}^+ \! / \partial^+
\right)
\left(
F_{\, \perp}^{\nu +} + i \widetilde F_{\, \perp}^{\nu +}
\right)
\, , \nonumber\\
{}[ \bar{\cal T}^{gg}_{jl} ]^{\mu\nu}
\!\!\!&=&\!\!\!
{\rm tr} \,
\left(
F^{+ \mu}_{\phantom{+} \, \perp} - i \widetilde F^{+ \mu}_{\phantom{+} \, \perp}
\right) \big( i \partial^+ \big)^{l - 1} C_{j - 1}^{5/2}
\left(
\stackrel{\leftrightarrow}{\cal D}{\!}^+ \! / \partial^+
\right)
\left(
F_{\, \perp}^{\nu +} - i \widetilde F_{\, \perp}^{\nu +}
\right) \, .
\label{g-maxhel}
\end{eqnarray}
\end{itemize}
Notice that we have split the maximal-helicity gluon operator (\ref{max-hel}) into
negative and positive total helicity components, ${\cal T}^{gg}$ and $\bar{\cal
T}^{gg}$, respectively. The Bose statistics of the gluon fields implies that the
conformal operators ${\cal O}^{gg}_{jl}$/$\widetilde {\cal O}^{gg}_{jl}$ vanish
for even/odd conformal spins $j$. To make these properties explicit we inserted
the signature factors into their definition.

\subsection{Two-scalar operators}

We recall that the six real scalar fields of the ${\cal N} = 4$ model are
combined into a complex antisymmetric field $\phi^{AB}$ which transforms with
respect to $\bit{6}$ of the $SU(4)$ flavor group and carries the conformal spin
$j_s=1/2$. Thus, the product of two representations for scalars gives $\bit{6}
\otimes \bit{\bar 6} = \bit{1} \oplus \bit{15} \oplus \bit{20}$, so that the
Clebsch-Gordan decomposition reads
\begin{eqnarray*}
\bar\phi_{AB} \phi^{CD}
\!\!\!&=&\!\!\! \ft{1}{12} \left( \delta_A^C \delta_B^D - \delta_A^D \delta_B^C
\right) \bar\phi_{FG} \phi^{FG}
\nonumber\\
&+&\!\!\! \ft12 \left\{ \delta_A^C {}[ P_{\bit{\scriptstyle 15}} ]^{DE}_{BF} -
\delta_A^D {}[ P_{\bit{\scriptstyle 15}} ]^{CE}_{BF} + \delta_B^D {}[
P_{\bit{\scriptstyle 15}} ]^{CE}_{AF} - \delta_B^C {}[ P_{\bit{\scriptstyle 15}}
]^{DE}_{AF} \right\}
\bar\phi_{EG} \phi^{FG}
\nonumber\\
&+&\!\!\! \ft{1}{12} {}[ P_{\bit{\scriptstyle 20}} ]^{CD;EF}_{AB;GH}
\bar\phi_{EF} \phi^{GH}
\, ,
\end{eqnarray*}
where the projector $[ P_{\bit{\scriptstyle 15}} ]^{DE}_{BF}$ was defined in
(\ref{Proj15}) and
\begin{equation}
{}[ P_{\bit{\scriptstyle 20}} ]^{CD;EF}_{AB;GH} \equiv \delta_G^C \delta_H^D
\varepsilon_{ABIJ} \varepsilon^{EFIJ} + \delta_H^D \varepsilon_{ABGJ}
\varepsilon^{CEFJ} + \delta_G^C \varepsilon_{ABHJ} \varepsilon^{DEFJ} +
\varepsilon_{ABGH} \varepsilon^{CDEF} \, .
\end{equation}
Applying (\ref{TwoParticleConformalOp}) for $j_1 = j_2 = 1/2$ and taking into
account that $P_j^{(0,0)} \sim C_j^{1/2}$, we introduce the following two-scalar
conformal operators:
\begin{itemize}
\item the singlet $SU(4)$ representation
\begin{equation}
\label{BiScalar}
{\cal O}_{jl}^{ss}
=
\ft12\sigma_j\, {\rm tr} \,
\bar\phi_{AB}
\big( i \partial^+ \big)^{l + 1} C_{j + 1}^{1/2}
\left(
\stackrel{\leftrightarrow}{\cal D}{\!}^+ \! / \partial^+
\right)
\phi^{AB} \, ,
\end{equation}
\item the non-singlet $\bit{15}$ and $\bit{20}$ $SU(4)$ representations
\begin{eqnarray}
{}[ {\cal O}_{jl}^{ss, \bit{\scriptstyle 15}} ]_A{}^B
\!\!\!&\equiv&\!\!\!
\ft12\sigma_{j+1}\,{\rm tr} \,
{}[ P_{\bit{\scriptstyle 15}} ]^{BC}_{AD} \,
\bar\phi_{CE}
\big( i \partial^+ \big)^{l + 1} C_{j + 1}^{1/2}
\left(
\stackrel{\leftrightarrow}{\cal D}{\!}^+ \! / \partial^+
\right)
\phi^{DE}
\, , \\
\label{SUSY-prim}
{}[ {\cal O}_{jl}^{ss, \bit{\scriptstyle 20}} ]_{AB}^{CD}
\!\!\!&\equiv&\!\!\!
\ft12\sigma_j\,{\rm tr} \,
{}[ P_{\bit{\scriptstyle 20}} ]^{CD;EF}_{AB;GH} \,
\bar\phi_{EF}
\big( i \partial^+ \big)^{l + 1} C_{j + 1}^{1/2}
\left(
\stackrel{\leftrightarrow}{\cal D}{\!}^+ \! / \partial^+
\right)
\phi^{GH} \, .
\end{eqnarray}
\end{itemize}
The Bose statistics implies that the operators ${\cal O}_{jl}^{ss}$ and
$[ {\cal O}_{jl}^{ss, \bit{\scriptstyle 20}} ]_A{}^B$ vanish for even $j$
whereas $[ {\cal O}_{jl}^{ss, \bit{\scriptstyle 15}} ]_A{}^B$ vanishes for
odd $j$.

\subsection{Mixed bosonic operators}

The last in a row of bosonic operators are mixed gluon-scalar operators,
which are
\begin{eqnarray}
{}[ {\cal T}^{sg}_{jl} ]^{\mu AB}
\!\!\!&=&\!\!\!
\frac{i}{\sqrt{2}} \,
{\rm tr} \,
\left(
F^{+ \mu}_{\phantom{+} \, \perp} - i \widetilde F^{+ \mu}_{\phantom{+} \, \perp}
\right)
\big( i \partial^+ \big)^l P_j^{(2, 0)}
\left( \stackrel{\leftrightarrow}{\cal D}{\!}^+ / \partial^+ \right)
\phi^{AB}
\, , \nonumber\\
{}[ \bar{\cal T}^{sg}_{jl} ]^\mu_{AB}
\!\!\!&=&\!\!\!
\frac{i}{\sqrt{2}} \,
{\rm tr} \,
\left(
F^{+ \mu}_{\phantom{+} \, \perp} + i \widetilde F^{+ \mu}_{\phantom{+} \, \perp}
\right) \big( i \partial^+ \big)^l P_j^{(2, 0)}
\left( \stackrel{\leftrightarrow}{\cal D}{\!}^+ / \partial^+ \right)
\bar\phi_{AB}
\, .
\end{eqnarray}
They do not possess a definite parity. The indices of the Jacobi polynomials are
defined by the conformal spins of the operators,
Eq.~(\ref{TwoParticleConformalOp}).

\subsection{Fermionic operators}

Finally, we define operators with fermionic quantum numbers. There are obviously
two types: constructed from the gaugino and the field strength and from the
gaugino and the scalar. In the latter case, the product of gaugino and scalar
fields is decomposed into $\bit{6} \otimes \bit{\bar 4} = \bit{4} \oplus
\bit{\overline{20\!}\,}$. The projection onto irreducible components is
accomplished with the Clebsch-Gordan decomposition
\begin{eqnarray*}
\phi^{AB} \bar\lambda_C = \ft13 \left( \delta_C^B \delta_D^A - \delta_C^A
\delta_D^B \right) \phi^{DE}
\bar\lambda_E
+ \ft16 [ P_{\bit{\scriptstyle \overline{20\!}\,}} ]^{AB;F}_{C;DE} \, \phi^{DE}
\bar\lambda_F \, ,
\end{eqnarray*}
where
\begin{equation}
{}[ P_{\bit{\scriptstyle \overline{20\!}\,}} ]^{AB;F}_{C;DE} = \left( \delta^F_C
\delta^I_G + \delta^F_G \delta^I_C \right) \varepsilon^{ABGH} \varepsilon_{DEIH}
\, .
\end{equation}
Analogously, one decomposes the product $\bar\phi_{AB}\lambda^C$ as $\bit{\bar 6}
\otimes \bit{4} = \bit{\overline{4}} \oplus \bit{{20\!}\,}$ with the projector
\begin{equation}
{}[ P_{\bit{\scriptstyle 20}} ]^{C;DE}_{AB;F} = \left( \delta_F^C \delta_I^G +
\delta_F^G \delta_I^C \right) \varepsilon_{ABGH} \varepsilon^{DEIH} \, .
\end{equation}
To complete the basis we introduce the operators built from the gaugino and the
gluon. Thus we have
\begin{itemize}
\item the gaugino-scalar $\bit{4}$ and $\bit{\bar 4}$
\begin{eqnarray}
{}[ \bar{\mit\Omega}_{jl}^{sq} ]^{\dot\alpha A}
\!\!\!&=&\!\!\! i \sqrt{2} \,
{\rm tr} \,
\phi^{AB} \big( i \partial^+ \big)^{l + 1} P_{j + 1}^{(0, 1)}
\left(
\stackrel{\leftrightarrow}{\cal D}{\!}^+ / \partial^+
\right)
\bar\lambda^{\dot\alpha}_B
\, , \nonumber\\
{}[ {\mit\Omega}^{sq}_{jl} ]_{\alpha A}
\!\!\!&=&\!\!\! i \sqrt{2} \,
{\rm tr} \,
\bar\phi_{AB}
\big( i \partial^+ \big)^{l + 1} P_{j + 1}^{(0, 1)}
\left(
\stackrel{\leftrightarrow}{\cal D}{\!}^+ / \partial^+
\right)
\lambda_\alpha^B \,
,
\end{eqnarray}
\item the gaugino-scalar $\bit{\overline{20\!}\,}$ and $\bit{20}$
\begin{eqnarray}
{}[
\bar{\mit\Omega}_{jl}^{sq, \bit{\scriptstyle \overline{20\!}\,}}
{}]^{\dot\alpha}{}^{AB}_C
\!\!\!&\equiv&\!\!\!
i \sqrt{2} \, {\rm tr} \,
{}[
P_{\bit{\scriptstyle \overline{20\!}\,}} ]^{AB;F}_{C;DE} \,
\phi^{DE}
\big( i \partial^+ \big)^{l + 1} P_{j + 1}^{(0, 1)}
\left(
\stackrel{\leftrightarrow}{\cal D}{\!}^+ / \partial^+
\right)
\bar\lambda^{\dot\alpha}_F
\, , \nonumber\\
{}[ {\mit\Omega}_{jl}^{sq, \bit{\scriptstyle 20}} ]_\alpha{}_{AB}^C
\!\!\!&\equiv&\!\!\!
i \sqrt{2} \, {\rm tr} \,
{}[ P_{\bit{\scriptstyle 20}} ]_{AB;F}^{C;DE} \,
\bar\phi_{DE}
\big( i \partial^+ \big)^{l + 1} P_{j + 1}^{(0, 1)}
\left(
\stackrel{\leftrightarrow}{\cal D}{\!}^+ / \partial^+
\right)
\lambda_\alpha^F
\, ,
\end{eqnarray}
\item the gluon-gaugino operators
\begin{eqnarray}
\label{GluonGaugeOper}
{}[{^\pm\!\bar{\mit\Omega}}^{gq}_{jl}]^{\mu A}_\alpha
\!\!\!&=&\!\!\!
\frac{1}{2} \,
{\rm tr} \,
\left(
F^{+ \mu}_{\phantom{+} \, \perp} \pm i \widetilde F^{+ \mu}_{\phantom{+} \, \perp}
\right)
\big( i \partial^+ \big)^l P_j^{(2, 1)}
\left( \stackrel{\leftrightarrow}{\cal D}{\!}^+ / \partial^+ \right)
\lambda_\alpha^A
\, , \nonumber\\
{}[{^\pm\!{\mit\Omega}}^{gq}_{jl}]^{\dot\alpha \mu}_A
\!\!\!&=&\!\!\!
\frac{1}{2} \,
{\rm tr} \,
\left(
F^{+ \mu}_{\phantom{+} \, \perp} \pm i \widetilde F^{+ \mu}_{\phantom{+} \, \perp}
\right)
\big( i \partial^+ \big)^l P_j^{(2, 1)}
\left(
\stackrel{\leftrightarrow}{\cal D}{\!}^+ / \partial^+
\right)
\bar\lambda^{\dot\alpha}_A
\, .
\end{eqnarray}
\end{itemize}
It is convenient to introduce into our consideration a projection of the
latter four operators in Eq.\ (\ref{GluonGaugeOper}) onto Pauli matrices.
Namely, one defines
\begin{equation}
{}[ {\mit\Omega}^{gq}_{jl} ]_{\alpha A}
\equiv
\bar\sigma^\perp_{\,\mu}{}_{\alpha \dot\beta}
{}[{^+\!{\mit\Omega}}^{gq}_{jl}]^{\dot\beta \mu}_A
\, , \qquad
{}[ \bar{\mit\Omega}_{jl}^{gq} ]^{\dot\alpha A}
\equiv
\sigma^\perp_{\,\mu}{}^{\dot\alpha \beta}
{}[{^-\!\bar{\mit\Omega}}^{gq}_{jl}]^{\mu A}_\beta \, .
\end{equation}
The remaining two projections of the maximal-helicity fermionic operators
vanish identically;
\begin{equation}
\bar\sigma^\perp_{\,\mu}{}_{\alpha \dot\beta}
{}[{^-\!{\mit\Omega}}^{gq}_{jl}]^{\dot\beta \mu}_A = 0
\, , \qquad
\sigma^\perp_{\,\mu}{}^{\dot\alpha \beta}
{}[{^+\!\bar{\mit\Omega}}^{gq}_{jl}]^{\mu A}_\beta = 0 \, .
\end{equation}

To summarize, in this section we used the collinear $SL(2)$ subgroup of the full
$SO(4,2)$ conformal group together with the internal $SU(4)$ symmetry to classify
all possible twist-two operators with respect to irreducible representations of
the two. The operators with the same $SU(4)$ charge and conformal spin mix under
renormalization and our goal is to diagonalize the corresponding mixing matrix
making use of supersymmetry.

\section{Building the supermultiplet}
\label{SuperMultiplet}

In the previous section, we have exploited the conformal invariance to
construct the complete set of twist-two operators with a definite conformal
spin $j$. Going from QCD to its supersymmetric extensions, one can
derive constraints on the properties of these operators. The reason for
this is that supersymmetry transformations relate elementary primary fields
of different conformal spins and, therefore, they lead to relations between
various conformal operators. As was already explained, multiparticle
operators of minimal twist are constructed solely from the ``good'' field
components. Therefore, out of the full supersymmetry algebra of the
${\cal N} = 4$ super-Yang-Mills, described in appendix \ref{N=4Lagrangian},
the analysis of such operators requires only a subalgebra for components
projected on the light cone. This implies that for multiparticle operators
of the lowest twist, the total superconformal group $SU(2,2|4)$ is reduced
down to its subgroup $SU(1,1|4) \sim SL(2|4)$, which can be regarded as
a supersymmetric extension of the collinear $SL(2)$-subgroup of the
four-dimensional conformal group.

In the $\mathcal{N}=4$ model, the supersymmetry transformations
(\ref{N=4SUSYrules}) mix ``good'' and ``bad'' components of the
primary fields. To project the transformation of the ``good" components
only, one has to impose the following constraint on the fermionic
transformation parameter
\begin{equation}
\label{AuxCond}
\xi_{+ \alpha}
\equiv
\ft12 \bar\sigma^-{}_{\alpha\dot\beta} \, \sigma^{+ \; \dot\beta\gamma}
\xi_\gamma
\equiv
0
\, , \qquad
\bar\xi_+^{\dot\alpha}
=
\ft12 \sigma^{- \; \dot\alpha\beta} \, \bar\sigma^+{}_{\beta\dot\gamma}
\bar\xi^{\dot\gamma}
=
0
\, ,
\end{equation}
which corresponds to $\ft12 \gamma^- \gamma^+ \xi = 0$ in four-component
notations so that $\xi = \xi_-$. From Eqs.\ (\ref{N=4SUSYrules}) we find
for the ``good" components of fields, in the $SU(4)$ covariant form, the
following rules
\begin{eqnarray}
\label{SUSYgood}
&&\delta A^\mu_\perp
=
- i \xi^{\alpha \; A} \bar\sigma^\mu_\perp{}_{\alpha\dot\beta}
\bar\lambda^{\dot\beta}_{+A}
- i \bar\xi_{\dot\alpha \; A} \sigma_\perp^{\mu\; \dot\alpha\beta}
\lambda_{+\beta}^{A}
\, , \nonumber\\
&&\delta \phi^{AB}
=
- i \sqrt{2}
\left\{
\xi^{\alpha \; A} \lambda_{+ \alpha}^B
-
\xi^{\alpha \; B} \lambda_{+ \alpha}^A
-
\varepsilon^{ABCD} \bar\xi_{\dot\alpha \; C} \bar\lambda^{\dot\alpha}_{+ D}
\right\}
\, , \nonumber\\
&&\delta \lambda^A_{+ \alpha}
=
- F^{+\perp}_{\phantom{+}\, \mu}
\bar\sigma^-{}_{\alpha\dot\beta} \, \sigma_\perp^{\mu \; \dot\beta\gamma}
\xi_\gamma^A - \sqrt{2} \left( {\partial}^+ \phi^{AB} \right)
\bar\sigma^-{}_{\alpha\dot\beta}
\bar\xi^{\dot\beta}_B
\, , \nonumber\\
&&\delta \bar\lambda_{+ A}^{\dot\alpha}
=
- F^{+\perp}_{\phantom{+}\, \mu}
\sigma^{- \; \dot\alpha\beta} \, \bar\sigma^{\mu}_{\perp \; \beta\dot\gamma}
\bar\xi^{\dot\gamma}_A
+ \sqrt{2} \left( {\partial}^+ \bar\phi_{AB} \right) \sigma^{- \;
\dot\alpha\beta} \xi_{\beta}^B
\, .
\end{eqnarray}
Note that $\delta A^+ = 0$ due to Eq.\ (\ref{AuxCond}) so that $\delta
F^{+\perp}_{\phantom{+}\, \mu} = {\cal D}^+ \delta A^\perp_\mu$. The following
comments are in order.

In the light-cone gauge $A^+ = 0$, the transformations (\ref{SUSYgood})
form the off-shell supersymmetry algebra, i.e., without gauge transformations,
needed in a generic case (\ref{N=4SUSYrules}) to bring the multiplet back
to the Wess-Zumino gauge, and without the equations of motion. Namely,
\begin{equation}
[\delta_2, \delta_1] \, X = 2 i \left\{ \xi_2^{\alpha A}
\bar\sigma^-{}_{\alpha \dot\beta}
\bar\xi^{\dot\beta}_{1A}
+
\bar\xi_{2 \dot\alpha A}
\sigma^{- \, \dot\alpha \beta}
\xi^A_{1 \beta}
\right\}
\partial^+ X
\, ,
\end{equation}
for $X = \{ A^\perp_\mu , \lambda_{+\alpha}^A ,
\bar\lambda^{\dot\alpha}_{+A} , \phi^{AB} \}$. In case the light-cone
gauge is lifted, one has to replace $\partial^+ A^\mu_\perp \to
F^+{}^\mu_\perp$ and $\partial^+ \lambda_{+\alpha}^A \to {\cal D}^+
\lambda^A_{+\alpha}$ with identical substitution rules for
$\bar\lambda^{\dot\alpha}_{+A}$ and $\phi^{AB}$.

Since the light-cone supersymmetry transformations are linear --- they do not
increase the number of fields --- the set of quasipartonic operators is closed.
Therefore, examining the transformation properties of such operators under
the transformations (\ref{SUSYgood}) one can
construct supermultiplets belonging to an irreducible representation of the
light-cone super-algebra. A unique property of conformal operators entering the
supermultiplets is that they diagonalize the dilatation operator and, therefore,
have an autonomous scale dependence.

The procedure of constructing supermultiplets of conformal operators is
straightforward \cite{BukFroKurLip84,BukFroKurLip85,BelMulSch98,BelMul00} once an
appropriate component of the supermultiplet is chosen. The right choice would be
any operator which renormalizes autonomously. Applying the light-cone
supersymmetric transformations (\ref{SUSYgood}) to such an operator, one can
reconstruct the remaining entries of the supermultiplet and, therefore, deduce
automatically the combinations of conformal operators which are the
eigenfunctions of the anomalous dimension matrix (see Ref.\ \cite{Bei02} for
the BNM sector).

There are many candidates for this component. This can be judged on the grounds
of unique quantum numbers either with respect to the Lorentz or isotopic groups.
For definiteness we choose the conformal operator ${\cal O}^{ss, \bit{\scriptstyle
20}}_{jl}$ defined in (\ref{SUSY-prim}), but one could equally have taken instead
any of the operators given in Eq.\ (\ref{AutoRenBoson}) or (\ref{AutoRenFermion})
below. We emphasize that the construction of the supermultiplet works to all
orders in the coupling constant provided the superconformal symmetry is not
broken on the quantum level. This is expected to be the case in ${\cal N} = 4$
super-Yang-Mills theory, however, potential complications with the explicit
implementation of regularization procedures in perturbative computations will be
postponed until section \ref{ADsupermultiplet}.

\subsection{Components of the supermultiplet}
\label{ComponentsSupermultiplet}

\begin{figure}[t]
\begin{center}
\mbox{
\begin{picture}(0,145)(240,0)
\put(0,0){\insertfig{16.5}{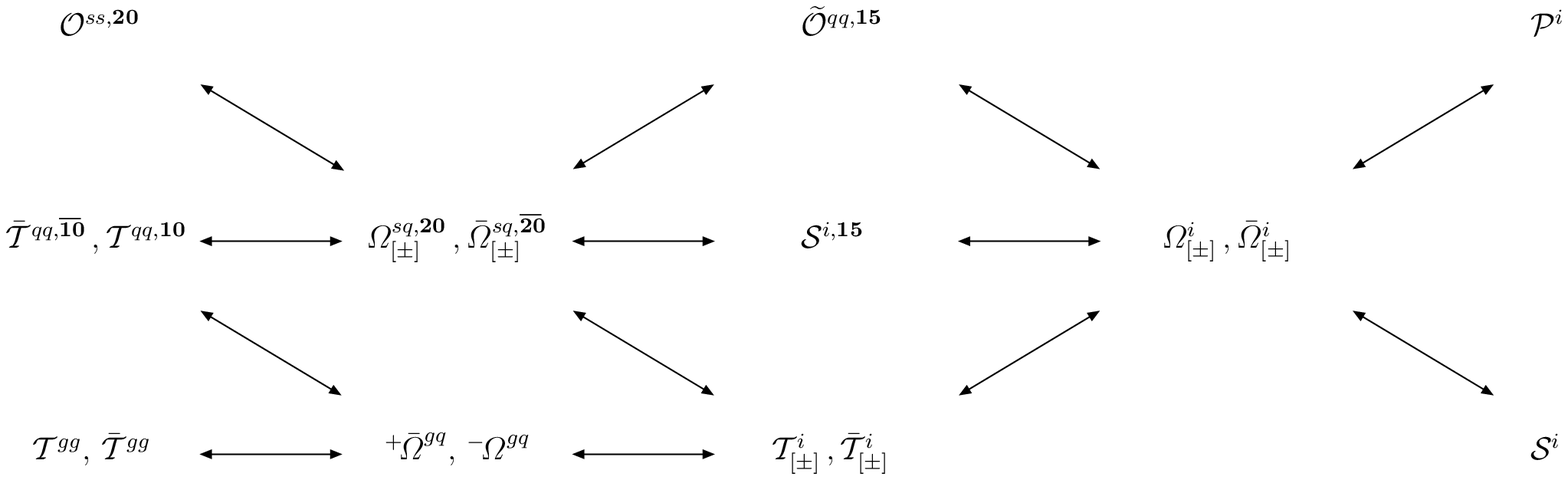}}
\end{picture}
}
\end{center}
\caption{\label{SUSY} A diagram representing the supermultiplet of
twist-two conformal operators in ${\cal N} = 4$ super-Yang-Mills theory.
The Lorentz, flavor and conformal-spin indices are omitted for brevity.
For explicit transformation rules, see appendix \ref{SUSYofComponents}.}
\end{figure}

The ${\cal N} = 4$ super-light-cone algebra (\ref{SUSYgood}) is closed in the
basis spanned by the bilinears introduced in section \ref{CountingOperators}.
As we just said, choosing ${\cal O}^{ss, \bit{\scriptstyle 20}}_{jl}$ as a
supersymmetric primary, the remaining components of the multiplet are deduced
as its descendants. This is demonstrated by an explicit calculation in appendix
\ref{SUSYtrans}. The result of our analysis is represented by the diagram in
Fig.\ \ref{SUSY}, where the arrows indicate the super-variation (\ref{SUSYgood})
of the corresponding conformal operator, and the operators within one step of
the diagram arise as its supersymmetric descendants.

All components of the supermultiplet can be separated into bosonic and fermionic
operators. Within each set, the operators are classified according to their $SU(4)$
charge. In addition, one has to consider separately the cases when the conformal
spin $j$ is even/odd: therefore, in what follows the $[\pm]$ subscript denotes
even ($+$) or odd ($-$) $j$'s.
\begin{itemize}
\item The bosonic components of the supermultiplet are the following:
\begin{eqnarray}
\label{Singlets}
&&{\cal S}^1_{jl}
=
6 {\cal O}_{jl}^{gg}
+
\frac{j}{4} {\cal O}_{jl}^{qq}
+
\frac{j (j + 1)}{4} {\cal O}_{jl}^{ss} \, ,
\nonumber\\
&&{\cal S}^2_{jl}
=
6 {\cal O}_{jl}^{gg}
-
\frac{1}{4} {\cal O}_{jl}^{qq}
-
\frac{(j + 1)(j + 2)}{12} {\cal O}_{jl}^{ss}
\, ,
\nonumber\\
&&{\cal S}^3_{jl}
=
6 {\cal O}_{jl}^{gg}
-
\frac{j + 3}{2} {\cal O}_{jl}^{qq}
+
\frac{(j + 2)(j + 3)}{4} {\cal O}_{jl}^{ss}
\, ,
\nonumber\\
&&{\cal P}^1_{jl}
=
6 \widetilde {\cal O}_{jl}^{gg}
+
\frac{j}{4} \widetilde {\cal O}_{jl}^{qq}
\, ,
\nonumber\\
&&{\cal P}^2_{jl}
=
6 \widetilde {\cal O}_{jl}^{gg}
-
\frac{j + 3}{4} \widetilde {\cal O}_{jl}^{qq}
\, ,
\end{eqnarray}
for the $SU(4)$ singlets \footnote{Notice that scalars do not contribute to the
odd operators ${\cal P}_{jl}^i$ since scalar operators have the opposite parity.},
and
\begin{eqnarray}
&&[ \bar{\cal T}^1_{[+] jl} ]^\mu_{AB}
=
2 (j + 1)
{}[ \bar{\cal T}^{sg}_{jl} ]^\mu_{AB}
-
{}[ {\cal T}^{qq, \bit{\scriptstyle \bar 6}}_{jl} ]^\mu_{AB}
\, ,
\nonumber\\
&&[ \bar{\cal T}^2_{[+] jl} ]^\mu_{AB} = 2 (j + 2) {}[ \bar{\cal
T}^{sg}_{jl} ]^\mu_{AB} + {}[ {\cal T}^{qq,\bit{\scriptstyle \bar 6}}_{jl}
]^\mu_{AB} \, ,
\nonumber\\
&&[ \bar{\cal T}^3_{[-] jl} ]^\mu_{AB}
=
{}[ \bar{\cal T}^{sg}_{jl} ]^\mu_{AB}
\, , \nonumber\\
&&[ {\cal T}^1_{[+] jl} ]^{\mu AB}
=
2 (j + 1)
{}[ {\cal T}^{sg}_{jl} ]^{\mu AB}
+
{}[ \bar {\cal T}^{qq, \bit{\scriptstyle 6}}_{jl} ]^{\mu AB}
\, ,
\nonumber\\
&&[ {\cal T}^2_{[+] jl} ]^{\mu AB}
=
2 (j + 2)
{}[ {\cal T}^{sg}_{jl} ]^{\mu AB}
-
{}[ \bar {\cal T}^{qq, \bit{\scriptstyle 6}}_{jl} ]^{\mu AB}
\, , \nonumber\\
&&[ {\cal T}^3_{[-] jl} ]^{\mu AB}
=
{}[ {\cal T}^{sg}_{jl} ]^{\mu AB}
\, ,
\label{conf-ex}
\end{eqnarray}
for the $\bit{6}$ and $\bit{\bar 6}$, and
\begin{eqnarray}
&&[{\cal S}^{1, \bit{\scriptstyle 15}}_{jl}]_A{}^B
=
(j + 2)
{}[ {\cal O}_{jl}^{ss, \bit{\scriptstyle 15}} ]_A{}^B
-
{}[ {\cal O}_{jl}^{qq, \bit{\scriptstyle 15}} ]_A{}^B
\, ,
\nonumber\\
&&[{\cal S}^{2, \bit{\scriptstyle 15}}_{jl}]_A{}^B
=
(j + 1)
{}[ {\cal O}_{jl}^{ss, \bit{\scriptstyle 15}} ]_A{}^B
+
{}[ {\cal O}_{jl}^{qq, \bit{\scriptstyle 15}} ]_A{}^B
\, ,
\end{eqnarray}
for the $\bit{15}$. Finally, the remaining bosonic operators, which have unique
quantum numbers and, as a consequence, renormalize auto\-no\-mous\-ly, are
\begin{equation}
\label{AutoRenBoson}
{}[ {\cal T}^{gg}_{jl} ]^{\mu\nu}
\, , \qquad
{}[ \bar {\cal T}^{gg}_{jl} ]^{\mu\nu}
\, , \qquad
{}[ {\cal T}^{qq, \bit{\scriptstyle 10}}_{jl} ]^{\mu AB}
\, , \qquad
{}[ \bar {\cal T}^{qq, \bit{\scriptstyle \overline{10 \!} \,}}_{jl} ]^\mu_{AB}
\, , \qquad
{}[ \widetilde {\cal O}_{jl}^{qq, \bit{\scriptstyle 15}} ]_A{}^B
\, , \qquad
{}[ {\cal O}_{jl}^{ss, \bit{\scriptstyle 20}} ]_{AB}^{CD}
\, .
\end{equation}

\item The fermionic components of the supermultiplet read as:
\begin{eqnarray}
&&[ {\mit\Omega}^1_{[+] jl} ]_{\alpha A}
=
3 [ {\mit\Omega}^{gq}_{jl} ]_{\alpha A}
+
{}[ {\mit\Omega}^{sq}_{jl} ]_{\alpha A}
\, , \nonumber\\
&&[ {\mit\Omega}^2_{[+] jl} ]_{\alpha A}
=
(j + 3) [ {\mit\Omega}^{gq}_{jl} ]_{\alpha A}
-
(j + 1) [ {\mit\Omega}^{sq}_{jl} ]_{\alpha A}
\, , \nonumber\\
&&[ {\mit\Omega}^1_{[-] jl} ]_{\alpha A}
=
3 (j + 3) [ {\mit\Omega}^{gq}_{jl} ]_{\alpha A}
+
(j + 1) [ {\mit\Omega}^{sq}_{jl} ]_{\alpha A}
\, , \nonumber\\
&&[ {\mit\Omega}^2_{[-] jl} ]_{\alpha A}
=
{}[ {\mit\Omega}^{gq}_{jl} ]_{\alpha A}
-
{}[ {\mit\Omega}^{sq}_{jl} ]_{\alpha A}
\, ,
\end{eqnarray}
and
\begin{eqnarray}
&&[ \bar{\mit\Omega}^1_{[+] jl} ]^{\dot\alpha A}
=
3 [ \bar{\mit\Omega}^{gq}_{jl} ]^{\dot\alpha A}
-
{}[ \bar{\mit\Omega}^{sq}_{jl} ]^{\dot\alpha A}
\, , \nonumber\\
&&[ \bar{\mit\Omega}^2_{[+] jl} ]^{\dot\alpha A}
=
(j + 3) [ \bar{\mit\Omega}^{gq}_{jl} ]^{\dot\alpha A}
+
(j + 1) [ \bar{\mit\Omega}^{sq}_{jl} ]^{\dot\alpha A}
\, , \nonumber\\
&&[ \bar{\mit\Omega}^1_{[-] jl} ]^{\dot\alpha A}
=
3 (j + 3) [ \bar{\mit\Omega}^{gq}_{jl} ]^{\dot\alpha A}
-
(j + 1) [ \bar{\mit\Omega}^{sq}_{jl} ]^{\dot\alpha A}
\, , \nonumber\\
&&[ \bar{\mit\Omega}^2_{[-] jl} ]^{\dot\alpha A}
=
{}[ \bar{\mit\Omega}^{gq}_{jl} ]^{\dot\alpha A}
+
{}[ \bar{\mit\Omega}^{sq}_{jl} ]^{\dot\alpha A}
\, ,
\end{eqnarray}
together with the remaining operators, which renormalize independently;
\begin{equation}
\label{AutoRenFermion}
{}[{^+\!\bar{\mit\Omega}}^{gq}_{jl}]^{\mu A}_\alpha
\, , \qquad
{}[{^-\!{\mit\Omega}}^{gq}_{jl}]^{\dot\alpha \mu}_A
\, , \qquad
{}[
\bar{\mit\Omega}_{jl}^{sq, \bit{\scriptstyle \overline{20\!}\,}}
{}]^{\dot\alpha}{}^{AB}_C
\, , \qquad
{}[ {\mit\Omega}_{jl}^{sq, \bit{\scriptstyle 20}} ]_\alpha{}_{AB}^C
\, .
\end{equation}
\end{itemize}

The following comments are in order: The result (\ref{Singlets}) for the
operators ${\cal S}^k_{jj}$ and ${\cal P}^k_{jj}$ coincides, modulo
multiplicative factors in front of ${\cal O}^{ss}_{jl}$, with the ones
of Ref.\ \cite{KotLip02}. To perform the comparison, one has to take the
``forward limit'' of the Wilson operators. This amounts to neglecting
contributions involving total derivatives and leads to
\begin{eqnarray*}
n_j^{-1} {\cal O}^{qq}_{jj}
\!\!\!&\to&\!\!\!
\sigma_j \, {\rm tr} \,
\bar\lambda_{\dot\alpha A}
\sigma^{+ \, \dot\alpha\beta} (i {\cal D}^+)^j \lambda^A_\beta
\, , \nonumber\\
n_j^{-1} {\cal O}^{gg}_{jj}
\!\!\!&\to&\!\!\!
\frac{j}{6} \, {\rm tr} \,
F_{\phantom{+} \mu}^{+ \perp} g^{\mu\nu}_\perp (i {\cal D}^+)^{j - 1} F_{\,
\nu}^{\perp +}
\, , \nonumber\\
n_j^{-1} {\cal O}^{ss}_{jj}
\!\!\!&\to&\!\!\!
\frac{2}{j + 1} \, {\rm tr} \,
\bar\phi_{AB} (i {\cal D}^+)^{j + 1} \phi^{AB}
\, ,
\end{eqnarray*}
where $n_j \equiv {\mit\Gamma} (2j + 2)/{\mit\Gamma}^2 (j + 1)$. Analogous
relations hold for the parity-odd sector.

The operators introduced in this subsection carry the same charges $j$ and
$l$ which define their transformation properties with respect to the collinear
$SL(2)$ subgroup. Note that for given $j$ and $l$ the bosonic and fermionic
components of the supermultiplet have different conformal spins: $J = j + 2$
and $J = j + 5/2$, respectively. The same holds true for their canonical
dimensions. For instance, the canonical dimension of the operators
${\cal O}_{jl}^{ss, \bit{\scriptstyle 20}}$ and
${\mit\Omega}^{sq, \bit{\scriptstyle 20}}_{[-] jl}$ is $j + 3$ and $j + 7/2$,
respectively.

\subsection{Moving along the multiplet}

The relations between various twist-two conformal operators entering the
supermultiplet in ${\cal N} = 4$ theory can be deduced from Fig.\ \ref{SUSY}.
As an example of a specific path which allows one to move along the multiplet
from, say, the left upper corner of the graph to its right bottom corner, is
achieved in four steps, in accordance with the number of the light-cone
supersymmetries,
\begin{equation}
{\cal O}_{jl}^{ss, \bit{\scriptstyle 20}}
\longrightarrow
{\mit\Omega}^{sq, \bit{\scriptstyle 20}}_{[-] jl}
\longrightarrow
\bar{\cal T}^2_{[+] jl}
\longrightarrow
{\mit\Omega}^2_{[+] jl}
\longrightarrow
{\cal S}^1_{jl} \, .
\end{equation}
The corresponding sequence of supersymmetric transformations looks like
\begin{eqnarray}
&&\delta [ {\cal O}_{jl}^{ss, \bit{\scriptstyle 20}}]_{AB}^{CD}
=
- \frac{1}{3} \frac{j + 2}{2j + 3}
{}[P_{\bit{\scriptstyle 20}}]_{AB;GH}^{CD;EF}
\,
\xi^{\alpha [G}
{}[
{\mit\Omega}_{[-] jl}^{sq, \bit{\scriptstyle 20}}
{}]_\alpha{}_{EF}^{H]}
+
\dots
\, , \nonumber\\
&&\delta [{\mit\Omega}^{sq, \bit{\scriptstyle 20}}_{[-] jl}]_\alpha{}^C_{AB}
=
\frac{1}{4} \frac{1}{j + 2}
[P_{\bit{\scriptstyle 20}}]_{AB;F}^{C;DE}
\,
[ \bar{\cal T}^2_{[+] j + 1l + 1} ]^\mu_{DE}
\bar\sigma^\perp_{\mu \, \alpha\dot\beta}
\sigma^{- \dot\beta\gamma}
\xi^F_\gamma
+
\dots
\, , \nonumber\\
&&\delta [ \bar{\cal T}^2_{[+] jl} ]^\mu_{AB}
=
- 2 \frac{j + 2}{2 j + 3}
\bar\xi_{\dot\alpha [A} \sigma^{\mu \, \dot\alpha\beta}_{\, \perp}
\, {}[ {\mit\Omega}^2_{[+] jl} ]_{\beta B]}
+
\dots
\, , \nonumber\\
&&\delta [ {\mit\Omega}^2_{[+] jl}]_{\alpha A}
=
- \frac{1}{j + 2}
{\cal S}^1_{j + 1l + 1}
\bar\sigma^-{}_{\alpha\dot\beta} \bar\xi^{\dot\beta}_A
+ \dots \, ,
\label{seq}
\end{eqnarray}
where the ellipses denote contributions of other conformal operators; see
appendix \ref{SUSYofComponents} for details. Notice that the parameter of the
supersymmetric transformations has the scaling dimension $(-1/2)$ and, therefore,
the operators entering both sides of these relations have scaling dimensions
which differ by the same amount. The generators of supersymmetric transformations
do not commute with the Casimir of the collinear subgroup $\bit{\cal L}^2$,
Eq.\ (\ref{PrimaryField}), and, as a consequence, the conformal spin of the
operators entering the right-hand side of (\ref{seq}) differ by $\pm 1/2$.

Examining the diagram shown in Fig.\ \ref{SUSY}, one deduces the following
remarkable property of the ${\cal N} = 4$ super-Yang-Mills theory. The fact
that the diagram is simply connected implies that {\sl all} two-particle
(quasipartonic) conformal operators are unified into a {\sl single}
supermultiplet. This should be compared with the ${\cal N} = 1$ super-Yang-Mills
theory \cite{BukFroKurLip85,BelMul99,BukFroKurLip84,BelMulSch98,BelMul00}. In
that case, a similar diagram has two disconnected components and, as a consequence,
the two-particle conformal operators form two different supermultiplets. One of
these supermultiplets comprises the operators with aligned helicities of
particles \cite{BukFroKurLip85,BelMul99} and, therefore, it inherits integrability
properties discovered in QCD \cite{BraDerKorMan99,Bel99a,Bel99b}. The last two
equations in (\ref{seq}) explicitly demonstrate a distinguished feature of the
${\cal N} = 4$ super-Yang-Mills. As was schematically illustrated in Eq.\
(\ref{AlignedNonaligned}) of the introduction, the ${\cal N} = 4$ supersymmetry
relates the operators of aligned helicities, ${}[\bar{\cal T}^2_{[+] jl}
]^\mu_{AB} \sim {}[ {\cal T}^{qq, \bit{ \scriptstyle \bar 6}}_{jl} ]^\mu_{AB}$,
to the ones built from fields of opposite helicities, ${\cal S}^1_{jl}\sim {\cal
O}^{qq}_{jl}$. Since the operators entering the supermultiplet have the same
properties with respect to the dilatation transformations, we conclude that
going from  QCD to ${\cal N} = 1$ and, finally, to ${\cal N} = 4$ theory,
the integrability in the subsector of maximal-helicity operators gets extended
to the entire sector of quasipartonic operators.

\section{Anomalous dimension of the supermultiplet}
\label{ADsupermultiplet}

If the superconformal symmetry is not broken by radiative corrections, then the
components of the supermultiplet have an autonomous scale dependence and their
anomalous dimensions are defined by a single function of the conformal spin $j$.
To calculate this function, one has to regularize the theory in such a way that
its symmetries are preserved. Unfortunately, a practical implementation of such
perturbative regularization procedure does not exist. The most prominent
candidate --- the dimensional reduction \cite{Sie78} --- preserves supersymmetry
to rather high orders of perturbation theory, but breaks conformal boosts starting
from two loops \cite{BelMul98,BelMulSch98,BelMul00}. The reason for this is that
the coupling constant acquires a nontrivial scaling dimension in the space-time
with $d \neq 4$ leading to a modification of the beta-function $\beta_d(g) =
\frac12 (d - 4)g + \beta(g)$. Thus, away from four dimensions, the conformal
invariance is destroyed even when $\beta(g) = 0$. As a consequence, the subtraction
of divergences via the conventional minimal subtraction procedure induces the
mixing of conformal operators of different conformal spin starting from two loops,
but still the conformal symmetry can be restored by performing a {\sl finite}
renormalization of conformal operators \cite{BelMul98}. This transformation does
not affect the eigenvalues of the mixing matrix, or equivalently, the anomalous
dimension of the supermultiplet, but it induces a finite scheme transformation
of its eigenstates.

The analysis of the present section relies on a tacit assumption that there
exists a subtraction scheme which preserves the superconformal covariance. Then
within this superconformal scheme the anomalous dimensions of components of the
supermultiplet are equal to each other. Note that the supersymmetry transformation
induces a shift $\pm 1/2$ in the conformal spin as we climb/descend the
supersymmetric tower of states --- here, conformal operators with different
quantum numbers. As a consequence, the anomalous dimensions of the components
are given by the same function with the argument shifted at most by four units
of the conformal spin.

Let us denote the anomalous dimension of the operator ${\cal O}^{ss,
\bit{\scriptstyle 20}}_{jl}$ as $\gamma(j)$. Then, going through the diagram
in Fig.\ \ref{SUSY} and inspecting the corresponding supersymmetric
transformations presented in appendix \ref{SUSYofComponents}, one uniquely
fixes the anomalous dimensions of all components of the supermultiplet (see
Table \ref{Table}).
\begin{table}[t]
\begin{center}
\begin{tabular}{|l||l|l|l|l|l|}
weight $\backslash$ ``level''
&
0
&
1
&
2
&
3
&
4 \\
\hline\hline
$\gamma (j - 2)$
&
&
&
&
&
${\cal S}^1_{jl}$
\\
$\gamma (j - 1)$
&
&
&
$\bar{\cal T}^2_{[+] jl}
\quad
{\cal S}^{2 [\bit{\scriptstyle 15}]}_{jl}
$
&
${\mit\Omega}^2_{[+] jl}$
&
${\cal P}^1_{jl}$
\\
$\gamma (j)$
&
${\cal O}^{s [\bit{\scriptstyle 20}]}_{jl}$
&
${\mit\Omega}^{[\bit{\scriptstyle 20}]}_{[-] jl}$
&
$\bar{\cal T}_{[-] jl}
\quad
\widetilde {\cal O}^{q [\bit{\scriptstyle 15}]}_{jl}
\quad
{\cal T}^{q [\bit{\scriptstyle 10}]}_{jl}
$
&
${\mit\Omega}^1_{[-] jl}
\quad
{^+\!\bar{\mit\Omega}^g}_{[-] jl}
$
&
${\cal S}^2_{jl}
\quad
{\cal T}^{gg}_{[-]jl}$
\\
$\gamma (j + 1)$
&
&
${\mit\Omega}^{[\bit{\scriptstyle 20}]}_{[+] jl}$
&
$\bar{\cal T}^1_{[+] jl}
\quad
{\cal S}^{1 [\bit{\scriptstyle 15}]}_{jl}
$
&
${\mit\Omega}^1_{[+] jl}
\quad
{^+\!\bar{\mit\Omega}^g}_{[+] jl}
$
&
${\cal P}^2_{jl}$
\\
$\gamma (j + 2)$
&
&
&
&
${\mit\Omega}^2_{[-] jl}$
&
${\cal S}^3_{jl}$
\\
\end{tabular}
\end{center}
\caption{\label{Table} Anomalous dimensions of the components of the supermultiplet
of conformal operators in $\mathcal{N}=4$ SYM. }
\end{table}

\subsection{Fixing the anomalous dimension of the supermultiplet}
\label{AnomalousDimension}

To determine the anomalous dimensions of the supermultiplet it suffices to
calculate the anomalous dimension of one of its components. The simplest
choice would be to consider either one of the operators displayed in Eqs.\
(\ref{AutoRenBoson}) and (\ref{AutoRenFermion}), or the operators $[
{\cal T}^3_{[-] jl} ]^{\mu AB}$ and $[ \bar{\cal T}^3_{[-] jl} ]^\mu_{AB}$
for odd $j$'s. Actually, for two such operators --- the maximal-helicity
gluonic or quark operators, Eqs.~(\ref{g-maxhel}) and (\ref{q-maxhel}),
respectively --- one can immediately borrow corresponding one-loop QCD
results to fix $\gamma(j)$. This would merely require an adjustment of
color factors with no additional calculation of Feynman diagrams. Beyond
leading order one has to add a few extra graphs involving scalars propagating
in loops to the existing QCD calculations and transform them to the
supersymmetry preserving dimensional reduction scheme.

However, in order to explicitly demonstrate that the components of the
supermultiplet have an autonomous scale dependence, we will compute the one-loop
anomalous dimension mixing matrix for the operators $[{\cal T}_{jl}^{qq,
\bit{\scriptstyle \bar 6}}]^\mu_{AB}$ and $[\bar{\cal T}^{sg}_{jl} ]^\mu_{AB}$.
We will show that the eigenstates of this matrix give rise to the components
$[\bar{\cal T}^1_{[+] jl}]^\mu_{AB}$ and $[\bar{\cal T}^2_{[+] jl}]^\mu_{AB}$
of the superconformal operator for even $j$, Eq.\ (\ref{conf-ex}), while for
odd $j$ we get the operator $[\bar{\cal T}^3_{[-]jl} ]^\mu_{AB}$ which
renormalizes autonomously.

Instead of dealing with an infinite tower of local conformal operators, we will
work with non-local light-cone operators
\begin{eqnarray}
&&{}
\bar {\cal T}^{sg}_{[\pm]}
\left( \xi_1, \xi_2 \right)
\equiv
\frac{i}{2\sqrt{2}} \, {\rm tr} \,
\left\{
\left(
F^{+ \mu}_{\phantom{+} \, \perp} + i \widetilde F^{+ \mu}_{\phantom{+} \, \perp}
\right) (\xi_2)\,
\bar\phi_{AB}
(\xi_1) \pm \left( \xi_1 \leftrightarrow \xi_2 \right) \right\}
\, , \label{non-local1}
\\
&&{}
{\cal T}^{qq, \bit{\scriptstyle \bar 6}}
\left( \xi_1, \xi_2 \right)
\equiv
\varepsilon_{ABCD} \, {\rm tr} \,
\left\{
\lambda^{\alpha C} (\xi_2)
\bar\sigma^+{}_{\alpha\dot\beta}
\sigma^\mu_{\,\perp}{}^{\dot\beta \gamma} \lambda_\gamma^D (\xi_1)
\right\} \, ,
\label{non-local2}
\end{eqnarray}
where we have suppressed the Lorentz and isotopic indices on the left-hand side
and have tacitly assumed the light-cone gauge $A^+ = 0$ and, therefore, neglected
the light-like Wilson lines stretched between the elementary fields. These nonlocal
operators serve as generating functions of local conformal operators $[{\cal
T}_{jl}^{qq, \bit{\scriptstyle \bar 6}}]^\mu_{AB}$ and $[\bar{\cal T}^{sg}_{jl}
]^\mu_{AB}$ when expanded in the Taylor series (\ref{NonlocalExpansionConformal}).

The computation of the dilatation operator in the basis (\ref{non-local1}) and
(\ref{non-local2}) has a number of advantages. Firstly, the renormalization
acquires a very concise form and has a transparent meaning in the coordinate
space; secondly, the conformal properties of the dilatation operator become
manifest \cite{BraMulKor03}.

\begin{figure}[t]
\begin{center}
\mbox{
\begin{picture}(0,65)(195,0)
\put(0,0){\insertfig{13}{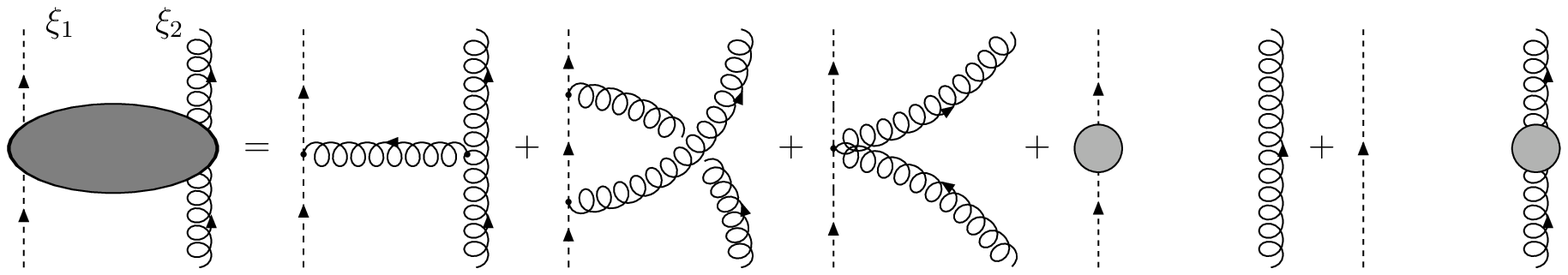}}
\end{picture}
}
\end{center}
\caption{\label{GStoGS} One loop diagrams contributing to the anomalous
dimension of $[ \bar {\cal T}^{sg}_{jl} ]^\mu_{AB}$ in the light-cone
gauge. The self-energy diagrams [the last two graphs] are multiplied
by the symmetry factor $1/2$. The dashed and wiggly lines represent
the scalar and gluon fields, respectively.}
\end{figure}

\begin{figure}[b]
\begin{center}
\mbox{
\begin{picture}(0,80)(130,0)
\put(0,0){\insertfig{9}{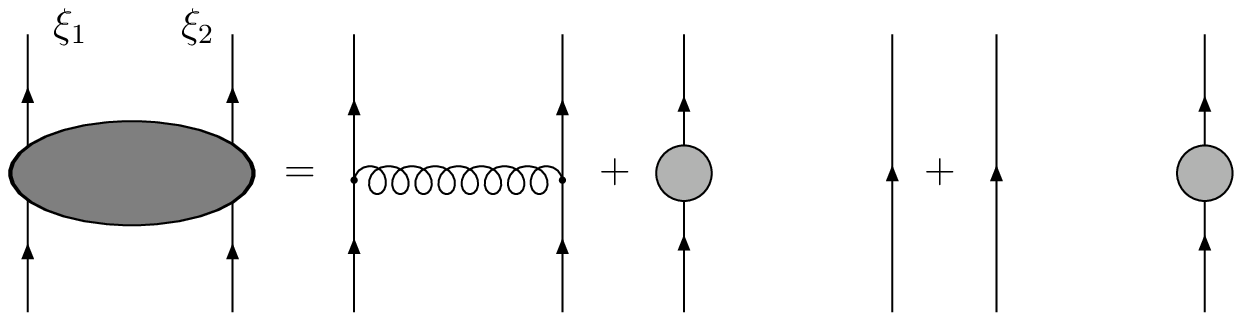}}
\end{picture}
}
\end{center}
\caption{\label{QQtoQQ} One-loop diagrams for diagonal ${\cal T}^{qq,
\bit{\scriptstyle \bar 6}} \to {\cal T}^{qq, \bit{\scriptstyle \bar 6}}$
transitions. The self-energy diagrams are multiplied by the combinatoric
factor $1/2$. The solid line represent the gaugino.}
\end{figure}

\begin{figure}[t]
\begin{center}
\mbox{
\begin{picture}(0,60)(140,0)
\put(0,0){\insertfig{9.5}{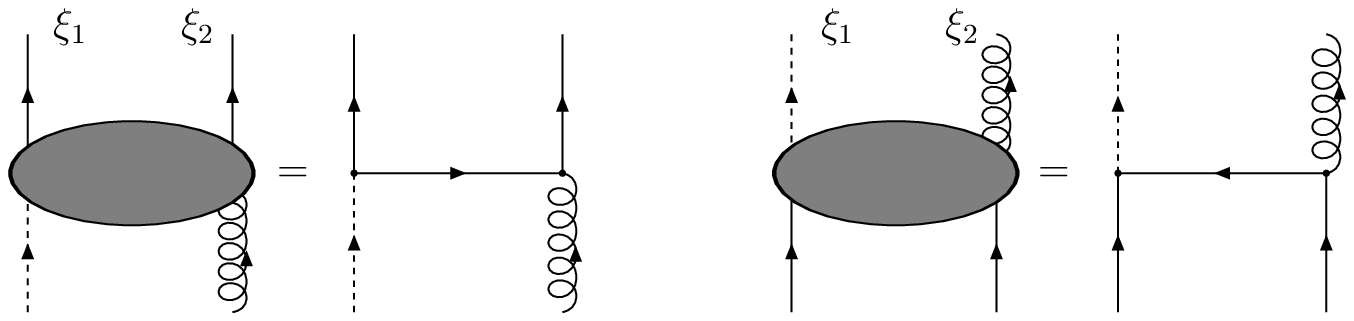}}
\end{picture}
}
\end{center}
\caption{\label{GStoQQ} One-loop transitions changing the particle
content of operators and contributing to the off-diagonal elements
of the mixing matrix.}
\end{figure}

The non-local operators (\ref{non-local1}) and (\ref{non-local2}) possess additional
singularities due to the light-light nature of separations between fields. To
regularize them one uses the dimensional regularization and subtracts divergences
in the $\overline{\rm MS}$-scheme. Due to the different parity under
$\xi_1\leftrightarrow\xi_2$, the operator ${\cal T}^{sg}_{[-]} \left( \xi_1,
\xi_2 \right)$ cannot mix with the remaining two operators and, therefore,
evolves independently. It is convenient to combine the parity-even operators
$\bar {\cal T}^{sg}_{[+]}$ and ${\cal T}^{qq, \bit{\scriptstyle\bar 6}}$ into
a two-vector
\begin{equation}
\label{ThetaTwoVector}
\bit{\bar{\cal T}}_{[+]}
\equiv
\left(
\begin{array}{c}
{\cal T}^{qq, \bit{\scriptstyle \bar 6}}
\\
\bar {\cal T}^{sg}_{[+]}
\end{array}
\right) \, .
\end{equation}
Let us discuss both cases in turn.

\medskip

\noindent $\bullet\, $ \textbf{Parity-even operators}: The one-loop mixing matrix
for the operators (\ref{ThetaTwoVector}) is given by Feynman diagrams displayed
in Figs.\ \ref{GStoGS}, \ref{QQtoQQ} and \ref{GStoQQ}. A computation, whose
detailed account can be found in appendix \ref{OneLoopEvolution}, leads to the
following result for the $\overline{\rm MS}$-subtracted operator%
\footnote{E.g., the operator, which generates finite Green functions.}
\begin{eqnarray}
\label{non-local-K}
\bit{\bar{\cal T}}_{[+]}^R \left( \xi_1, \xi_2 \right) \!\!\!&=&\!\!\!
\int_0^1 d y \int_0^1 d z \, \theta (1 - y - z)
\\
&\times&\!\!\! \left( \delta (y) \delta (z) \1 - \frac{g^2 N_c}{8 \pi^2}
\frac{S_\varepsilon}{\varepsilon} \bit{\mathbb{K}} (y, z) \right)
\bit{\bar{\cal T}}_{[+]}  \left( \bar y \xi_1 + y \xi_2 , z \xi_1 +
\bar z \xi_2 \right) \, , \nonumber
\end{eqnarray}
with $S_\varepsilon \equiv \exp (\varepsilon (\gamma_{\rm E} - \ln 4 \pi))$. The
integral kernel of the one-loop dilatation operator is given by
\begin{equation}
\bit{\mathbb{K}}(y, z) = \left(
\begin{array}{cc}
\bar y \left[ 1/y \right]_+
&
4 \, \bar y
\\
1/4
&
(1 + y) \left[ 1/y \right]_+
\end{array}
\right)
\delta (z)
+
\left(
\begin{array}{cc}
\bar z \left[ 1/z \right]_+
&
4 \, \bar z
\\
1/4
&
\bar z^2 \left[ 1/z \right]_+
\end{array}
\right)
\delta (y)
\, ,
\end{equation}
with $\bar y \equiv 1 - y$ and analogously for other variables. Here and below
the plus-regularization is defined conventionally as $\left[ 1/y \right]_+ \equiv
1/y - \delta (y) \int_0^1 d y'/y'$.

Equation (\ref{non-local-K}) has a simple physical meaning: under the
renormalization group flow the fields are displaced on the light-cone in the
direction toward each other. One can verify that (\ref{non-local-K}) is invariant
under the projective transformations on the light-cone (\ref{sl2}) and, as a
consequence, the operator $\mathbb{K}$ commutes with the generators of the
collinear $SL(2)$ subgroup. The above integral representation for the one-loop
dilatation operator is diagonalized in the basis $\bit{\bar{\cal T}}_{[+]}^R
\left( \xi_1, \xi_2 \right) \sim (\xi_1 - \xi_2)^j$. Going from the
coordinate to the momentum representation, one can show (see Eq.\
(\ref{FourierTransformDO})) that the dilatation operator admits the conformal
partial-wave expansion (\ref{PartialWave}) making the diagonalization property
manifest.

Let us expand the non-local light-cone operator $\bit{\bar{\cal T}}_{[+]}^R
\left( \xi_1, \xi_2 \right)$ in the Taylor series over local conformal operators
(\ref{NonlocalExpansionConformal}) and, then, substitute it into (\ref{non-local-K}).
Due to the positive parity, the expansion will involve only conformal operators
with even conformal spins and each term of the expansion will diagonalize the
operator $\mathbb{K}$. In this way, one obtains the renormalization group equation
\begin{equation}
\frac{d}{d \ln \mu}
[\bit{\bar {\cal T}}^R_{[+] jl}]^\mu_{AB}
=
- \frac{g^2 N_c}{8 \pi^2}
\bit{\gamma}^{(0)}_j
[\bit{\bar {\cal T}}^R_{[+] jl}]^\mu_{AB}
+
{\cal O} (g^4)
\, ,
\end{equation}
where we have extracted the color factor from the anomalous dimension
matrix, cf.\ (\ref{ADinPT}),
\begin{equation}
\bit{\gamma}^{(0)}_j =
\left(
\begin{array}{cc}
4 \psi (j + 2) + 4 \gamma_{\rm E}
&
- 8
\\
-
\frac{2}{(j + 1)(j + 2)}
&
4 \psi (j + 2) + 4 \gamma_{\rm E} - \frac{4}{(j + 1)(j + 2)}
\end{array}
\right)
\, .
\end{equation}
The eigenvectors of this matrix define two conformal operators
$[\bar{\cal T}^1_{[+] jl}]^\mu_{AB}$ and $[\bar{\cal T}^2_{[+]
jl}]^\mu_{AB}$, Eq.~(\ref{conf-ex}), and the corresponding eigenvalues determine
their anomalous dimensions
\begin{eqnarray}
{}[\bar{\cal T}^1_{[+] jl}]^\mu_{AB} : \quad \gamma^{(0)}_{j,{\rm I}}
\!\!\!&\equiv&\!\!\! \gamma^{(0)} (j + 1) = 4 \psi (j + 3) + 4 \gamma_{\rm E}
\, , \nonumber\\
{}[\bar{\cal T}^2_{[+] jl}]^\mu_{AB} : \quad \gamma^{(0)}_{j, {\rm II}}
\!\!\!&\equiv&\!\!\! \gamma^{(0)} (j - 1) = 4 \psi (j + 1) + 4 \gamma_{\rm E} \,
,
\label{Res1}
\end{eqnarray}
respectively.

\medskip

\noindent $\bullet\, $ \textbf{Parity-odd operators}: The non-local light-cone
operator $[ \bar {\cal T}^{sg}_{[-]} ]^\mu_{AB} \left( \xi_1, \xi_2 \right)$
has an autonomous scale dependence. It satisfies an evolution equation analogous
to (\ref{non-local-K}) with the evolution kernel given by
\begin{equation}
{\cal \mathbb{K}} (y, z) =
\bar y
\left[ 1/y \right]_+ \delta (z)
+
\bar z^2
\left[ 1/z \right]_+ \delta (y)
\, .
\end{equation}
Analogous to the previous consideration, one expands $\bar {\cal T}^{sg}_{[-]}
\left( \xi_1, \xi_2 \right)$ over local conformal operators and calculates
their anomalous dimension with the result
\begin{equation}
\label{Res2}
{}[\bar{\cal T}^3_{[-] jl}]^\mu_{AB} : \quad \gamma^{(0)}_{j, {\rm III}}
\equiv
\gamma^{(0)} (j)
=
4 \psi (j + 2) + 4 \gamma_{\rm E}
\, .
\end{equation}
Here we used the notation for the component of the multiplet (\ref{conf-ex}).
The obtained eigensystem (\ref{Res1}) and (\ref{Res2}) is in perfect
agreement with Table \ref{Table}.

To summarize, we demonstrated in this section that the anomalous dimensions of
different components of the supermultiplet are determined by the same function of
the conformal spin $j$. To one-loop accuracy, this function $\gamma^{(0)} (j)$ is
expressed in terms of the Euler $\psi$-function, $\psi (j + 2) = \sum_{n = 1}^{j
+ 1} \ft{1}{n} - \gamma_{\rm E}$~\cite{Osborn}. Making use of the recent two-loop
calculation of the anomalous dimensions of $SU(4)$-singlet operators
\cite{KotLip02} one can fix the
two-loop correction to the anomalous dimension of the supermultiplet%
\begin{equation}
\label{2-loops}
\gamma (j)
\equiv
\frac{g^2 N_c}{8 \pi^2} \gamma^{(0)} (j)
+
\left( \frac{g^2 N_c}{8 \pi^2} \right)^2 \gamma^{(1)} (j)
+
{\cal O} (g^6)
\, ,
\end{equation}
where
\begin{eqnarray}
\gamma^{(0)} (j)
\!\!\!&=&\!\!\!
4 \sum_{n = 1}^{j + 1} \frac{1}{n}
\, , \nonumber\\
\gamma^{(1)} (j)
\!\!\!&=&\!\!\!
- 4 \sum_{n = 1}^{j + 1}
\Bigg\{
\frac{\sigma_n}{n^3}
+
\frac{2}{n} \sum_{m = 1}^{j + 1}
\frac{1}{m^2}
\Big(
1 + (- 1)^m \theta_{n + 1, m}
\Big)
\Bigg\} \, .
\end{eqnarray}
Here $\sigma_n$ stands the signature factor (\ref{SignatureFactor}) and the
step-function is $\theta_{n,m} = \{ 1 , n > m ; 0 , n \leq m \}$.

\section{Discussion and conclusions}
\label{Discussion}

In this paper, we have studied the constraints imposed by supersymmetry on
renormalization properties of twist-two operators. These operators involve only
``good'' components of the fundamental fields and, as a consequence, they form
a closed set with respect to the action of the collinear $SL(2)$ subgroup of
the full conformal group. In conformal theories, the operators that carry the
same conformal spin mix under renormalization, and diagonalizing the corresponding
mixing matrix one constructs their linear combinations which have an autonomous
scale dependence. Being solutions to characteristic (polynomial) equations, the
anomalous dimensions $\gamma(j)$ are, in general, {\sl multi-valued} functions of
the conformal spin $j$ in a Yang-Mills theory. These properties are generic and
do not rely on supersymmetry.

Supersymmetry has the following remarkable consequences. Firstly, it allows one
to classify all conformal operators with respect to the action of superconformal
transformations of the collinear $SL(2|4)$ subgroup of the full $SU (2,2|4)$
superconformal group of the $\mathcal{N}=4$ SYM theory. As was shown in section
\ref{ComponentsSupermultiplet}, the coefficients defining relative weights of
conformal operators in the expressions for the $\mathcal{N}=4$ superconformal
operators have a very simple form. They can be interpreted as Clebsch-Gordan
coefficients in the tensor product of two irreducible representations of $SL(2|4)$.
Secondly, the superconformal operators entering the same supermultiplet have the
same scaling dimension $\gamma(j)$. In distinction with a general case of
conformal Yang-Mills theory, it is a {\sl single-valued}, meromorphic function of
the conformal spin $j$. In the $\mathcal{N}=4$ theory, the one-loop anomalous
dimension is given by the Euler $\psi$-function of the conformal Casimir while
the two-loop corrections involve its generalizations, Eq.~(\ref{2-loops}).
Thirdly, we have demonstrated that a unique feature of the $\mathcal{N}=4$ SYM is
that {\sl all} quasipartonic operators with different $SU(4)$ quantum numbers
fall into a single supermultiplet. Among them there is a subsector of the
operators of maximal helicity, Eqs.\ (\ref{q-maxhel}) and (\ref{g-maxhel}), which
plays a special role. Independently on the presence of supersymmetry, the
dilatation operator in conformal Yang-Mills theory exhibits a hidden symmetry in
this sector in the multi-color limit --- it is equivalent to a Hamiltonian of
integrable $SL(2)$ Heisenberg spin
magnet~\cite{BraDerMan98,BraDerKorMan99,Bel99a,Bel99b,DerKorMan00}. In the
$\mathcal{N}=4$ SYM theory, this symmetry is extended to the whole supermultiplet
of quasipartonic operators with the only difference that the collinear $SL(2)$
group is replaced by its supersymmetric extension $SL(2|4)$, which is a subsector
of the $PSU (2, 2 | 4)$-symmetric one-loop dilatation operator defined in Refs.\
\cite{Bei03a,BeiSta03}.

So far we have discussed twist-two conformal operators --- two-particle
blocks of multiparticle operators. The above consideration can be repeated
to construct multi-particle superconformal operators involving ``good''
components but calculations become rather tedious. Our findings can be
re-expressed in a concise form if one combines ``good'' components of fields
into a light-cone superfield ${\mit\Phi} (\xi, \theta^A)$. To obtain the
supermultiplet of conformal operators constructed in section
\ref{ComponentsSupermultiplet}, one has to examine the operator expansion
of the product of two superfields ${\mit\Phi} (\xi,\theta^A) {\mit\Phi}
(0,0)$ and expand it over irreducible components similar to
(\ref{NonlocalExpansionConformal}). Each component of the operator product
expansion evolves autonomously under dilatations and has the anomalous
dimension $\gamma(j)$ with $j$ defined by quadratic Casimir of the $SL(2|4)$
group. As a consequence, the one-loop dilatation operator in the
$\mathcal{N}=4$ super-Yang-Mills acts on the product of two superfield as
a two-particle Hamiltonian of the $SL(2|4)$ Heisenberg spin magnet. As the
next step we consider the product of superfields $\prod_{k=1}^L {\mit\Phi}
(\xi_k,\theta^A_k)$. For this subclass of operators the particle creation
and annihilation is absent, so that the number of sites $L$ is conserved.
In the multi-color limit, the dilatation operator has a two-particle
structure and defines a Hamiltonian of the integrable $SL(2|4)$ spin
chain with $L$ sites. The multi-particle superconformal operators and their
anomalous dimensions can be read from the energy spectrum of this spin chain.

Integrability of particle interactions implies stringent restrictions of the
spectrum of anomalous dimensions in gauge theories at weak coupling. It is
expected that integrability is a genuine property of the $\mathcal{N}=4$
super-Yang-Mills and, therefore, it has also to reveal itself at strong
coupling. Within the gauge/string correspondence, the operators with large
conformal spin are dual to semiclassical stringy states with a large angular
momentum on $AdS_5$. This allows to establish a correspondence between
asymptotic behavior of anomalous dimensions of two-particle operators with
large conformal spin at strong coupling and the energy of a long folded
revolving string \cite{GubKlePol02,TseFro02,Sta03}. For multiparticle operators
the spectrum of the anomalous dimensions at strong coupling was calculated in
\cite{BelGorKor03} using their intrinsic relation with a cusp anomaly of
Wilson loops. This correspondence has been employed in earlier analyses
of twist-two operators in Ref.\ \cite{Kru02}. It was conjectured in
\cite{BelGorKor03} that the string configuration, corresponding to higher-twist
operators, possesses string junctions, e.g., a three-particle operator
is expressed by a three-string, to match them to conserved charges of the
gauge theory \cite{BelGorKor03}. A more precise identification of string
configurations still awaits deeper exploration.

\vspace{1cm}

A.B.\ would like to express his gratitude to Edward Witten for the warm
hospitality extended to him at the Institute for Advanced Studies and for
numerous enlightening and inspiring discussions. He thanks Arkady Vainshtein
for an instructive conversation. The present work was supported in part by
the US Department of Energy under contract DE-FG02-93ER40762 (A.B.), by Sofya
Kovalevskaya programme of the Alexander von Humboldt Foundation (A.M.),
by the NATO Fellowship (A.M.\ and S.D.) and in part by the grant 00-01-005-00
from the Russian Foundation for Fundamental Research (A.M.\ and S.D.).

\appendix

\setcounter{section}{0}
\setcounter{equation}{0}
\renewcommand{\theequation}{\Alph{section}.\arabic{equation}}

\section{${\cal N} = 4$ super-Yang-Mills}
\label{N=4Lagrangian}

In order to set our conventions, we perform a textbook exercise of reducing
the ten-dimensional ${\cal N} = 1$ super-Yang-Mills down to the
four-dimensional Minkowski space with the result of getting the ${\cal N} = 4$
supersymmetric theory \cite{GliSchOli77,BriSchSch77}. We will adopt the $SU(4)$
covariant form of Refs.\ \cite{DorHolKhoMat02,BelVanNie00} adjusted to the
presently used nomenclature of the Dirac algebra in four dimensions which is
discussed in the subsequent appendix \ref{DiracAlgebra}. One starts with
\begin{equation}
{\cal L}_{10}
= {\rm tr}
\left\{
- \ft{1}{2} F_{MN} F^{MN}
+
i \, \bar{\mit\Psi} {\mit\Gamma}_M {\cal D}^M {\mit\Psi}
\right\}
\, ,
\end{equation}
where the field strength is $F_{MN} = \partial_M A_N - \partial_N A_M
- i g [A_M, A_N]$ and the covariant derivative correspondingly is
${\cal D}_M = \partial_M - i g [A_M, \cdot]$. All fields are matrix-valued
in the $SU (N)$ gauge group, i.e., ${\mit\Phi} \equiv {\mit\Phi}^a t^a$
for any ${\mit\Phi} = \{ A_M , {\mit\Psi}\}$, with the generators
normalized as ${\rm tr} \, t^a t^b = \ft12 \delta^{ab}$. Here ${\mit\Psi}$
is the Majorana-Weyl spinor, i.e., satisfying the conditions ${\mit\Psi}^T
C_{10} = {\mit\Psi}^\dagger {\mit\Gamma}^0 \equiv \bar{\mit\Psi}$ and
${\mit\Gamma}^{11} {\mit\Psi} = {\mit\Psi}$, and the solution to them has
the form
\begin{equation}
{\mit\Psi}
=
\left(
\begin{array}{c}
1
\\
0
\end{array}
\right)
\otimes
\left(
\begin{array}{c}
\lambda_\alpha^A
\\
0
\end{array}
\right)
+
\left(
\begin{array}{c}
0
\\
1
\end{array}
\right)
\otimes
\left(
\begin{array}{c}
0
\\
\bar\lambda_A^{\dot\alpha}
\end{array}
\right)
\, ,
\end{equation}
with obviously $\left( \lambda^A_\alpha \right)^\ast = \bar\lambda_{\dot\alpha
\, A}$. The Lagrangian is a density with respect to supersymmetric transformation
rules
\begin{equation}
\label{N=1SUSYrules}
\delta {\mit\Psi} = \ft{i}2 \, F^{MN} {\mit\Gamma}_{MN} \xi
\, , \qquad
\delta A_M = - i \, \bar\xi {\mit\Gamma}_M {\mit\Psi}
\, ,
\end{equation}
i.e., $\delta {\cal L} = \partial_M \Delta^M$ with $\Delta_M = {\rm tr} \,
\bar\xi \left\{ 2 i F_{MN} {\mit\Gamma}^N + \ft12 F^{NL} {\mit\Gamma}_{NL}
{\mit\Gamma}_M \right\} {\mit\Psi}$, where we introduced a matrix
${\mit\Gamma}^{MN} \equiv \ft{i}2 \left( {\mit\Gamma}^M {\mit\Gamma}^N -
{\mit\Gamma}^N {\mit\Gamma}^M \right)$.

Reducing the ten-dimensional Lagrangian to four dimensions we get the
standard maximally supersymmetric theory,
\begin{eqnarray}
{\cal L}_{{\cal N} = 4} = {\rm tr} \,
\bigg\{
\!\!\!&-&\!\!\! \ft12
F_{\mu\nu} F^{\mu\nu}
+
\ft12
\left( {\cal D}_\mu \phi^{AB} \right)
\left( {\cal D}^\mu \bar\phi_{AB} \right)
+
\ft{1}8 g^2
[\phi^{AB}, \phi^{CD}] [\bar\phi_{AB}, \bar\phi_{CD}]
\nonumber\\
&+&\!\!\!
2 i \bar\lambda_{\dot\alpha A}
\sigma^{\dot\alpha \beta}_\mu {\cal D}^\mu
\lambda^A_\beta
-
\sqrt{2} g
\lambda^{\alpha A}
[\bar\phi_{AB}, \lambda_\alpha^B]
+
\sqrt{2} g
\bar\lambda_{\dot\alpha A}
[\phi^{AB}, \bar\lambda^{\dot\alpha}_B]
\bigg\}
\, ,
\end{eqnarray}
where we have introduced the complex scalar field $\phi^{AB}$ related to
the six real components of the ten-dimensional gauge field $A^a$ via
\begin{equation}
\phi^{AB} = \frac{1}{\sqrt{2}} {\mit\Sigma}^{a \, AB} A^a
\, , \qquad
\bar\phi_{AB}
=
\left( \phi^{AB} \right)^\ast
=
\ft12 \varepsilon_{ABCD} \phi^{CD}
=
\frac{1}{\sqrt{2}} \bar{\mit\Sigma}^a_{AB} A^a
\, ,
\end{equation}
where $\varepsilon_{ABCD} = \varepsilon^{ABCD}$. Here we have split the
ten-dimensional index $M = \{ \mu, a \}$ into four-dimensional $\mu =
0,1,2,3$ and six-dimensional $a = 1, \dots, 6$. The Lagrangian
${\cal L}_{{\cal N} = 4}$ is invariant under the transformation rules
deduced from (\ref{N=1SUSYrules}),
\begin{eqnarray}
\label{N=4SUSYrules}
&&\delta A^\mu
=
- i \xi^{\alpha \; A} \bar\sigma^\mu{}_{\alpha\dot\beta}
\bar\lambda^{\dot\beta}_A
- i \bar\xi_{\dot\alpha \; A} \sigma^{\mu\; \dot\alpha\beta}
\lambda_{\beta}^A
\, , \nonumber\\
&&\delta \phi^{AB}
=
- i \sqrt{2}
\left\{
\xi^{\alpha \; A} \lambda_\alpha^B - \xi^{\alpha \; B} \lambda_\alpha^A
-
\varepsilon^{ABCD} \bar\xi_{\dot\alpha \; C} \bar\lambda^{\dot\alpha}_D
\right\}
\, , \nonumber\\
&&\delta \lambda^A_\alpha
=
\ft{i}2 F_{\mu\nu} \sigma^{\mu\nu}{}_\alpha{}^\beta \xi_\beta^A
-
\sqrt{2} \left( {\cal D}_\mu \phi^{AB} \right)
\bar\sigma^\mu{}_{\alpha\dot\beta} \bar\xi^{\dot\beta}_B
+
i g [\phi^{AB} , \bar\phi_{BC}] \xi_\alpha^C
\, , \nonumber\\
&&\delta \bar\lambda_A^{\dot\alpha}
=
\ft{i}2 F_{\mu\nu} \bar\sigma^{\mu\nu}{}^{\dot\alpha}{}_{\dot\beta}
\bar\xi^{\dot\beta}_A
+
\sqrt{2} \left( {\cal D}_\mu \bar\phi_{AB} \right)
\sigma^{\mu \, \dot\alpha\beta} \xi_\beta^B
+
i g[\bar\phi_{AB} , \phi^{BC}] \bar\xi^{\dot\alpha}_C
\, .
\end{eqnarray}

\setcounter{equation}{0}
\renewcommand{\theequation}{\Alph{section}.\arabic{equation}}

\section{Dirac algebra in various dimensions}
\label{DiracAlgebra}

In this appendix we introduce representations of Clifford algebras in
four, six and ten dimensions. Let us introduce them in turn.

$\bullet$ $D = 3 + 1$:
The four-component spinor is composed from two Weyl spinors, transforming
with respect to conjugated factors of the Lorentz group $L_+^\uparrow = SO
(3,1) = SO (4, {\bf C})_{\downarrow R} \approx \left( SL (2, {\bf C})
\otimes SL (2, {\bf C}) \right)_{\downarrow R}$ which are labeled by a pair
$\left( j_1, j_2 \right)$ of eigenvalues $j_i(j_i + 1)$ of the $SL (2)$
Casimir operators $\mbox{\boldmath$J$}{}^2_i$,
\begin{equation}
\psi
=
\left(
\begin{array}{c}
\lambda_\alpha
\\
\bar\lambda^{\dot\alpha}
\end{array}
\right)
\, ,
\end{equation}
with $\lambda_\alpha \sim (\ft12, 0)$ and $\bar\lambda^{\dot\alpha}
\sim (0 , \ft12)$. The Dirac matrices admit the form
\begin{equation}
\label{4Dmatrices}
\gamma^\mu
=
\left(
\begin{array}{cc}
0 & \bar\sigma^\mu{}_{\alpha \dot\beta}
\\
\sigma^\mu{}^{\; \dot\alpha \beta} & 0
\end{array}
\right)
\, , \qquad
\gamma^5
=
i \gamma^0 \gamma^1 \gamma^2 \gamma^3
=
\left(
\begin{array}{rr}
1 & 0
\\
0 & - 1
\end{array}
\right)
\, ,
\end{equation}
where $\sigma^\mu = (1 , \bit{\sigma})$ and $\bar\sigma^\mu = (1 , -
\bit{\sigma})$ with the conventional vector of Pauli matrices $\bit{\sigma}$.
The Clifford algebra is $\{ \gamma^\mu , \gamma^\nu \} = 2 g^{\mu\nu}$, with
the metric of signature $g^{\mu\nu} = {\rm diag} (+, -, -, -)$ and reads for
two-dimensional matrices
\begin{equation}
\bar\sigma^\mu{}_{\alpha\dot\gamma} \sigma^{\nu \; \dot\gamma\beta}
+
\bar\sigma^\nu{}_{\alpha\dot\gamma} \sigma^{\mu \; \dot\gamma\beta}
= 2 g^{\mu\nu} \delta_\alpha^\beta
\, ,
\qquad
\sigma^{\mu \; \dot\alpha\gamma} \bar\sigma^\nu{}_{\gamma\dot\beta}
+
\sigma^{\nu \; \dot\alpha\gamma} \bar\sigma^\mu{}_{\gamma\dot\beta}
= 2 g^{\mu\nu} \delta_{\dot\alpha}^{\dot\beta}
\, .
\end{equation}
The charge conjugation matrix is
\begin{equation}
C_4
=
i \gamma^2 \gamma^0
=
\left(
\begin{array}{rr}
- \varepsilon^{\alpha\beta} & 0
\\
0 & - \varepsilon_{\dot\alpha \dot\beta}
\end{array}
\right)
\, ,
\end{equation}
where $\varepsilon^{\alpha\beta} = - \varepsilon_{\dot\alpha \dot\beta}$
and $\varepsilon^{12} = \varepsilon_{12} = - \varepsilon_{\dot 1 \dot 2}
= - \varepsilon^{\dot 1 \dot 2} = 1$. The raising and lowering of Weyl
spinor indices is accomplished by the following set of rules
\begin{equation}
\lambda^\alpha = \varepsilon^{\alpha \beta} \lambda_\beta
\, , \qquad
\bar\lambda_{\dot\alpha}
= \varepsilon_{\dot\alpha \dot\beta} \bar\lambda^{\dot\beta}
\, , \qquad
\lambda_\alpha = \lambda^\beta \varepsilon_{\beta\alpha}
\, , \qquad
\bar\lambda^{\dot\alpha}
= \bar\lambda_{\dot\beta} \varepsilon^{\dot\beta \dot\alpha}
\, ,
\end{equation}
corresponding to conventions of Ref.\ \cite{Soh85}. While for the
$\sigma$-matrices we have
\begin{equation}
\sigma^{\mu \; \dot\alpha\beta}
=
\varepsilon^{\beta\gamma}
\bar\sigma^\mu{}_{\gamma\dot\delta}
\varepsilon^{\dot\delta\dot\alpha}
\, , \qquad
\bar\sigma^\mu{}_{\alpha\dot\beta}
=
\varepsilon_{\dot\beta\dot\gamma}
\sigma^{\mu \; \dot\gamma\delta}
\varepsilon_{\delta\alpha}
\, .
\end{equation}
The charge conjugation matrix obeys the relations $C_4^T = - C_4$, $C_4^2 = -1$
and $C_4 \gamma^\mu = - \left( \gamma^\mu \right){\!}^T C_4$. The Majorana
condition for a four-component spinor $\psi^T C_4 = \psi^\dagger \gamma^0 \equiv
\bar\psi$ reads in terms of the Weyl components $\left( \bar\lambda^{\dot\alpha}
\right)^\ast = \lambda^\alpha$ and $\left( \lambda_\alpha \right )^\ast =
\bar\lambda_{\dot\alpha}$. The sigma-matrices transform under the complex
conjugation as follows:
\begin{equation}
\left( \sigma^{\mu \, \dot\alpha\beta} \right)^\ast
=
\sigma^{\mu \, \dot\beta\alpha}
\, , \qquad
\left( \bar\sigma^\mu{}_{\alpha\dot\beta} \right)^\ast
=
\bar\sigma^\mu{}_{\beta\dot\alpha}
\, .
\end{equation}
Note that $\left( \varepsilon_{\dot\alpha\dot\beta} \right)^\ast = -
\varepsilon_{\alpha\beta}$. Any tensor can be expanded in the following
basis of two-dimensional matrices $\{ \1, \sigma^\mu, \bar\sigma^\mu ,
\sigma^{\mu\nu} , \bar\sigma^{\mu\nu} \}$, where we introduced
\begin{equation}
\sigma^{\mu\nu}{}_\alpha{}^\beta
\equiv
\ft{i}{2}
\left[
\bar\sigma^\mu{}_{\alpha\dot\gamma} \sigma^{\nu \; \dot\gamma\beta}
-
\bar\sigma^\nu{}_{\alpha\dot\gamma} \sigma^{\mu \; \dot\gamma\beta}
\right]
\, , \qquad
\bar\sigma^{\mu\nu}{\;}^{\dot\alpha}{}_{\dot\beta}
\equiv
\ft{i}{2}
\left[
\sigma^{\mu \; \dot\alpha\gamma} \bar\sigma^\nu{}_{\gamma\dot\beta}
-
\sigma^{\nu \; \dot\alpha\gamma} \bar\sigma^\mu{}_{\gamma\dot\beta}
\right]
\, ,
\end{equation}
which satisfy the following ``self-duality'' conditions
\begin{equation}
\ft{i}2 \varepsilon^{\mu\nu\rho\sigma} \sigma_{\rho\sigma}
=
\sigma^{\mu\nu}
\, , \qquad
\ft{i}2 \varepsilon^{\mu\nu\rho\sigma} \bar\sigma_{\rho\sigma}
=
- \bar\sigma^{\mu\nu}
\, .
\end{equation}
Note that $\sigma^{\mu\nu}{\,}_\alpha{}^\beta = \varepsilon^{\beta\gamma}
\sigma^{\mu\nu}{\,}_\gamma{}^\delta \varepsilon_{\delta\alpha}$ and
analogously for $\bar\sigma$ with undotted indices replaced by dotted.
Then, the products of two Weyl spinors obey the following Fierz identities
\begin{eqnarray}
\label{FierzWeyl}
\xi_\alpha \zeta^\beta
\!\!\!&=&\!\!\!
-
\ft12 \, \delta_\alpha^\beta \left( \xi^{\gamma} \zeta_\gamma \right)
+
\ft18 \, \sigma^{\mu\nu}{\,}_\alpha{}^\beta
\left( \xi^\gamma
\sigma_{\mu\nu}{\,}_\gamma{}^\delta
\zeta_\delta
\right)
\, , \nonumber\\
\bar\xi^{\dot\alpha} \bar\zeta_{\dot\beta}
\!\!\!&=&\!\!\!
-
\ft12 \, \delta^{\dot\alpha}_{\dot\beta}
\left( \bar\xi_{\dot\gamma} \bar\zeta^{\dot\gamma} \right)
+
\ft18 \, \bar\sigma^{\mu\nu}{\,}^{\dot\alpha}{}_{\dot\beta}
\left( \bar\xi_{\dot\gamma}
\bar\sigma_{\mu\nu}{\,}^{\dot\gamma}{}_{\dot\delta}
\bar\zeta^{\dot\delta}
\right)
\, , \nonumber\\
&&
\bar\xi^{\dot\alpha} \xi^\beta
=
\ft12 \, \sigma^{\mu \, \dot\alpha\beta}
\left(
\bar\xi_{\dot\gamma}
\sigma_\mu{}^{\dot\gamma\delta}
\xi_\delta
\right)
\, ,
\end{eqnarray}
with the following traces used in the derivation
\begin{eqnarray}
&&\!\!\!\!\!\ft12 {\rm tr}
\left\{
\,
\bar\sigma^\mu \sigma^\nu
\right\}
=
g^{\mu\nu}
\, , \nonumber\\
\ft12 {\rm tr}
\left\{
\,
\bar\sigma^\mu \sigma^\nu \bar\sigma^\rho \sigma^\sigma
\right\}
\!\!\!&=&\!\!\!
g^{\mu \nu} g^{\rho \sigma}
+
g^{\mu \sigma} g^{\nu \rho}
-
g^{\mu \rho} g^{\nu \sigma}
-
i \varepsilon^{\mu\nu\rho\sigma}
\, , \\
\ft12 {\rm tr}
\left\{
\,
\sigma^\mu \bar\sigma^\nu \sigma^\rho \bar\sigma^\sigma
\right\}
\!\!\!&=&\!\!\!
g^{\mu \nu} g^{\rho \sigma}
+
g^{\mu \sigma} g^{\nu \rho}
-
g^{\mu \rho} g^{\nu \sigma}
+
i \varepsilon^{\mu\nu\rho\sigma}
\, ,
\end{eqnarray}
where the normal position of spinor indices as in Eq.\ (\ref{4Dmatrices})
is implied. The complex conjugation of products of Weyl spinors includes
the reversal of ordering, e.g.,
\begin{equation}
\left( \xi^\alpha \zeta_\alpha \right)^\ast
\equiv
\left( \zeta_\alpha \right)^\ast \left( \xi^\alpha \right)^\ast
=
\bar\zeta_{\dot\alpha} \bar\xi^{\dot\alpha}
\, .
\end{equation}

The following relations are useful to perform the Dirac algebra for the
product of three matrices:
\begin{equation}
\label{GtoEpsilon}
\sigma_\perp^{\mu \; \dot\alpha\gamma}
\left(
\delta^\beta_\gamma
-
\bar\sigma^+{}_{\gamma\dot\delta}
\sigma^{- \; \dot\delta\beta}
\right)
=
i \varepsilon^{\mu\nu}_\perp
\sigma_\nu^{\perp \; \dot\alpha\beta}
\, , \qquad
\bar\sigma_\perp^{\mu}{}_{\alpha\dot\gamma}
\left(
\delta_{\dot\beta}^{\dot\gamma}
-
\sigma^{+ \; \dot\gamma\delta}
\bar\sigma^-{}_{\delta\dot\beta}
\right)
=
- i \varepsilon^{\mu\nu}_\perp
\bar\sigma^\perp_{\nu \; \alpha\dot\beta}
\, ,
\end{equation}
for contraction of transverse Lorentz indices
\begin{equation}
\label{ThreeTransverse}
\sigma_\mu^{\perp \, \dot\alpha\beta}
\bar\sigma^\perp_{\, \nu \, \beta\dot\gamma}
\sigma_\perp^{\mu \, \dot\gamma\delta}
=
\bar\sigma^\perp_{\, \mu \, \alpha\dot\beta}
\sigma_{\, \nu}^{\perp \, \dot\beta\gamma}
\bar\sigma^\mu_{\perp \, \gamma\dot\delta}
= 0 \, ,
\end{equation}
and for the antisymmetric tensors
\begin{equation}
\label{SigmaToOne}
\sigma^{\, \mu\nu}_{\perp\perp}{\, }_\alpha{}^\beta
=
- \varepsilon_\perp^{\mu\nu}
\left\{
\delta_\alpha^\beta
-
\bar\sigma^-{}_{\alpha\dot\gamma}
\sigma^{+ \, \dot\gamma\beta}
\right\}
\, , \qquad
\bar\sigma^{\, \mu\nu}_{\perp\perp}{\, }^{\dot\alpha}{}_{\dot\beta}
=
\varepsilon_\perp^{\mu\nu}
\left\{
\delta^{\dot\alpha}_{\dot\beta}
-
\sigma^{- \, \dot\alpha\gamma}
\bar\sigma^+{}_{\gamma\dot\beta}
\right\}
\, .
\end{equation}

$\bullet$ $D = 6$:
The Dirac matrices in the six-dimensional Euclidean space are taken
in the form
\begin{equation}
\hat\gamma^a
=
\left(
\begin{array}{cc}
0 & {\mit\Sigma}^a{}^{\, AB}
\\
\bar{\mit\Sigma}^a{}_{AB} & 0
\end{array}
\right)
\, , \qquad
\hat\gamma^7
=
i
\hat\gamma^1 \hat\gamma^2 \hat\gamma^3 \hat\gamma^4 \hat\gamma^5 \hat\gamma^6
=
\left(
\begin{array}{rr}
1 & 0
\\
0 & - 1
\end{array}
\right)
\, ,
\end{equation}
where ${\mit\Sigma}^a{}^{\, AB} = (\eta_{i AB}, i \bar\eta_{i AB})$ and
$\bar{\mit\Sigma}^a{}_{\, AB} = (\eta_{i AB}, - i \bar\eta_{i AB})$ are
expressed in terms of 't Hooft symbols \cite{Hoo78}. They obey the
Clifford algebra $\{ \hat\gamma^a , \hat\gamma^b \} = - 2 \delta^{ab}$,
where $\delta^{ab} = {\rm diag} (+, \dots, +)$, with the normalization
chosen with regards to the fact that it will be a part of the
ten-dimensional Clifford algebra. 't Hooft symbols obey the following
relations
\begin{eqnarray}
&&
\eta_{iAB}
\equiv
\varepsilon_{iAB}
+
\delta_{iA} \, \delta_{4B}
-
\delta_{iB} \, \delta_{4A}
\, , \nonumber\\
&&
\bar\eta_{iAB}
\equiv
\varepsilon_{iAB}
-
\delta_{iA} \, \delta_{4B}
+
\delta_{iB} \, \delta_{4A}
\, ,
\end{eqnarray}
and $\bar\eta_{iAB} = (-1)^{\delta_{4A} + \delta_{4B}} \eta_{iAB}$. They
form a basis of anti-symmetric $4 \times 4$ matrices and are (anti-) selfdual
in vector indices ($\varepsilon_{1234} = \varepsilon^{1234} = 1$)
\begin{equation}
\eta_{iAB}
=
\ft12 \varepsilon_{ABCD} \, \eta_{iCD}
\, , \qquad
\bar\eta_{iAB}
=
- \ft12 \varepsilon_{ABCD} \, \bar\eta_{iCD}
\, .
\end{equation}
The $\eta$-symbols obey the following relations
\begin{eqnarray}
\label{eta-eta1}
&&
\varepsilon_{ijk} \, \eta_{jAB} \, \eta_{kCD}
=
\delta_{AC} \, \eta_{iBD}
+
\delta_{BD} \, \eta_{iAC}
-
\delta_{AD} \, \eta_{iBC}
-
\delta_{BC} \, \eta_{iAD}
\, , \nonumber\\
&&
\eta_{iAB} \, \eta_{iCD}
=
\delta_{AC} \, \delta_{BD}
-
\delta_{AD} \, \delta_{BC}
+
\varepsilon_{ABCD}
\, , \nonumber\\
\label{eta-eta3}
&&
\eta_{iAB} \, \eta_{jAC}
=
\delta_{ij} \, \delta_{BC}
+
\varepsilon_{ijk} \, \eta_{kBC}
\, , \nonumber\\
&&
\varepsilon_{ABCE} \, \eta_{iDE}
=
\delta_{DA} \, \eta_{iBC}
+
\delta_{DC} \, \eta_{iAB}
-
\delta_{DB} \, \eta_{iAC}
\, , \\
&&
\eta_{iAB} \, \eta_{iAB}
=
12
\, , \nonumber\\
&&
\eta_{iAB} \, \eta_{jAB}
=
4 \delta_{ij}
\, , \nonumber\\
&&
\eta_{iAB} \, \eta_{iAC}
=
3 \delta_{BC}
\, . \nonumber
\end{eqnarray}
The same holds for $\bar\eta$ except for
\begin{equation}
\bar\eta_{iAB} \, \bar\eta_{iCD}
=
\delta_{AC} \, \delta_{BD}
-
\delta_{AD} \, \delta_{BC}
-
\varepsilon_{ABCD}
\, .
\end{equation}
Obviously $\eta_{iAB} \, \bar\eta_{jAB} = 0$ due to their different
duality properties.

The six-dimensional charge conjugation matrix is
\begin{equation}
C_6
=
\hat\gamma^1 \hat\gamma^2 \hat\gamma^3
=
\left(
\begin{array}{cc}
0 & \delta_A{}^B
\\
\delta^A{}_B & 0
\end{array}
\right)
\, ,
\end{equation}
with properties $C_6^T = C_6$ and $C_6 \hat\gamma^a = - \left( \hat\gamma^a
\right){\!}^T C_6$.

$\bullet$ $D = 9 + 1$:
We use the representations for the four- and six-dimensional Clifford algebra
elaborated above in order to construct a representation for the ten-dimensional
matrices. Namely,
\begin{equation}
{\mit\Gamma}^M
=
\left( \1 \otimes \gamma^\mu , \hat\gamma^a \otimes \gamma^5 \right)
\, .
\end{equation}
They obey the algebra $\{ {\mit\Gamma}^M, {\mit\Gamma}^N \} = 2 g^{MN}$
with the Minkowski metric $g^{MN} = {\rm diag} (+, - , \dots , -)$. The
chiral and the charge conjugation matrices are defined as
\begin{equation}
{\mit\Gamma}^{11} = \hat\gamma^7 \otimes \gamma^5
\, , \qquad
C_{10} = C_6 \otimes C_4
\, ,
\end{equation}
respectively. Obviously, the latter satisfies the conditions $C_{10}^T = - C_{10}$
and $C_{10} {\mit\Gamma}^M = - \left( {\mit\Gamma}^M \right){\!}^T C_{10}$.

\setcounter{equation}{0}
\renewcommand{\theequation}{\Alph{section}.\arabic{equation}}

\section{Building the supermultiplet: transformation of ...}
\label{SUSYtrans}

In this appendix we give the complete result of the supersymmetric
transformations of conformal operators in ${\cal N} = 4$ Yang-Mills
theory, which allows to construct their supermultiplet. In the final
subsection \ref{SUSYofComponents} we give the result for the transformation
of the components of the multiplet. Throughout this section the
symmetrization and antisymmetrization operations are defined as follows:
\begin{equation}
\label{SymAsym}
T^{[AB]} \equiv T^{AB} - T^{BA}
\, , \qquad
T^{\{AB\}} \equiv T^{AB} + T^{BA}
\, .
\end{equation}

\subsection{... bosonic operators}

$SU(4)$ singlet operators first:
For the parity-even operators we get
\begin{eqnarray}
\delta {\cal O}_{jl}^{qq}
\!\!\!&=&\!\!\!
- \sigma_j
\frac{j + 2}{2j + 3}
\nonumber\\
&\times&\!\!\!
\bigg\{
(j + 3)
\left(
\xi^{\alpha A} [ {\mit\Omega}^{gq}_{jl} ]_{\alpha A}
+
\bar\xi_{\dot\alpha A} [ \bar{\mit\Omega}_{jl}^{gq} ]^{\dot\alpha A}
\right)
-
(j + 1)
\left(
\xi^{\alpha A} [ {\mit\Omega}^{gq}_{j - 1 l} ]_{\alpha A}
+
\bar\xi_{\dot\alpha A} [ \bar{\mit\Omega}_{j - 1 l}^{gq} ]^{\dot\alpha A}
\right)
\nonumber\\
&&\!\!\!\!-
(j + 1)
\left(
\xi^{\alpha A} [ {\mit\Omega}^{sq}_{jl} ]_{\alpha A}
-
\bar\xi_{\dot\alpha A} [ \bar{\mit\Omega}_{jl}^{sq} ]^{\dot\alpha A}
\right)
+
(j + 1)
\left(
\xi^{\alpha A} [ {\mit\Omega}^{sq}_{j - 1 l} ]_{\alpha A}
-
\bar\xi_{\dot\alpha A} [ \bar{\mit\Omega}_{j - 1 l}^{sq} ]^{\dot\alpha A}
\right)
\bigg\}
\, , \\
\delta {\cal O}_{jl}^{gg}
\!\!\!&=&\!\!\!
\sigma_j
\frac{(j + 2) (j + 3)}{12 (2j + 3)}
\nonumber\\
&\times&\!\!\!
\bigg\{
j \left(
\xi^{\alpha A} [ {\mit\Omega}^{gq}_{jl} ]_{\alpha A}
+
\bar\xi_{\dot\alpha A} [ \bar{\mit\Omega}_{jl}^{gq} ]^{\dot\alpha A}
\right)
+
(j + 1)
\left(
\xi^{\alpha A} [ {\mit\Omega}^{gq}_{j - 1 l} ]_{\alpha A}
+
\bar\xi_{\dot\alpha A} [ \bar{\mit\Omega}_{j - 1 l}^{gq} ]^{\dot\alpha A}
\right)
\bigg\}
\, , \\
\delta {\cal O}_{jl}^{ss}
\!\!\!&=&\!\!\!
- \sigma_j \frac{2}{2j + 3}
\nonumber\\
&\times&\!\!\!
\bigg\{
(j + 2) \left(
\xi^{\alpha A} [ {\mit\Omega}^{sq}_{jl} ]_{\alpha A}
-
\bar\xi_{\dot\alpha A} [ \bar{\mit\Omega}_{jl}^{sq} ]^{\dot\alpha A}
\right)
+
(j + 1)
\left(
\xi^{\alpha A} [ {\mit\Omega}^{sq}_{j - 1 l} ]_{\alpha A}
-
\bar\xi_{\dot\alpha A} [ \bar{\mit\Omega}_{j - 1 l}^{sq} ]^{\dot\alpha A}
\right)
\bigg\}
\, .
\end{eqnarray}

Analogously, for the parity-odd operators we get
\begin{eqnarray}
\delta \widetilde{\cal O}_{jl}^{qq}
\!\!\!&=&\!\!\!
- \sigma_{j + 1}
\frac{j + 2}{2j + 3}
\nonumber\\
&\times&\!\!\!
\bigg\{
- (j + 3)
\left(
\xi^{\alpha A} [ {\mit\Omega}^{gq}_{jl} ]_{\alpha A}
-
\bar\xi_{\dot\alpha A} [ \bar{\mit\Omega}_{jl}^{gq} ]^{\dot\alpha A}
\right)
+
(j + 1)
\left(
\xi^{\alpha A} [ {\mit\Omega}^{gq}_{j - 1 l} ]_{\alpha A}
-
\bar\xi_{\dot\alpha A} [ \bar{\mit\Omega}_{j - 1 l}^{gq} ]^{\dot\alpha A}
\right)
\nonumber\\
&&\
- \, (j + 1)
\left(
\xi^{\alpha A} [ {\mit\Omega}^{sq}_{jl} ]_{\alpha A}
+
\bar\xi_{\dot\alpha A} [ \bar{\mit\Omega}_{jl}^{sq} ]^{\dot\alpha A}
\right)
+
(j + 1)
\left(
\xi^{\alpha A} [ {\mit\Omega}^{sq}_{j - 1 l} ]_{\alpha A}
+
\bar\xi_{\dot\alpha A} [ \bar{\mit\Omega}_{j - 1 l}^{sq} ]^{\dot\alpha A}
\right)
\bigg\}
\, , \\
\delta \widetilde{\cal O}_{jl}^{gg}
\!\!\!&=&\!\!\!
- \sigma_{j + 1}
\frac{(j + 2) (j + 3)}{12 (2j + 3)}
\nonumber\\
&\times&\!\!\!
\left\{
j
\left(
\xi^{\alpha A} [ {\mit\Omega}^{gq}_{jl} ]_{\alpha A}
-
\bar\xi_{\dot\alpha A} [ \bar{\mit\Omega}_{jl}^{gq} ]^{\dot\alpha A}
\right)
+ (j + 1)
\left(
\xi^{\alpha A} [ {\mit\Omega}^{gq}_{j - 1 l} ]_{\alpha A}
-
\bar\xi_{\dot\alpha A} [ \bar{\mit\Omega}_{j - 1 l}^{gq} ]^{\dot\alpha A}
\right)
\right\}
\, .
\end{eqnarray}
Here we have used Eqs.\ (\ref{GtoEpsilon}) and relations between Jacobi
polynomials of different order \cite{PruBryMar83}.

The variation of $SU(4)$ non-singlet bosonic operators reads for
scalar bilinears,
\begin{eqnarray}
\delta [ {\cal O}_{jl}^{ss, \bit{\scriptstyle 15}}]_A{}^B
\!\!\!&=&\!\!\!
- \sigma_{j + 1} \frac{2}{3 (2j + 3)}
{}[P_{\bit{\scriptstyle 15}}]_{AC}^{BD}
\bigg\{
(j + 2)
\left(
\xi^{\alpha C}
{}[ {\mit\Omega}^{sq}_{jl} ]_{\alpha D}
+
\bar\xi_{\dot\alpha D}
{}[ \bar{\mit\Omega}_{jl}^{sq} ]^{\dot\alpha C}
\right)
\\
&&\qquad\qquad\qquad\qquad\quad \
+ \,
(j + 1)
\left(
\xi^{\alpha C}
{}[ {\mit\Omega}^{sq}_{j - 1 l} ]_{\alpha D}
+
\bar\xi_{\dot\alpha D}
{}[ \bar{\mit\Omega}_{j - 1 l}^{sq} ]^{\dot\alpha C}
\right)
\bigg\}
\nonumber\\
&&\!\!\!
+ \sigma_{j + 1} \frac{1}{6 (2j + 3)}
\bigg\{
(j + 2)
\left(
\xi^{\alpha C}
{}[ {\mit\Omega}_{jl}^{sq, \bit{\scriptstyle 20}} ]_\alpha{}_{AC}^B
+
\bar\xi_{\dot\alpha C}
{}[
\bar{\mit\Omega}_{jl}^{sq, \bit{\scriptstyle \overline{20\!}\,}}
{}]^{\dot\alpha}{}^{BC}_A
\right)
\nonumber\\
&&\qquad\qquad\qquad
+\, (j + 1)
\left(
\xi^{\alpha C}
{}[ {\mit\Omega}_{j - 1 l}^{sq, \bit{\scriptstyle 20}} ]_\alpha{}_{AC}^B
+
\bar\xi_{\dot\alpha C}
{}[
\bar{\mit\Omega}_{j - 1 l}^{sq, \bit{\scriptstyle \overline{20\!}\,}}
{}]^{\dot\alpha}{}^{BC}_A
\right)
\bigg\}
\, , \nonumber
\end{eqnarray}
and quark operators
\begin{eqnarray}
\delta [ {\cal O}_{jl}^{qq, \bit{\scriptstyle 15}}]_A{}^B
\!\!\!&=&\!\!\!
\sigma_{j + 1}
\frac{j + 2}{3 (2j + 3)}
{}[P_{\bit{\scriptstyle 15}}]_{AC}^{BD}
\\
&\times&\!\!\!
\bigg\{
3 (j + 3)
\left(
\xi^{\alpha C} [ {\mit\Omega}^{gq}_{jl} ]_{\alpha D}
-
\bar\xi_{\dot\alpha D} [ \bar{\mit\Omega}_{jl}^{gq} ]^{\dot\alpha C}
\right)
-
3 (j + 1)
\left(
\xi^{\alpha C} [ {\mit\Omega}^{gq}_{j - 1 l} ]_{\alpha D}
-
\bar\xi_{\dot\alpha D} [ \bar{\mit\Omega}_{j - 1 l}^{gq} ]^{\dot\alpha C}
\right)
\nonumber\\
&&\!
- (j + 1)
\left(
\xi^{\alpha C} [ {\mit\Omega}^{sq}_{jl} ]_{\alpha D}
+
\bar\xi_{\dot\alpha D} [ \bar{\mit\Omega}_{jl}^{sq} ]^{\dot\alpha C}
\right)
+
(j + 1)
\left(
\xi^{\alpha C} [ {\mit\Omega}^{sq}_{j - 1 l} ]_{\alpha D}
+
\bar\xi_{\dot\alpha D} [ \bar{\mit\Omega}_{j - 1 l}^{sq} ]^{\dot\alpha C}
\right)
\bigg\}
\nonumber\\
&-&
\!\!\!\sigma_{j + 1} \frac{(j + 1) (j + 2)}{6 (2j + 3)}
\nonumber\\
&\times&\!\!\!
\bigg\{
\left(
\xi^{\alpha C}
{}[ {\mit\Omega}_{jl}^{sq, \bit{\scriptstyle 20}} ]_\alpha{}_{AC}^B
+
\bar\xi_{\dot\alpha C}
{}[
\bar{\mit\Omega}_{jl}^{sq, \bit{\scriptstyle \overline{20\!}\,}}
{}]^{\dot\alpha}{}^{BC}_A
\right)
-
\left(
\xi^{\alpha C}
{}[ {\mit\Omega}_{j - 1l}^{sq, \bit{\scriptstyle 20}} ]_\alpha{}_{AC}^B
+
\bar\xi_{\dot\alpha C}
{}[
\bar{\mit\Omega}_{j - 1l}^{sq, \bit{\scriptstyle \overline{20\!}\,}}
{}]^{\dot\alpha}{}^{BC}_A
\right)
\bigg\}
\, , \nonumber\\
\delta [ \widetilde {\cal O}_{jl}^{qq, \bit{\scriptstyle 15}}]_A{}^B
\!\!\!&=&\!\!\!
- \sigma_j
\frac{j + 2}{3 (2j + 3)}
{}[P_{\bit{\scriptstyle 15}}]_{AC}^{BD}
\\
&\times&\!\!\!
\bigg\{
3 (j + 3)
\left(
\xi^{\alpha C} [ {\mit\Omega}^{gq}_{jl} ]_{\alpha D}
+
\bar\xi_{\dot\alpha D} [ \bar{\mit\Omega}_{jl}^{gq} ]^{\dot\alpha C}
\right)
-
3 (j + 1)
\left(
\xi^{\alpha C} [ {\mit\Omega}^{gq}_{j - 1 l} ]_{\alpha D}
+
\bar\xi_{\dot\alpha D} [ \bar{\mit\Omega}_{j - 1 l}^{gq} ]^{\dot\alpha C}
\right)
\nonumber\\
&&\!
+ (j + 1)
\left(
\xi^{\alpha C} [ {\mit\Omega}^{sq}_{jl} ]_{\alpha D}
-
\bar\xi_{\dot\alpha D} [ \bar{\mit\Omega}_{jl}^{sq} ]^{\dot\alpha C}
\right)
-
(j + 1)
\left(
\xi^{\alpha C} [ {\mit\Omega}^{sq}_{j - 1 l} ]_{\alpha D}
-
\bar\xi_{\dot\alpha D} [ \bar{\mit\Omega}_{j - 1 l}^{sq} ]^{\dot\alpha C}
\right)
\bigg\}
\nonumber\\
&-&
\!\!\!\sigma_j \frac{(j + 1) (j + 2)}{6 (2j + 3)}
\nonumber\\
&\times&\!\!\!
\bigg\{
\left(
\xi^{\alpha C}
{}[ {\mit\Omega}_{jl}^{sq, \bit{\scriptstyle 20}} ]_\alpha{}_{AC}^B
-
\bar\xi_{\dot\alpha C}
{}[
\bar{\mit\Omega}_{jl}^{sq, \bit{\scriptstyle \overline{20\!}\,}}
{}]^{\dot\alpha}{}^{BC}_A
\right)
-
\left(
\xi^{\alpha C}
{}[ {\mit\Omega}_{j - 1l}^{sq, \bit{\scriptstyle 20}} ]_\alpha{}_{AC}^B
-
\bar\xi_{\dot\alpha C}
{}[
\bar{\mit\Omega}_{j - 1l}^{sq, \bit{\scriptstyle \overline{20\!}\,}}
{}]^{\dot\alpha}{}^{BC}_A
\right)
\bigg\}
\, . \nonumber
\end{eqnarray}
Finally, the $\bit{20}$ transforms via
\begin{eqnarray}
\label{Os20}
\delta [ {\cal O}_{jl}^{ss, \bit{\scriptstyle 20}}]_{AB}^{CD}
\!\!\!&=&\!\!\!
- \frac{\sigma_j}{6 (2j + 3)}
{}[P_{\bit{\scriptstyle 20}}]_{AB;GH}^{CD;EF}
\bigg\{
(j + 2)
\left(
\xi^{\alpha [G}
{}[
{\mit\Omega}_{jl}^{sq, \bit{\scriptstyle 20}}
{}]_\alpha{}_{EF}^{H]}
-
\bar\xi_{\dot\alpha [E}
{}[
\bar{\mit\Omega}_{jl}^{sq, \bit{\scriptstyle \overline{20\!}\,}}
{}]^{\dot\alpha}{}^{GH}_{F]}
\right)
\nonumber\\
&&\qquad\qquad\qquad\qquad\ \ \,
+ \,
(j + 1)
\left(
\xi^{\alpha [G}
{}[
{\mit\Omega}_{j - 1l}^{sq, \bit{\scriptstyle 20}}
{}]_\alpha{}_{EF}^{H]}
-
\bar\xi_{\dot\alpha [E}
{}[
\bar{\mit\Omega}_{j - 1l}^{sq, \bit{\scriptstyle \overline{20\!}\,}}
{}]^{\dot\alpha}{}^{GH}_{F]}
\right)
\bigg\}
\, .
\end{eqnarray}

\subsection{... maximal-helicity bosonic operators}

For antisymmetric representation
\begin{eqnarray}
\delta [ {\cal T}^{qq, \bit{\scriptstyle 6}}_{jl} ]^{\mu AB}
\!\!\!&=&\!\!\!
- \sigma_{j + 1}
\frac{2 (j + 2)}{2 j + 3}
\xi^{\alpha [A}
\left\{
(j + 3)
{}[{^+\!\bar{\mit\Omega}}^{gq}_{jl}]^{\mu B]}_\alpha
-
(j + 1)
{}[{^+\!\bar{\mit\Omega}}^{gq}_{j - 1 l}]^{\mu B]}_\alpha
\right\}
\nonumber\\
&&\!\!\!
+ \sigma_{j + 1}
\frac{(j + 1)(j + 2)}{12 (2 j + 3)}
\bar\xi_{\dot\alpha C} \sigma^{\mu \, \dot\alpha\beta}_{\, \perp}
\bigg\{
8
\varepsilon^{ABCD}
\left(
[ {\mit\Omega}_{jl}^{sq} ]_{\beta D}
-
[ {\mit\Omega}_{j - 1 l}^{sq} ]_{\beta D}
\right)
\nonumber\\
&&\qquad\qquad\qquad\qquad\qquad\qquad
-
\varepsilon^{CDE[A}
\left(
[
{\mit\Omega}_{jl}^{sq, \bit{\scriptstyle 20}}
]_\beta{}^{B]}_{DE}
-
[
{\mit\Omega}_{j - 1 l}^{sq, \bit{\scriptstyle 20}}
]_\beta{}^{B]}_{DE}
\right)
\bigg\}
\, , \\
\delta [ \bar{\cal T}^{qq, \bit{\scriptstyle \bar 6}}_{jl} ]^\mu_{AB}
\!\!\!&=&\!\!\!
- \sigma_{j + 1}
\frac{2 (j + 2)}{2 j + 3}
\bar\xi_{\dot\alpha [A}
\left\{
(j + 3)
{}[{^-\! {\mit\Omega}}^{gq}_{jl}]^{\dot\alpha \mu}_{B]}
-
(j + 1)
{}[{^-\! {\mit\Omega}}^{gq}_{j - 1l}]^{\dot\alpha \mu}_{B]}
\right\}
\nonumber\\
&&\!\!\!
- \sigma_{j + 1}
\frac{(j + 1)(j + 2)}{12 (2 j + 3)}
\xi^{\alpha C} \bar\sigma^\mu_{\perp \alpha\dot\beta}
\bigg\{
8
\varepsilon_{ABCD}
\left(
[ \bar{\mit\Omega}_{jl}^{sq} ]^{\dot\beta D}
-
[ \bar{\mit\Omega}_{j - 1 l}^{sq} ]^{\dot\beta D}
\right)
\nonumber\\
&&\qquad\qquad\qquad\qquad\qquad\qquad
-
\varepsilon_{CDE[A}
\left(
{}[
{\mit\Omega}_{jl}^{sq, \bit{\scriptstyle \overline{20\!}\,}}
]^{\dot\beta}{}^{DE}_{B]}
-
{}[
{\mit\Omega}_{j - 1l}^{sq, \bit{\scriptstyle \overline{20\!}\,}}
]^{\dot\beta}{}^{DE}_{B]}
\right)
\bigg\}
\, ,
\end{eqnarray}
and for symmetric representation
\begin{eqnarray}
\label{T10}
\delta [ {\cal T}^{qq, \bit{\scriptstyle 10}}_{jl} ]^{\mu AB}
\!\!\!&=&\!\!\!
- \sigma_j
\frac{2 (j + 2)}{2 j + 3}
\xi^{\alpha \{A}
\left\{
(j + 3)
{}[{^+\!\bar{\mit\Omega}}^{gq}_{jl}]^{\mu B\}}_\alpha
-
(j + 1)
{}[{^+\!\bar{\mit\Omega}}^{gq}_{j - 1 l}]^{\mu B\}}_\alpha
\right\}
\nonumber\\
&&\!\!\!
- \sigma_j
\frac{(j + 1)(j + 2)}{12 (2 j + 3)}
\bar\xi_{\dot\alpha C} \sigma^{\mu \, \dot\alpha\beta}_{\, \perp}
\varepsilon^{CDE\{A}
\left(
[
{\mit\Omega}_{jl}^{sq, \bit{\scriptstyle 20}}
]_\beta{}^{B\}}_{DE}
-
[
{\mit\Omega}_{j - 1 l}^{sq, \bit{\scriptstyle 20}}
]_\beta{}^{B\}}_{DE}
\right)
\bigg\}
\, , \\
\label{T10bar}
\delta [ \bar{\cal T}^{qq, \bit{\scriptstyle \overline{10\!}\,}}_{jl} ]^\mu_{AB}
\!\!\!&=&\!\!\!
- \sigma_j
\frac{2 (j + 2)}{2 j + 3}
\bar\xi_{\dot\alpha \{A}
\left\{
(j + 3)
{}[{^-\! {\mit\Omega}}^{gq}_{jl}]^{\dot\alpha \mu}_{B\}}
-
(j + 1)
{}[{^-\! {\mit\Omega}}^{gq}_{j - 1l}]^{\dot\alpha \mu}_{B\}}
\right\}
\nonumber\\
&&\!\!\!
+ \sigma_j
\frac{(j + 1)(j + 2)}{12 (2 j + 3)}
\xi^{\alpha C} \bar\sigma^\mu_{\perp \alpha\dot\beta}
\varepsilon_{CDE\{A}
\left(
{}[
{\mit\Omega}_{jl}^{sq, \bit{\scriptstyle \overline{20\!}\,}}
]^{\dot\beta}{}^{DE}_{B\}}
-
{}[
{\mit\Omega}_{j - 1l}^{sq, \bit{\scriptstyle \overline{20\!}\,}}
]^{\dot\beta}{}^{DE}_{B\}}
\right)
\, ,
\end{eqnarray}
Here the square and curly brackets on indices denote antisymmetrization
and symmetrization as defined in Eq.\ (\ref{SymAsym}).

The maximal-helicity gluonic operators transform into the maximal-helicity
gauge-gaugino operators via
\begin{eqnarray}
\delta [ {\cal T}^{gg}_{jl} ]^{\mu\nu}
\!\!\!&=&\!\!\!
\frac{(j + 2)(j + 3)}{3 (2 j + 3)}
\bigg\{
\bar\xi_{\dot\alpha A}
\sigma^{\nu \, \dot\alpha\beta}_\perp
\left(
j
{}[{^+\!\bar{\mit\Omega}}^{gq}_{jl}]^{\mu A}_\beta
+
(j + 1)
{}[{^+\!\bar{\mit\Omega}}^{gq}_{j - 1l}]^{\mu A}_\beta
\right)
\nonumber\\
&&\qquad\quad
- (- 1)^j
\bar\xi_{\dot\alpha A}
\sigma^{\mu \, \dot\alpha\beta}_\perp
\left(
j
{}[{^+\!\bar{\mit\Omega}}^{gq}_{jl}]^{\nu A}_\beta
+
(j + 1)
{}[{^+\!\bar{\mit\Omega}}^{gq}_{j - 1 l}]^{\nu A}_\beta
\right)
\bigg\}
\, , \\
\delta [ \bar {\cal T}^{gg}_{jl} ]^{\mu\nu}
\!\!\!&=&\!\!\!
\frac{(j + 2)(j + 3)}{3 (2 j + 3)}
\bigg\{
\xi^{\alpha A}
\bar\sigma^{\nu}_{\perp \alpha\dot\beta}
\left(
j
{}[{^-\! {\mit\Omega}}^{gq}_{jl}]^{\dot\beta \mu}_A
+
(j + 1)
{}[{^-\! {\mit\Omega}}^{gq}_{j - 1l}]^{\dot\beta \mu}_A
\right)
\nonumber\\
&&\qquad\quad
- (- 1)^j
\xi^{\alpha A}
\bar\sigma^{\mu}_{\perp \alpha\dot\beta}
\left(
j
{}[{^-\! {\mit\Omega}}^{gq}_{jl}]^{\dot\beta \nu}_A
+
(j + 1)
{}[{^-\! {\mit\Omega}}^{gq}_{j - 1l}]^{\dot\beta \nu}_A
\right)
\bigg\}
\, .
\end{eqnarray}

Notice that in the ${\cal N} = 1$ case, ${\cal T}$'s form a separate
supermultiplet and do not enter the one with the chiral-even operators
(\ref{BiLambda}), (\ref{BiGauge}) and (\ref{BiScalar}).

\subsection{... mixed bosonic operators}

\begin{eqnarray}
\delta [ {\cal T}^{sg}_{jl} ]^{\mu AB}
\!\!\!&=&\!\!\!
\frac{1}{3 (2 j + 3)}
\xi^{\alpha [A} \bar\sigma^\mu_{\, \perp \, \alpha\dot\beta}
\bigg\{
\left(
3 (j + 3)
{}[ \bar{\mit\Omega}^{gq}_{jl} ]^{\dot\alpha B]}
+
(- 1)^j (j + 1)
{}[ \bar{\mit\Omega}^{sq}_{jl} ]^{\dot\alpha B]}
\right)
\nonumber\\
&&\qquad\qquad\qquad\qquad \
+
(j + 2)
\left(
3
{}[ \bar{\mit\Omega}^{gq}_{jl} ]^{\dot\alpha B]}
+
(- 1)^j
{}[ \bar{\mit\Omega}^{sq}_{jl} ]^{\dot\alpha B]}
\right)
\bigg\}
\nonumber\\
&-&\!\!\!
\frac{(- 1)^j}{6(2 j + 3)}
\xi^{\alpha C} \bar\sigma^\mu_{\, \perp \, \alpha\dot\beta}
\left\{
(j + 1)
[
\bar{\mit\Omega}_{jl}^{sq, \bit{\scriptstyle \overline{20\!}\,}}
]^{\dot\beta}{}^{AB}_C
+
(j + 2)
[
\bar{\mit\Omega}_{j - 1l}^{sq, \bit{\scriptstyle \overline{20\!}\,}}
]^{\dot\beta}{}^{AB}_C
\right\}
\nonumber\\
&-&\!\!\!
\frac{2}{2 j + 3}
\varepsilon^{ABCD}
\bar\xi_{\dot\alpha C}
\left\{
(j + 3)
{}[{^-\!{\mit\Omega}}^{gq}_{jl}]^{\dot\alpha \mu}_D
+
(j + 2)
{}[{^-\!{\mit\Omega}}^{gq}_{j - 1 l}]^{\dot\alpha \mu}_D
\right\}
\, ,
\end{eqnarray}
and
\begin{eqnarray}
\delta [ \bar {\cal T}^{sg}_{jl} ]^\mu_{AB}
\!\!\!&=&\!\!\!
- \frac{1}{3 (2 j + 3)}
\bar\xi_{\dot\alpha [A} \sigma^{\mu \, \dot\alpha\beta}_{\, \perp}
\bigg\{
\left(
3 (j + 3)
{}[ {\mit\Omega}^{gq}_{jl} ]_{\alpha B]}
-
(- 1)^j (j + 1)
{}[ {\mit\Omega}^{sq}_{jl} ]_{\alpha B]}
\right)
\nonumber\\
&&\qquad\qquad\qquad\qquad \
+
(j + 2)
\left(
3
{}[ {\mit\Omega}^{gq}_{j - 1 l} ]_{\alpha B]}
-
(- 1)^{j}
{}[ {\mit\Omega}^{sq}_{j - 1 l} ]_{\alpha B]}
\right)
\bigg\}
\nonumber\\
&-&\!\!\!
\frac{(- 1)^j}{6(2 j + 3)}
\bar\xi_{\dot\alpha C} \sigma^{\mu \, \dot\alpha\beta}_{\, \perp}
\left\{
(j + 1)
[
{\mit\Omega}_{jl}^{sq, \bit{\scriptstyle 20}}
]_{\beta}{}_{AB}^C
+
(j + 2)
[
{\mit\Omega}_{j - 1l}^{sq, \bit{\scriptstyle 20}}
]_{\beta}{}_{AB}^C
\right\}
\nonumber\\
&+&\!\!\!
\frac{2}{2 j + 3}
\varepsilon_{ABCD}
\xi^{\alpha C}
\left\{
(j + 3)
{}[{^+\! \bar{\mit\Omega}}^{gq}_{jl}]^{\mu D}_\alpha
+
(j + 2)
{}[{^+\! \bar{\mit\Omega}}^{gq}_{j - 1 l}]^{\mu D}_\alpha
\right\}
\, .
\end{eqnarray}

\subsection{... fermionic operators}

Transformation of gluon-gaugino operators
\begin{eqnarray}
\delta [ {\mit\Omega}^{gq}_{j l} ]_{\alpha A}
\!\!\!&=&\!\!\!
-
\frac{1}{(j + 2)(j + 3)}
\bigg\{
\left(
6 {\cal O}^{gg}_{j l + 1}
-
\ft18 (j + 3) {\cal O}^{qq}_{j l + 1}
\right)
-
\left(
6 \widetilde {\cal O}^{gg}_{j l + 1}
-
\ft18 (j + 3) \widetilde {\cal O}^{qq}_{j l + 1}
\right)
\\
&&\qquad
+
\left(
6 {\cal O}^{gg}_{j + 1 l + 1}
+
\ft18 (j + 1) {\cal O}^{qq}_{j + 1 l + 1}
\right)
-
\left(
6 \widetilde {\cal O}^{gg}_{j + 1 l + 1}
+
\ft18 (j + 1) \widetilde {\cal O}^{qq}_{j + 1 l + 1}
\right)
\bigg\}
\bar\sigma^-{}_{\alpha\dot\beta}
\bar\xi^{\dot\beta}_A
\nonumber\\
&&\!\!\!-
\frac{1}{2 (j + 2)(j + 3)}
\bigg\{
(j + 1)
\left(
{}[ \widetilde {\cal O}_{j + 1 l + 1}^{qq, \bit{\scriptstyle 15}} ]_A{}^B
-
{}[ {\cal O}_{j + 1 l + 1}^{qq, \bit{\scriptstyle 15}} ]_A{}^B
\right)
\nonumber\\
&&\qquad\qquad\qquad\qquad\qquad\qquad\qquad\quad \ \
- (j + 3)
\left(
{}[ \widetilde {\cal O}_{j l + 1}^{qq, \bit{\scriptstyle 15}} ]_A{}^B
-
{}[ {\cal O}_{j l + 1}^{qq, \bit{\scriptstyle 15}} ]_A{}^B
\right)
\bigg\}
\bar\sigma^-{}_{\alpha\dot\beta}
\bar\xi^{\dot\beta}_B
\nonumber\\
&&\!\!\!
-
\frac{j + 1}{2 (j + 2)}
\left\{
[\bar {\cal T}^{sg}_{j l + 1} ]^\mu_{AB}
+
[ \bar {\cal T}^{sg}_{j + 1 l + 1} ]^\mu_{AB}
\right\}
\bar\sigma^\perp_{\, \mu \, \alpha\dot\beta}
\sigma^{- \, \dot\beta\gamma}
\xi^B_\gamma
\, , \nonumber\\
\delta [ \bar{\mit\Omega}_{jl}^{gq} ]^{\dot\alpha A}
\!\!\!&=&\!\!\!
-
\frac{1}{(j + 2)(j + 3)}
\bigg\{
\left(
6 {\cal O}^{gg}_{j l + 1}
-
\ft18 (j + 3) {\cal O}^{qq}_{j l + 1}
\right)
+
\left(
6 \widetilde {\cal O}^{gg}_{j l + 1}
-
\ft18 (j + 3) \widetilde {\cal O}^{qq}_{j l + 1}
\right)
\\
&&\qquad
+
\left(
6 {\cal O}^{gg}_{j + 1 l + 1}
+
\ft18 (j + 1) {\cal O}^{qq}_{j + 1 l + 1}
\right)
+
\left(
6 \widetilde {\cal O}^{gg}_{j + 1 l + 1}
+
\ft18 (j + 1) \widetilde {\cal O}^{qq}_{j + 1 l + 1}
\right)
\bigg\}
\sigma^{- \, \dot\alpha\beta}
\xi^A_\beta
\nonumber\\
&&\!\!\!
-
\frac{1}{2 (j + 2)(j + 3)}
\bigg\{
(j + 1)
\left(
{}[ \widetilde {\cal O}_{j + 1 l + 1}^{qq, \bit{\scriptstyle 15}} ]_B{}^A
+
{}[ {\cal O}_{j + 1 l + 1}^{qq, \bit{\scriptstyle 15}} ]_B{}^A
\right)
\nonumber\\
&&\qquad\qquad\qquad\qquad\qquad\qquad\qquad\quad \ \
- (j + 3)
\left(
{}[ \widetilde {\cal O}_{j l + 1}^{qq, \bit{\scriptstyle 15}} ]_B{}^A
+
{}[ {\cal O}_{j l + 1}^{qq, \bit{\scriptstyle 15}} ]_B{}^A
\right)
\bigg\}
\sigma^{- \, \dot\alpha\beta}
\xi^B_\beta
\nonumber\\
&&\!\!\!
+
\frac{j + 1}{2 (j + 2)}
\left\{
{}[ {\cal T}^{sg}_{j l + 1} ]^{\mu AB}
+
{}[ {\cal T}^{sg}_{j + 1 l + 1} ]^{\mu AB}
\right\}
\sigma^{\perp \, \dot\alpha\beta}_{\, \mu}
\bar\sigma^-{}_{\beta\dot\gamma}
\bar\xi^{\dot\gamma}_B
\, , \nonumber
\end{eqnarray}
and for scalar-gaugino operators
\begin{eqnarray}
\delta [ {\mit\Omega}^{sq}_{jl} ]_{\alpha A}
\!\!\!&=&\!\!\!
\frac{1}{8 (j + 2)}
\bigg\{
3 \widetilde {\cal O}^{qq}_{j + 1 l + 1}
-
3 \widetilde {\cal O}^{qq}_{j l + 1}
\\
&&\quad
+
\left(
3 {\cal O}^{qq}_{j + 1 l + 1}
+
2 (j + 2) {\cal O}^{ss}_{j + 1 l + 1}
\right)
-
\left(
3 {\cal O}^{qq}_{j l + 1}
-
2 (j + 2) {\cal O}^{ss}_{j l + 1}
\right)
\bigg\}
\bar\sigma^-{}_{\alpha\dot\beta}
\bar\xi^{\dot\beta}_A
\nonumber\\
&+&\!\!\!
\frac{1}{2 (j + 2)}
\bigg\{
{}[ \widetilde {\cal O}_{j l + 1}^{qq, \bit{\scriptstyle 15}} ]_A{}^B
-
{}[ \widetilde {\cal O}_{j + 1 l + 1}^{qq, \bit{\scriptstyle 15}} ]_A{}^B
+
\left( [ {\cal O}_{j l + 1}^{qq, \bit{\scriptstyle 15}} ]_A{}^B
+
2 (j + 2) [ {\cal O}_{j l + 1}^{ss, \bit{\scriptstyle 15}} ]_A{}^B
\right)
\nonumber\\
&&\qquad\qquad\qquad\qquad \
-
\left( [ {\cal O}_{j + 1 l + 1}^{qq, \bit{\scriptstyle 15}} ]_A{}^B
-
2 (j + 2) {}[ {\cal O}_{j + 1 l + 1}^{ss, \bit{\scriptstyle 15}} ]_A{}^B
\right)
\!
\bigg\}
\bar\sigma^-{}_{\alpha\dot\beta}
\bar\xi^{\dot\beta}_B
\nonumber\\
&+&\!\!\!
\frac{1}{2 (j + 2)}
\Big\{
\left(
{}[ {\cal T}^{qq, \bit{\scriptstyle \bar 6}}_{j l + 1} ]^\mu_{AB}
+
(- 1)^j (j + 1)
{}[ \bar {\cal T}^{sg}_{j l + 1} ]^\mu_{AB}
\right)
\nonumber\\
&&\qquad\qquad\qquad\qquad\
-
\left(
{}[ {\cal T}^{qq, \bit{\scriptstyle \bar 6}}_{j + 1 l + 1} ]^\mu_{AB}
+
(- 1)^j (j + 3)
{}[ \bar {\cal T}^{sg}_{j + 1 l + 1} ]^\mu_{AB}
\right)
\Big\}
\bar\sigma^\perp_{\, \mu \, \alpha\dot\beta}
\sigma^{- \, \dot\beta\gamma}
\xi^B_\gamma
\, , \nonumber\\
\delta [ \bar{\mit\Omega}_{jl}^{sq} ]^{\dot\alpha A}
\!\!\!&=&\!\!\!
\frac{1}{8 (j + 2)}
\bigg\{
3 \widetilde {\cal O}^{qq}_{j + 1 l + 1}
-
3 \widetilde {\cal O}^{qq}_{j l + 1}
\\
&&\quad
-
\left(
3 {\cal O}^{qq}_{j + 1 l + 1}
+
2 (j + 2) {\cal O}^{ss}_{j + 1 l + 1}
\right)
+
\left(
3 {\cal O}^{qq}_{j l + 1}
-
2 (j + 2) {\cal O}^{ss}_{j l + 1}
\right)
\bigg\}
\sigma^{- \, \dot\alpha\beta}
\xi^A_\beta
\nonumber\\
&-&\!\!\!
\frac{1}{2 (j + 2)}
\bigg\{
{}[ \widetilde {\cal O}_{j l + 1}^{qq, \bit{\scriptstyle 15}} ]_B{}^A
-
{}[ \widetilde {\cal O}_{j + 1 l + 1}^{qq, \bit{\scriptstyle 15}} ]_B{}^A
-
\left( [ {\cal O}_{j l + 1}^{qq, \bit{\scriptstyle 15}} ]_B{}^A
+
2 (j + 2) [ {\cal O}_{j l + 1}^{ss, \bit{\scriptstyle 15}} ]_B{}^A
\right)
\nonumber\\
&&\qquad\qquad\qquad\qquad\
+
\left( [ {\cal O}_{j + 1 l + 1}^{qq, \bit{\scriptstyle 15}} ]_B{}^A
-
2 (j + 2) [ {\cal O}_{j + 1 l + 1}^{ss, \bit{\scriptstyle 15}} ]_B{}^A
\right)
\!
\bigg\}
\sigma^{- \, \dot\alpha\beta}
\xi^B_\beta
\nonumber\\
&-&\!\!\!
\frac{1}{2 (j + 2)}
\Big\{
\left(
{}[ \bar {\cal T}^{qq, \bit{\scriptstyle 6}}_{j l + 1} ]^{\mu AB}
-
(- 1)^j (j + 1)
{}[ {\cal T}^{sg}_{j l + 1} ]^{\mu AB}
\right)
\nonumber\\
&&\qquad\qquad\qquad\qquad\quad
-
\left(
{}[ \bar {\cal T}^{qq, \bit{\scriptstyle 6}}_{j + 1 l + 1} ]^{\mu AB}
-
(- 1)^j (j + 3)
{}[ {\cal T}^{sg}_{j + 1 l + 1} ]^{\mu AB}
\right)
\Big\}
\sigma_{\, \mu}^{\perp \, \dot\alpha\beta}
\sigma^-{}_{\beta\dot\gamma}
\bar\xi^{\dot\gamma}_B
\, ,
\nonumber
\end{eqnarray}
where we have used Fierz identities (\ref{FierzWeyl}) and the
properties (\ref{ThreeTransverse}) and (\ref{SigmaToOne}).

One can easily recognize from these variations the result for ${\cal N}
= 1$ supersymmetric Yang-Mills theory by replacing $\ft18 \to \ft12$ in
coefficients on the right-hand side of the variations of ${\mit\Omega}^{gq}_{jl}$.
From the latter one can easily read off the components of the supermultiplet
of conformal operators in this theory \cite{BukFroKurLip85,BelMulSch98,BelMul00}
which form the chiral superfield \cite{BelMul00}.

Transformation of the maximal-helicity fermion operators
\begin{eqnarray}
\delta
{}[{^+\! \bar{\mit\Omega}}^{gq}_{jl}]^{\mu A}_\alpha
\!\!\!&=&\!\!\!
\frac{1}{8(j + 2)(j + 3)}
\bigg\{
(j + 3)
\left(
\varepsilon^{ABCD}
\left(
2 (j + 1)
{}[ \bar {\cal T}^{sg}_{jl + 1} ]^\mu_{CD}
-
{}[ {\cal T}^{qq, \bit{\scriptstyle \bar 6}}_{jl + 1} ]^\mu_{CD}
\right)
+
2 [ {\cal T}^{qq, \bit{\scriptstyle 10}}_{j l + 1} ]^{\mu AB}
\right)
\nonumber\\
&&\quad
+ (j + 1)
\left(
\varepsilon^{ABCD}
\left(
2 (j + 3)
{}[ \bar {\cal T}^{sg}_{j + 1 l + 1} ]^\mu_{CD}
+
{}[ {\cal T}^{qq, \bit{\scriptstyle \bar 6}}_{j + 1l + 1} ]^\mu_{CD}
\right)
-
2 [ {\cal T}^{qq, \bit{\scriptstyle 10}}_{j  + 1 l + 1} ]^{\mu AB}
\right)
\bigg\}
\bar\sigma^-{}_{\alpha\dot\beta}
\bar\xi^{\dot\beta}_B
\nonumber\\
&-&\!\!\!
\frac{3}{2 (j + 2)(j + 3)}
\left\{
[ {\cal T}^{gg}_{j l + 1} ]^{\mu\nu}
+
[ {\cal T}^{gg}_{j + 1 l + 1} ]^{\mu\nu}
\right\}
\bar\sigma^\perp_{\, \nu \, \alpha\dot\beta}
\sigma^{- \, \dot\beta\gamma}
\xi_\gamma^A
\, ,
\end{eqnarray}
and
\begin{eqnarray}
\delta
{}[{^-\! {\mit\Omega}}^{gq}_{jl}]^{\mu \dot\alpha}_A
\!\!\!&=&\!\!\!
\frac{- 1}{8(j + 2)(j + 3)}
\bigg\{
(j + 3)
\left(
\varepsilon_{ABCD}
\left(
2 (j + 1)
{}[ {\cal T}^{sg}_{jl + 1} ]^{\mu CD}
+
{}[ \bar{\cal T}^{qq, \bit{\scriptstyle 6}}_{jl + 1} ]^{\mu CD}
\right)
-
2 [ \bar{\cal T}^{qq, \bit{\scriptstyle \overline{10\!}\,}}_{j l + 1} ]^\mu_{AB}
\right)
\nonumber\\
&&\quad
+ (j + 1)
\left(
\varepsilon_{ABCD}
\left(
2 (j + 3)
{}[ {\cal T}^{sg}_{j + 1l + 1} ]^{\mu CD}
-
{}[ \bar{\cal T}^{qq, \bit{\scriptstyle 6}}_{j + 1l + 1} ]^{\mu CD}
\right)
+
2 [ \bar{\cal T}^{qq, \bit{\scriptstyle \overline{10\!}\,}}_{j + 1 l + 1} ]^\mu_{AB}
\right)
\bigg\}
\sigma^{- \, \dot\alpha\beta}
\xi^B_\beta
\nonumber\\
&-&\!\!\!
\frac{3}{2 (j + 2)(j + 3)}
\left\{
[ \bar{\cal T}^{gg}_{j l + 1} ]^{\mu\nu}
+
[ \bar{\cal T}^{gg}_{j + 1 l + 1} ]^{\mu\nu}
\right\}
\sigma^{\perp \, \dot\alpha\beta}_{\, \nu}
\bar\sigma^-{}_{\beta\dot\gamma}
\bar\xi_A^{\dot\gamma}
\, .
\end{eqnarray}

The $\bit{20}$ transforms as
\begin{eqnarray}
\delta
{}[\bar {\mit\Omega}^{sq, \bit{\scriptstyle \overline{20\!}\,}}_{jl}]^{\dot\alpha}{}^{AB}_C
\!\!\!&=&\!\!\!
\frac{1}{j + 2}
[P_{\bit{\scriptstyle \overline{20\!}\,}}]^{AB;F}_{C;DE}
\bigg\{
\left(
- [ {\cal O}_{j l + 1}^{qq, \bit{\scriptstyle 15}} ]_F{}^E
+
(j + 2)
{}[ {\cal O}_{j l + 1}^{ss, \bit{\scriptstyle 15}} ]_F{}^E
\right)
\\
&&\qquad
+
\left(
{}[ {\cal O}_{j + 1 l + 1}^{qq, \bit{\scriptstyle 15}} ]_F{}^E
+
(j + 2)
{}[ {\cal O}_{j + 1 l + 1}^{ss, \bit{\scriptstyle 15}} ]_F{}^E
\right)
+
{}[ \widetilde {\cal O}_{j l + 1}^{qq, \bit{\scriptstyle 15}} ]_F{}^E
-
{}[ \widetilde {\cal O}_{j + 1 l + 1}^{qq, \bit{\scriptstyle 15}} ]_F{}^E
\bigg\}
\sigma^{- \dot\alpha\beta}
\xi^D_\beta
\nonumber\\
&+&\!\!\!
\frac{1}{2}
\left(
{}[ {\cal O}_{j l + 1}^{ss, \bit{\scriptstyle 20}} ]^{AB}_{CD}
+
{}[ {\cal O}_{j + 1l + 1}^{ss, \bit{\scriptstyle 20}} ]^{AB}_{CD}
\right)
\sigma^{- \dot\alpha\beta}
\xi^D_\beta
\nonumber\\
&+&\!\!\!
\frac{3}{2 (j + 2)}
\varepsilon^{ABDE}
\left(
{}[ \bar{\cal T}^{qq, \bit{\scriptstyle \overline{10\!}\,}}_{jl + 1} ]^\mu_{CD}
-
{}[ \bar{\cal T}^{qq, \bit{\scriptstyle \overline{10\!}\,}}_{j + 1l + 1} ]^\mu_{CD}
\right)
\sigma^{\perp \, \dot\alpha\beta}_{\, \mu}
\bar\sigma^-{}_{\beta\dot\gamma}
\bar\xi_E^{\dot\gamma}
\nonumber\\
&+&\!\!\! \frac{(- 1)^j}{4(j + 2)}
[P_{\bit{\scriptstyle \overline{20\!}\,}}]^{AB;F}_{C;DE}
\bigg\{
\left(
2 (j + 1)
{}[ {\cal T}^{sg}_{jl + 1} ]^{\mu DE}
+
{}[ \bar{\cal T}^{qq, \bit{\scriptstyle 6}}_{jl + 1} ]^{\mu DE}
\right)
\nonumber\\
&&\qquad\qquad\qquad\quad
-
\left(
2 (j + 3)
{}[ {\cal T}^{sg}_{j + 1l + 1} ]^{\mu DE}
-
{}[ \bar{\cal T}^{qq, \bit{\scriptstyle 6}}_{j + 1l + 1} ]^{\mu DE}
\right)
\bigg\}
\sigma^{\perp \, \dot\alpha\beta}_{\, \mu}
\bar\sigma^-{}_{\beta\dot\gamma}
\bar\xi_F^{\dot\gamma}
\, , \nonumber\\
\delta
{}[{\mit\Omega}^{sq, \bit{\scriptstyle 20}}_{jl}]_\alpha{}^C_{AB}
\!\!\!&=&\!\!\!
\frac{1}{j + 2}
{}[P_{\bit{\scriptstyle 20}}]_{AB;F}^{C;DE}
\bigg\{
\left(
- [ {\cal O}_{j l + 1}^{qq, \bit{\scriptstyle 15}} ]_E{}^F
+
(j + 2)
{}[ {\cal O}_{j l + 1}^{ss, \bit{\scriptstyle 15}} ]_E{}^F
\right)
\\
&&\qquad
+
\left(
{}[ {\cal O}_{j + 1 l + 1}^{qq, \bit{\scriptstyle 15}} ]_E{}^F
+
(j + 2)
{}[ {\cal O}_{j + 1 l + 1}^{ss, \bit{\scriptstyle 15}} ]_E{}^F
\right)
-
{}[ \widetilde {\cal O}_{j l + 1}^{qq, \bit{\scriptstyle 15}} ]_E{}^F
+
{}[ \widetilde {\cal O}_{j + 1 l + 1}^{qq, \bit{\scriptstyle 15}} ]_E{}^F
\bigg\}
\bar\sigma^-{}_{\alpha\dot\beta}
\bar\xi_D^{\dot\beta}
\nonumber\\
&-&\!\!\!
\frac{1}{2}
\left(
{}[ {\cal O}_{j l + 1}^{ss, \bit{\scriptstyle 20}} ]_{AB}^{CD}
+
{}[ {\cal O}_{j + 1l + 1}^{ss, \bit{\scriptstyle 20}} ]_{AB}^{CD}
\right)
\bar\sigma^-{}_{\alpha\dot\beta}
\bar\xi_D^{\dot\beta}
\nonumber\\
&-&\!\!\!
\frac{3}{2 (j + 2)}
\varepsilon_{ABDE}
\left(
{}[ {\cal T}^{qq, \bit{\scriptstyle 10}}_{jl + 1} ]^{\mu CD}
-
{}[ {\cal T}^{qq, \bit{\scriptstyle 10}}_{j + 1l + 1} ]^{\mu CD}
\right)
\bar\sigma^\perp_{\mu \, \alpha\dot\beta}
\sigma^{- \dot\beta\gamma}
\xi^E_\gamma
\nonumber\\
&+&\!\!\! \frac{(- 1)^j}{4(j + 2)}
[P_{\bit{\scriptstyle 20}}]_{AB;F}^{C;DE}
\bigg\{
\left(
2 (j + 1)
{}[ \bar{\cal T}^{sg}_{jl + 1} ]^\mu_{DE}
-
{}[ {\cal T}^{qq, \bit{\scriptstyle \bar 6}}_{jl + 1} ]^\mu_{DE}
\right)
\nonumber\\
&&\qquad\qquad\qquad\quad
-
\left(
2 (j + 3)
{}[ \bar{\cal T}^{sg}_{j + 1l + 1} ]^\mu_{DE}
+
{}[ {\cal T}^{qq, \bit{\scriptstyle \bar 6}}_{j + 1l + 1} ]^\mu_{DE}
\right)
\bigg\}
\bar\sigma^\perp_{\mu \, \dot\alpha\beta}
\sigma^{- \dot\beta\gamma}
\xi^F_\gamma
\, . \nonumber
\end{eqnarray}

\subsection{... components of the supermultiplet}
\label{SUSYofComponents}

As we explained in the main text, in order to construct automatically
components of the supermultiplet, one has to choose the right primary.
To this end we use ${\cal O}^{ss, \bit{\scriptstyle 20}}_{jl}$. Then,
the first level descendants are, according to Eq.\ (\ref{Os20}), the
operators $[{\mit\Omega}^{sq, \bit{\scriptstyle 20}}_{[\pm] jl}]_\alpha{}^C_{AB}$
and its complex conjugated sibling. In turn their descendants are
\begin{eqnarray}
&&\delta
{}[{\mit\Omega}^{sq, \bit{\scriptstyle 20}}_{[+] jl}]_\alpha{}^C_{AB}
=
\frac{1}{2 (j + 2)}
\left\{
2
[P_{\bit{\scriptstyle 20}}]_{AB;F}^{C;DE}
\left(
[{\cal S}^{1, \bit{\scriptstyle 15}}_{jl + 1}]_E{}^F
+
[\widetilde {\cal O}^{qq, \bit{\scriptstyle 15}}_{j + 1l + 1}]_E{}^F
\right)
-
(j + 2)
[{\cal O}^{ss, \bit{\scriptstyle 20}}_{j + 1l + 1}]_{AB}^{CD}
\right\}
\bar\sigma^-_{\alpha\dot\beta} \bar\xi_D^{\dot\beta}
\nonumber\\
&&\qquad\qquad\qquad
+ \frac{1}{4 (j + 2)}
\Big\{
6 \varepsilon_{ABEF}
{}[ {\cal T}^{qq, \bit{\scriptstyle 10}}_{j + 1l + 1} ]^{\mu CE}
\\
&&\qquad\qquad\qquad\qquad\qquad
+
[P_{\bit{\scriptstyle 20}}]_{AB;F}^{C;DE}
\left(
[ \bar{\cal T}^1_{[+] jl + 1} ]^\mu_{DE}
-
2 (j + 3)
[ \bar{\cal T}^3_{[-] j + 1l + 1} ]^\mu_{DE}
\right)
\Big\}
\bar\sigma^\perp_{\mu \, \alpha\dot\beta}
\sigma^{- \dot\beta\gamma}
\xi^F_\gamma
\, , \nonumber\\
&&\delta
{}[{\mit\Omega}^{sq, \bit{\scriptstyle 20}}_{[-] jl}]_\alpha{}^C_{AB}
=
\frac{1}{2 (j + 2)}
\left\{
2
[P_{\bit{\scriptstyle 20}}]_{AB;F}^{C;DE}
\left(
[{\cal S}^{2, \bit{\scriptstyle 15}}_{j + 1l + 1}]_E{}^F
-
[\widetilde {\cal O}^{qq, \bit{\scriptstyle 15}}_{jl + 1}]_E{}^F
\right)
-
(j + 2)
[{\cal O}^{ss, \bit{\scriptstyle 20}}_{jl + 1}]_{AB}^{CD}
\right\}
\bar\sigma^-_{\alpha\dot\beta} \bar\xi_D^{\dot\beta}
\nonumber\\
&&\qquad\qquad\qquad
- \frac{1}{4 (j + 2)}
\Big\{
6 \varepsilon_{ABEF}
{}[ {\cal T}^{qq, \bit{\scriptstyle 10}}_{jl + 1} ]^{\mu CE}
\\
&&\qquad\qquad\qquad\qquad\qquad
-
[P_{\bit{\scriptstyle 20}}]_{AB;F}^{C;DE}
\left(
[ \bar{\cal T}^2_{[+] j + 1l + 1} ]^\mu_{DE}
-
2 (j + 1)
[ \bar{\cal T}^3_{[-] jl + 1} ]^\mu_{DE}
\right)
\Big\}
\bar\sigma^\perp_{\mu \, \alpha\dot\beta}
\sigma^{- \dot\beta\gamma}
\xi^F_\gamma
\, . \nonumber
\end{eqnarray}
When one combines the operators with same quantum numbers and conformal
spin they form the components of the supermultiplet, and, of course, the
eigenfunctions of the anomalous dimension matrix as was demonstrated
in section \ref{AnomalousDimension}. Evaluating supersymmetric descendants
of all arising bosonic components one finds the following fermionic
components; namely,
\begin{eqnarray}
&&
\delta [ \bar{\cal T}^1_{[+] jl} ]^\mu_{AB}
=
- \frac{2 (j + 1)}{3 (2 j + 3)}
\bar\xi_{\dot\alpha [A} \sigma^{\mu \, \dot\alpha\beta}_{\, \perp}
\left\{
(j + 3)
{}[ {\mit\Omega}^1_{[+] jl} ]_{\beta B]}
+
3 (j + 2)
{}[ {\mit\Omega}^2_{[-] j - 1l} ]_{\beta B]}
\right\}
\nonumber\\
&&\qquad\qquad\qquad
-
\frac{j + 1}{3}
\bar\xi_{\dot\alpha C} \sigma^{\mu \, \dot\alpha\beta}_{\, \perp}
{}[{\mit\Omega}^{sq, \bit{\scriptstyle 20}}_{[+] jl}]_\beta{}^C_{AB}
+
4 (j + 3)
\varepsilon_{ABCD}
\xi^{\alpha C}
{}[{^+\! \bar{\mit\Omega}}^{gq}_{[+] jl}]^{\mu D}_\beta
\, , \\
&&
\delta [ \bar{\cal T}^2_{[+] jl} ]^\mu_{AB}
=
- \frac{2 (j + 2)}{3 (2 j + 3)}
\bar\xi_{\dot\alpha [A} \sigma^{\mu \, \dot\alpha\beta}_{\, \perp}
\left\{
3
{}[ {\mit\Omega}^2_{[+] jl} ]_{\beta B]}
+
{}[ {\mit\Omega}^1_{[-] j - 1l} ]_{\beta B]}
\right\}
\nonumber\\
&&\qquad\qquad\qquad
-
\frac{j + 2}{3}
\bar\xi_{\dot\alpha C} \sigma^{\mu \, \dot\alpha\beta}_{\, \perp}
{}[{\mit\Omega}^{sq, \bit{\scriptstyle 20}}_{[-]j - 1l}]_\beta{}^C_{AB}
+
4 (j + 2)
\varepsilon_{ABCD}
\xi^{\alpha C}
{}[{^+\! \bar{\mit\Omega}}^{gq}_{[-]j - 1l}]^{\mu D}_\beta
\, , \\
\label{SUSYthetaMinus}
&&
\delta [ \bar{\cal T}^3_{[-] jl} ]^\mu_{AB}
=
- \frac{1}{3 (2 j + 3)}
\bar\xi_{\dot\alpha [A} \sigma^{\mu \, \dot\alpha\beta}_{\, \perp}
\left\{
{}[ {\mit\Omega}^1_{[-] jl} ]_{\beta B]}
+
(j + 2)
{}[ {\mit\Omega}^1_{[+] j - 1l} ]_{\beta B]}
\right\}
\nonumber\\
&&\qquad\qquad\qquad
+
\frac{1}{6 (2j + 3)}
\bar\xi_{\dot\alpha C} \sigma^{\mu \, \dot\alpha\beta}_{\, \perp}
\left\{
(j + 1)
{}[{\mit\Omega}^{sq, \bit{\scriptstyle 20}}_{[-] jl}]_\beta{}^C_{AB}
+
(j + 2)
{}[{\mit\Omega}^{sq, \bit{\scriptstyle 20}}_{[+] j - 1l}]_\beta{}^C_{AB}
\right\}
\nonumber\\
&&\qquad\qquad\qquad
+
\frac{2}{2j + 3}
\varepsilon_{ABCD}
\xi^{\alpha C}
\left\{
(j + 3)
{}[{^+\! \bar{\mit\Omega}}^{gq}_{[-] jl}]^{\mu D}_\alpha
+
(j + 2)
{}[{^+\! \bar{\mit\Omega}}^{gq}_{[+] j - 1l}]^{\mu D}_\alpha
\right\}
\, ,
\end{eqnarray}
for the maximal-helicity operators and analogous relations holding for
complex conjugated operators which can be read off from appendix
\ref{SUSYtrans}. Here and below we will not present these equations for
their redundancy. Next
\begin{eqnarray}
&&\delta
[\widetilde {\cal O}^{qq, \bit{\scriptstyle 15}}_{jl}]_A{}^B
=
- \frac{2 (j + 2)}{3 (2j + 3)}
[P_{\bit{\scriptstyle 15}}]_{AC}^{BD}
\bigg\{
\xi^{\alpha C}
\left(
[ {\mit\Omega}^1_{[-] jl}]_{\alpha D}
-
(j + 1)
[ {\mit\Omega}^1_{[+] j - 1l}]_{\alpha D}
\right)
\nonumber\\
&&\qquad\qquad\qquad\qquad\qquad\qquad\qquad \
+
\bar\xi_{\dot\alpha D}
\left(
[ \bar{\mit\Omega}^1_{[-] jl}]^{\dot\alpha C}
-
(j + 1)
[ \bar{\mit\Omega}^1_{[+] j - 1l}]^{\dot\alpha C}
\right)
\bigg\}
\nonumber\\
&&\qquad\qquad\qquad
-
\frac{(j + 1)(j + 2)}{3 (2j + 3)}
\bigg\{
\xi^{\alpha C}
\left(
{}[{\mit\Omega}^{sq, \bit{\scriptstyle 20}}_{[-] jl}]_\alpha{}^B_{AC}
-
{}[{\mit\Omega}^{sq, \bit{\scriptstyle 20}}_{[+] j - 1l}]_\alpha{}^B_{AC}
\right)
\nonumber\\
&&\qquad\qquad\qquad\qquad\qquad\qquad\quad
-
\bar\xi_{\dot\alpha C}
\left(
{}[
\bar{\mit\Omega}^{sq, \bit{\scriptstyle \overline{20\!}\,}}_{[-] jl}
]^{\dot\alpha}{}^{BC}_A
-
{}[
\bar{\mit\Omega}^{sq, \bit{\scriptstyle \overline{20\!}\,}}_{[+] j - 1l}
]^{\dot\alpha}{}^{BC}_A
\right)
\bigg\}
\, , \nonumber\\
&&\delta
[{\cal S}^{1, \bit{\scriptstyle 15}}_{jl}]_A{}^B
=
- \frac{2 (j + 2)}{3 (2j + 3)}
[P_{\bit{\scriptstyle 15}}]_{AC}^{BD}
\bigg\{
(j + 3)
\xi^{\alpha C}
[ {\mit\Omega}^1_{[+] jl}]_{\alpha D}
-
3 (j + 1)
\xi^{\alpha C}
[ {\mit\Omega}^2_{[-] j - 1l}]_{\alpha D}
\nonumber\\
&&\qquad\qquad\qquad\qquad\qquad\qquad\qquad
-
(j + 3)
\bar\xi_{\dot\alpha D}
[ \bar{\mit\Omega}^1_{[+] jl}]^{\dot\alpha C}
+
3 (j + 1)
\bar\xi_{\dot\alpha D}
[ \bar{\mit\Omega}^2_{[-] j - 1l}]^{\dot\alpha C}
\bigg\}
\nonumber\\
&&\qquad\qquad\qquad
+
\frac{j + 2}{3}
\left\{
\xi^{\alpha C}
{}[{\mit\Omega}^{sq, \bit{\scriptstyle 20}}_{[+] jl}]_\alpha{}^B_{AC}
+
\bar\xi_{\dot\alpha C}
{}[
\bar{\mit\Omega}^{sq, \bit{\scriptstyle \overline{20\!}\,}}_{[+] jl}
]^{\dot\alpha}{}^{BC}_A
\right\}
\, , \nonumber\\
&&\delta
[{\cal S}^{2, \bit{\scriptstyle 15}}_{jl}]_A{}^B
=
\frac{2}{3 (2j + 3)}
[P_{\bit{\scriptstyle 15}}]_{AC}^{BD}
\bigg\{
\xi^{\alpha C}
\left(
3 (j + 2)
[ {\mit\Omega}^2_{[+] jl}]_{\alpha D}
-
(j + 1)
[ {\mit\Omega}^1_{[-] j - 1l}]_{\alpha D}
\right)
\nonumber\\
&&\qquad\qquad\qquad\qquad\qquad\qquad\quad \;
-
\bar\xi_{\dot\alpha D}
\left(
3 (j + 2)
[ \bar{\mit\Omega}^2_{[+] jl}]^{\dot\alpha C}
-
(j + 1)
[ \bar{\mit\Omega}^1_{[-] j - 1l}]^{\dot\alpha C}
\right)
\bigg\}
\nonumber\\
&&\qquad\qquad\qquad
+
\frac{j + 1}{3}
\left\{
\xi^{\alpha C}
{}[{\mit\Omega}^{sq, \bit{\scriptstyle 20}}_{[-] j - 1l}]_\alpha{}^B_{AC}
+
\bar\xi_{\dot\alpha C}
{}[
\bar{\mit\Omega}^{sq, \bit{\scriptstyle \overline{20\!}\,}}_{[-] j - 1l}
]^{\dot\alpha}{}^{BC}_A
\right\}
\, ,
\end{eqnarray}
for the rest. The subsequent step gives
\begin{eqnarray}
&&\delta [ {\mit\Omega}^1_{[+] jl}]_{\alpha A}
=
- \frac{1}{(j + 2)(j + 3)}
\bigg\{
3
\left(
{\cal S}^2_{j + 1l + 1}
-
{\cal P}^2_{jl + 1}
\right)
\delta_A^B
-
(j + 3)
[{\cal S}^{1, \bit{\scriptstyle 15}}_{jl + 1}]_A{}^B
\nonumber\\
&&\qquad\qquad\qquad\qquad\qquad\qquad\qquad
\qquad\qquad\qquad\qquad\quad \,
+
(2j + 3)
[\widetilde {\cal O}^{qq, \bit{\scriptstyle 15}}_{j + 1l + 1}]_A{}^B
\bigg\}
\bar\sigma^-{}_{\alpha\dot\beta} \bar\xi^{\dot\beta}_B
\nonumber\\
&&\qquad\qquad\qquad
- \frac{1}{2 (j + 2)}
\left\{
{}[ \bar{\cal T}^1_{[+] jl + 1} ]^\mu_{AB}
-
2 (2j + 3)
{}[ \bar{\cal T}^3_{[-] j + 1l + 1} ]^\mu_{AB}
\right\}
\bar\sigma^\perp_{\mu \, \dot\alpha\beta}
\sigma^{- \dot\beta\gamma}
\xi_\gamma^B
\, , \nonumber\\
&&\delta [ {\mit\Omega}^1_{[-] jl}]_{\alpha A}
=
- \frac{1}{j + 2}
\bigg\{
3
\left(
{\cal S}^2_{jl + 1}
-
{\cal P}^1_{j + 1l + 1}
\right)
\delta_A^B
-
(j + 1)
[{\cal S}^{2, \bit{\scriptstyle 15}}_{j + 1l + 1}]_A{}^B
\nonumber\\
&&\qquad\qquad\qquad\qquad\qquad\qquad\qquad
\qquad\qquad\qquad\qquad\quad \,
-
(2j + 5)
[\widetilde {\cal O}^{qq, \bit{\scriptstyle 15}}_{jl + 1}]_A{}^B
\bigg\}
\bar\sigma^-{}_{\alpha\dot\beta} \bar\xi^{\dot\beta}_B
\nonumber\\
&&\qquad\qquad\qquad
- \frac{j + 1}{2 (j + 2)}
\left\{
{}[ \bar{\cal T}^2_{[+] j + 1l + 1} ]^\mu_{AB}
+
2 (2j + 5)
{}[ \bar{\cal T}^3_{[-] jl + 1} ]^\mu_{AB}
\right\}
\bar\sigma^\perp_{\mu \, \dot\alpha\beta}
\sigma^{- \dot\beta\gamma}
\xi_\gamma^B
\, , \nonumber\\
&&\delta [ {\mit\Omega}^2_{[+] jl}]_{\alpha A}
=
- \frac{1}{j + 2}
\left\{
\left(
{\cal S}^1_{j + 1l + 1}
-
{\cal P}^1_{jl + 1}
\right)
\delta_A^B
+
(j + 2)
[{\cal S}^{2, \bit{\scriptstyle 15}}_{jl + 1}]_A{}^B
\right\}
\bar\sigma^-{}_{\alpha\dot\beta} \bar\xi^{\dot\beta}_B
\nonumber\\
&&\qquad\qquad\qquad
- \frac{j + 1}{2 (j + 2)}
{}[ \bar{\cal T}^2_{[+] jl + 1} ]^\mu_{AB}
\sigma^\perp_{\mu \, \dot\alpha\beta}
\sigma^{- \dot\beta\gamma}
\xi_\gamma^B
\, , \nonumber\\
&&\delta [ {\mit\Omega}^2_{[-] jl}]_{\alpha A}
=
- \frac{1}{(j + 2)(j + 3)}
\left\{
\left(
{\cal S}^3_{jl + 1}
-
{\cal P}^2_{j + 1l + 1}
\right)
\delta_A^B
+
(j + 2)
[{\cal S}^{1, \bit{\scriptstyle 15}}_{j + 1l + 1}]_A{}^B
\right\}
\bar\sigma^-{}_{\alpha\dot\beta} \bar\xi^{\dot\beta}_B
\nonumber\\
&&\qquad\qquad\qquad
- \frac{1}{2 (j + 2)}
{}[ \bar{\cal T}^1_{[+] j + 1l + 1} ]^\mu_{AB}
\bar\sigma^\perp_{\mu \, \dot\alpha\beta}
\sigma^{- \dot\beta\gamma}
\xi_\gamma^B
\, ,
\end{eqnarray}
and for the maximal-helicity fermionic operators
\begin{eqnarray}
&&\delta [{^+\! \bar{\mit\Omega}}^{gq}_{[+] jl}]^{\mu \dot\alpha}_A
=
- \frac{1}{8 (j + 2)(j + 3)}
\Big\{
2 (j + 1)
{}[ {\cal T}^{qq, \bit{\scriptstyle 10}}_{j + 1l + 1} ]^{\mu AB}
\nonumber\\
&&\qquad\qquad\qquad\qquad\qquad
-
(j + 3)
\varepsilon^{ABCD}
\left(
{}[ \bar{\cal T}^1_{[+] jl + 1} ]^\mu_{CD}
+
2 (j + 1)
{}[ \bar{\cal T}^3_{[-] j + 1l + 1} ]^\mu_{CD}
\right)
\Big\}
\bar\sigma^-{}_{\alpha\dot\beta} \bar\xi^{\dot\beta}_B
\nonumber\\
&&\qquad\qquad\qquad
- \frac{3}{2 (j + 2)(j + 3)}
\left\{
{}[{\cal T}^{gg}_{[+] jl + 1}]^{\mu\nu}
+
{}[{\cal T}^{gg}_{[-] j + 1l + 1}]^{\mu\nu}
\right\}
\bar\sigma^\perp_{\nu \, \dot\alpha\beta}
\sigma^{- \dot\beta\gamma}
\xi_\gamma^A
\, , \nonumber\\
&&\delta [{^+\! \bar{\mit\Omega}}^{gq}_{[-] jl}]^{\mu \dot\alpha}_A
=
\frac{1}{8 (j + 2)(j + 3)}
\Big\{
2 (j + 3)
{}[ {\cal T}^{qq, \bit{\scriptstyle 10}}_{jl + 1} ]^{\mu AB}
\nonumber\\
&&\qquad\qquad\qquad\qquad\qquad
+
(j + 1)
\varepsilon^{ABCD}
\left(
{}[ \bar{\cal T}^2_{[+] j + 1l + 1} ]^\mu_{CD}
+
2 (j + 3)
{}[ \bar{\cal T}^3_{[-] jl + 1} ]^\mu_{CD}
\right)
\Big\}
\bar\sigma^-{}_{\alpha\dot\beta} \bar\xi^{\dot\beta}_B
\nonumber\\
&&\qquad\qquad\qquad
- \frac{3}{2 (j + 2)(j + 3)}
\left\{
{}[{\cal T}^{gg}_{[-] jl + 1}]^{\mu\nu}
+
{}[{\cal T}^{gg}_{[+] j + 1l + 1}]^{\mu\nu}
\right\}
\bar\sigma^\perp_{\nu \, \dot\alpha\beta}
\sigma^{- \dot\beta\gamma}
\xi_\gamma^A
\, , \nonumber\\
&&\delta [{^-\! {\mit\Omega}}^{gq}_{[+] jl}]^{\mu \dot\alpha}_A
=
- \frac{1}{8 (j + 2)(j + 3)}
\Big\{
2 (j + 1)
{}[ \bar{\cal T}^{qq, \bit{\scriptstyle \overline{10\!}\,}}_{j + 1l + 1} ]^\mu_{AB}
\nonumber\\
&&\qquad\qquad\qquad\qquad\qquad
+
(j + 3)
\varepsilon_{ABCD}
\left(
{}[ {\cal T}^1_{[+] jl + 1} ]^{\mu CD}
+
2 (j + 1)
{}[ {\cal T}^3_{[-] j + 1l + 1} ]^{\mu CD}
\right)
\Big\}
\sigma^{- \dot\alpha\beta} \xi_\beta^B
\nonumber\\
&&\qquad\qquad\qquad
- \frac{3}{2 (j + 2)(j + 3)}
\left\{
{}[\bar{\cal T}^{gg}_{[+] jl + 1}]^{\mu\nu}
+
{}[\bar{\cal T}^{gg}_{[-] j + 1l + 1}]^{\mu\nu}
\right\}
\sigma^{\perp \, \dot\alpha\beta}_{\, \nu}
\bar\sigma^-{}_{\beta\dot\gamma}
\bar\xi_A^{\dot\gamma}
\, , \nonumber\\
&&\delta [{^-\! {\mit\Omega}}^{gq}_{[-] jl}]^{\mu \dot\alpha}_A
=
\frac{1}{8 (j + 2)(j + 3)}
\Big\{
2 (j + 3)
{}[ \bar{\cal T}^{qq, \bit{\scriptstyle \overline{10\!}\,}}_{jl + 1} ]^\mu_{AB}
\nonumber\\
&&\qquad\qquad\qquad\qquad\qquad
-
(j + 1)
\varepsilon_{ABCD}
\left(
{}[ {\cal T}^2_{[+] j + 1l + 1} ]^{\mu CD}
+
2 (j + 3)
{}[ {\cal T}^3_{[-] jl + 1} ]^{\mu CD}
\right)
\Big\}
\sigma^{- \dot\alpha\beta} \xi_\beta^B
\nonumber\\
&&\qquad\qquad\qquad
- \frac{3}{2 (j + 2)(j + 3)}
\left\{
{}[\bar{\cal T}^{gg}_{[-] jl + 1}]^{\mu\nu}
+
{}[\bar{\cal T}^{gg}_{[+] j + 1l + 1}]^{\mu\nu}
\right\}
\sigma^{\perp \, \dot\alpha\beta}_{\, \nu}
\bar\sigma^-{}_{\beta\dot\gamma}
\bar\xi_A^{\dot\gamma}
\, .
\end{eqnarray}
Finally, one observes that the supersymmetric algebra closes by varying
the flavor-singlet bosonic components of the supermultiplet,
\begin{eqnarray}
&&\delta {\cal S}^1_{jl}
=
(j + 1)
\left\{
\xi^{\alpha A}
[ {\mit\Omega}^2_{[+] j - 1l} ]_{\alpha A}
+
\bar\xi_{\dot\alpha A}
[ \bar{\mit\Omega}^2_{[+] j - 1l} ]^{\dot\alpha A}
\right\}
\, , \nonumber\\
&&\delta {\cal S}^2_{jl}
=
\frac{j + 2}{6 (2j + 3)}
\Big\{
(2j + 1)
\left(
\xi^{\alpha A}
[ {\mit\Omega}^1_{[-] jl} ]_{\alpha A}
+
\bar\xi_{\dot\alpha A}
[ \bar{\mit\Omega}^1_{[-] jl} ]^{\dot\alpha A}
\right)
\nonumber\\
&&\qquad\qquad\qquad\quad
+
(j + 1) (2j + 5)
\left(
\xi^{\alpha A}
[ {\mit\Omega}^1_{[+] j - 1l} ]_{\alpha A}
+
\bar\xi_{\dot\alpha A}
[ \bar{\mit\Omega}^1_{[+] j - 1l} ]^{\dot\alpha A}
\right)
\Big\}
\, , \nonumber\\
&&\delta {\cal S}^3_{jl}
=
(j + 2)(j + 3)
\left\{
\xi^{\alpha A}
[ {\mit\Omega}^2_{[-] jl} ]_{\alpha A}
+
\bar\xi_{\dot\alpha A}
[ \bar{\mit\Omega}^2_{[-] jl} ]^{\dot\alpha A}
\right\}
\, , \nonumber\\
&&\delta {\cal P}^1_{jl}
=
- \frac{j + 2}{2 (2j + 3)}
\Big\{
j
\left(
\xi^{\alpha A}
[ {\mit\Omega}^2_{[+] jl} ]_{\alpha A}
-
\bar\xi_{\dot\alpha A}
[ \bar{\mit\Omega}^2_{[+] jl} ]^{\dot\alpha A}
\right)
\nonumber\\
&&\qquad\qquad\qquad\quad
+
(j + 1)
\left(
\xi^{\alpha A}
[ {\mit\Omega}^1_{[-] j - 1l} ]_{\alpha A}
-
\bar\xi_{\dot\alpha A}
[ \bar{\mit\Omega}^1_{[-] j - 1l} ]^{\dot\alpha A}
\right)
\Big\}
\, , \nonumber\\
&&\delta {\cal P}^2_{jl}
=
- \frac{(j + 1)(j + 2)(j + 3)}{2 (2j + 3)}
\Big\{
\left(
\xi^{\alpha A}
[ {\mit\Omega}^1_{[+] jl} ]_{\alpha A}
-
\bar\xi_{\dot\alpha A}
[ \bar{\mit\Omega}^1_{[+] jl} ]^{\dot\alpha A}
\right)
\nonumber\\
&&\qquad\qquad\qquad\qquad\qquad\qquad\quad
+
\left(
\xi^{\alpha A}
[ {\mit\Omega}^2_{[-] j - 1l} ]_{\alpha A}
-
\bar\xi_{\dot\alpha A}
[ \bar{\mit\Omega}^2_{[-] j - 1l} ]^{\dot\alpha A}
\right)
\Big\}
\, .
\end{eqnarray}
The remaining transformation laws which are no displayed above, i.e.,
for $[ {\cal T}^{qq, \bit{\scriptstyle 10}}_{jl} ]^{\mu AB}$,
$[ \bar{\cal T}^{qq, \bit{\scriptstyle \overline{10\!}\,}}_{jl} ]^\mu_{AB}$
and $[{\cal O}^{ss, \bit{\scriptstyle 20}}_{jl}]_{AB}^{CD}$ are given
in Eqs.\ (\ref{T10}), (\ref{T10bar}) and (\ref{Os20}), respectively,
where, for the uniformity of notations used in the present section, one has
to add the corresponding subscript $[\pm]$ depending on the $j$-parity.

\setcounter{equation}{0}
\renewcommand{\theequation}{\Alph{section}.\arabic{equation}}

\section{One-loop renormalization in light-cone gauge}
\label{OneLoopEvolution}

Let us give a few details of the derivation of anomalous dimensions which
fix the scaling weight of the supermultiplet. We perform the calculation
in the light-cone gauge $A^+ = 0$ which preserves supersymmetry of the
Lagrangian. The residual gauge freedom is fixed by imposing the antisymmetric
boundary condition on the gauge potential at the light-cone infinity, i.e.,
$A^\mu_\perp (\infty) + A^\mu_\perp (- \infty) = 0$, which results into the
principal value prescription on the spurious pole in the gluon density matrix.

\subsection{Propagators}

\begin{figure}[t]
\begin{center}
\mbox{
\begin{picture}(0,130)(180,0)
\put(0,0){\insertfig{13}{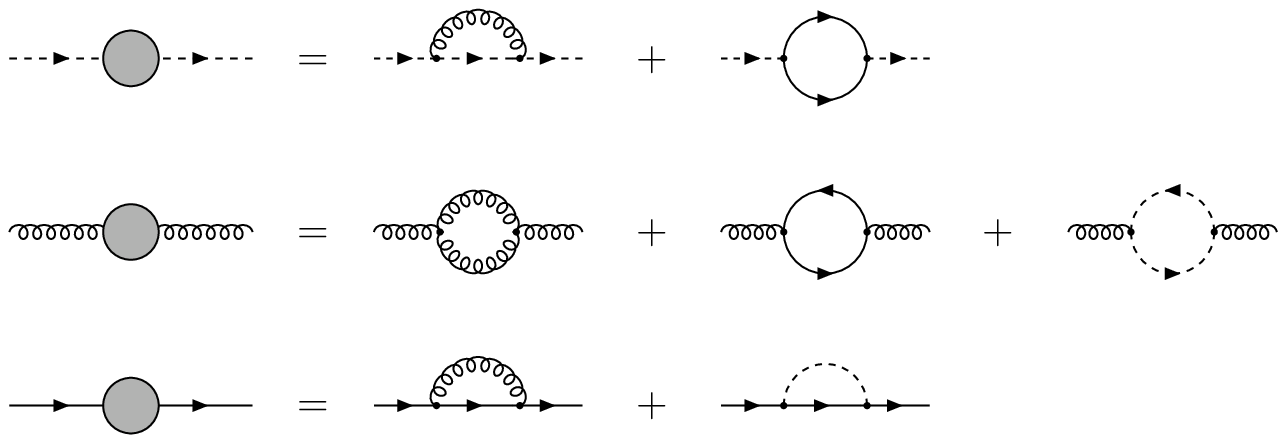}}
\end{picture}
}
\end{center}
\caption{\label{OneLoopProp} One-loop corrections to the scalar, gluon
and gaugino propagators.}
\end{figure}

Let us present first the propagators in ${\cal N} = 4$ Yang-Mills theory
fixed with the light-cone gauge with a principal value prescription.

$\bullet$ Scalar propagator:
\begin{equation}
\langle 0 | T \bar\phi^a_{AB} (z_1) \phi^{b \, CD} (z_2) | 0 \rangle
=
i \delta^{ab} \left( \delta_A^C \delta_B^D - \delta_A^D \delta_B^C \right)
\int \frac{d^4 k}{(2 \pi)^4}
{\rm e}^{- i k \cdot (z_1 - z_2)} \frac{{\cal Z} (k)}{k^2 + i 0}
\, .
\end{equation}

$\bullet$ Gluon propagator:
\begin{equation}
\langle 0 | T A^a_\mu (z_1) A^b_\nu (z_2) | 0 \rangle
=
- i \delta^{ab}
\int \frac{d^4 k}{(2 \pi)^4}
{\rm e}^{- i k \cdot (z_1 - z_2)}
U_{\mu\rho} (k) d^{\rho\sigma} (k) U_{\sigma\nu} (k)
\frac{{\cal Z} (k)}{k^2 + i 0}
\, .
\end{equation}
The spurious pole in the gluon density matrix
\begin{equation}
d_{\mu\nu} (k) = g_{\mu\nu} - \frac{k_\mu n_\nu + k_\nu n_\mu}{[k^+]}
\, ,
\end{equation}
is regularized via the principal value $1/[k^+] = \ft12 \Big[ 1/(k^+ \! + i 0)
+ 1/(k^+ \! - i 0) \Big]$. The presence of the divergent
tensor $U_{\mu\nu}$,
\begin{equation}
U_{\mu\nu} (k)
=
g_{\mu\nu}
-
\widetilde {\cal Z} (k) \frac{k_\mu n_\nu + k_\nu n_\mu}{k^+}
\, ,
\end{equation}
is an artifact of the Lorentz symmetry breaking effects by the gauge
fixing vector $n_\mu$ and the use of the principal value prescription
which leads to different renormalization constants for components of
``good'' and ``bad'' components of the tensor fields.

$\bullet$ Fermion propagator:
\begin{equation}
\langle 0 |
T \lambda^{a A}_\alpha (z_1) \bar\lambda^b_{\dot\beta B} (z_2)
| 0 \rangle
=
i \delta^{ab} \delta^A_B
\int \frac{d^4 k}{(2 \pi)^4}
{\rm e}^{- i k \cdot (z_1 - z_2)}
U_\alpha{}^\beta (k)
k_\mu \bar\sigma^\mu{}_{\beta\dot\gamma}
U^{\dot\gamma}{}_{\dot\beta} (k)
\frac{{\cal Z} (k)}{k^2 + i 0}
\, .
\end{equation}
Here the rotation matrix reads
\begin{equation}
U_\alpha{}^\beta (k)
=
\delta_\alpha^\beta
-
\frac{\widetilde {\cal Z} (k)}{2 k^+}
\bar\sigma^+{}_{\alpha\dot\gamma} k^\mu \sigma_\mu{}^{\dot\gamma\beta}
\, ,
\end{equation}
and its conjugate is
\begin{equation}
U^{\dot\beta}{}_{\dot\alpha}
=
\left( U_\alpha{}^\beta \right)^\ast
\, .
\end{equation}

Since the light-cone gauge fixing does not break the supersymmetry
at the Lagrangian level, the renormalization constants of all elementary
fields are equal. At one loop they are given by the expressions
\begin{equation}
\label{ZConst}
{\cal Z} (k)
=
1 + \frac{g^2 N_c}{4 \pi^2} \frac{S_\varepsilon}{\varepsilon}
\int d q^+ \frac{k^+}{q^+ \! - k^+ \!} \,
\vartheta_{11}^0 (q^+ , q^+ \! - k^+)
\, ,
\end{equation}
and
\begin{equation}
\label{ZtildeConst}
\widetilde {\cal Z} (k)
=
\frac{g^2 N_c}{8 \pi^2} \frac{S_\varepsilon}{\varepsilon}
\left\{
1
+
\int d q^+ \frac{k^+}{q^+ \! - k^+ \!} \,
\vartheta_{11}^0 (q^+ , q^+ \! - k^+)
\right\}
\, ,
\end{equation}
where $\varepsilon \equiv (4 - d)/2$ is the parameter of the dimensional
regularization. Note that the ``good" components of the fields are
renormalized by ${\cal Z}$ and are not affected by $\widetilde {\cal Z}$.
Here and below we introduced the functions
\begin{equation}
\vartheta^{k}_{\alpha_1 \dots \alpha_j} (x_1, \dots , x_j)
\equiv
\int \frac{d \beta}{2 \pi i} \beta^k
\prod_{\ell = 1}^j \left( x_\ell \beta - 1 + i 0 \right)^{- \alpha_\ell}
\, .
\end{equation}
The first few of them, which appear in lowest-order calculations, read
\begin{eqnarray}
&&\vartheta^0_{11} (x_1, x_2) = \frac{\theta (x_1) - \theta (x_2)}{x_1 - x_2}
\, , \\
&&\vartheta^0_{111} (x_1, x_2 , x_3)
=
\frac{x_2}{x_1 - x_2} \vartheta^0_{11} (x_2, x_3)
-
\frac{x_1}{x_1 - x_2} \vartheta^0_{11} (x_1, x_3)
\, .
\end{eqnarray}
They are expressed in terms of the step function $\theta (x) = \{ 1, x
\geq 0 ; 0, x < 0 \}$.

Note that the field renormalization constants (\ref{ZConst}) and
(\ref{ZtildeConst}) are infrared sensitive due to the non-causal choice
of the regularization of the spurious poles in the light-cone propagator.
Had we chosen the Mandelstam-Leibbrandt prescription, the field
renormalization constants are free from infrared singularities and
moreover they are identically zero, demonstrating the ultraviolet
finiteness of the ${\cal N} = 4$ super-Yang-Mills theory \cite{Man83}.

\subsection{Kernels}

Here we give the result for the one-loop dilatation operator for the
maximal-helicity bosonic operators in $\bit{\bar 6}$ of $SU(4)$.
The calculation of the renormalization group kernels of these
light-cone operators is most easily performed in the Fourier
transformed space, i.e.,
\begin{equation}
{\cal O} (x_1, x_2)
\equiv
\int \frac{d \xi_1}{2 \pi} \frac{d \xi_2}{2 \pi}
\,
{\rm e}^{ - i x_1 \xi_1 - i x_2 \xi_2}
\,
{\cal O} (\xi_1, \xi_2)
\, ,
\end{equation}
where ${\cal O} = \bar {\cal T}^{sg}, \dots$. The computation of the one-loop
diagrams in Figs.\ \ref{GStoGS}, \ref{QQtoQQ} and \ref{GStoQQ}, along the line of
Refs.\ \cite{BukKurLip83,Bel97}, results into the following matrix equation
\begin{equation}
[ \bit{\bar{\cal T}}(x_1, x_2) ]_{\rm 1-loop} = - \frac{g^2 N_c}{8 \pi^2}
\frac{S_\varepsilon}{\varepsilon} \int d y_1 d y_2 \, \delta (x_1 + x_2 - y_1 -
y_2) \bit{K} ( x_1, x_2 | y_1, y_2 ) [ \bit{\bar{\cal T}} (y_1, y_2) ]_{\rm
0-loop} \, ,
\end{equation}
with the two-vector
\begin{equation}
\bit{\bar{\cal T}}(x_1, x_2) = \left(
\begin{array}{c}
{} {\cal T}^{qq, \bit{\scriptstyle\bar 6}}
\\
\bar{\cal T}^{sg}
\end{array}
\right) (x_1, x_2)
\, ,
\end{equation}
evolving with the matrix kernel
\begin{equation}
\bit{K} =
\left(
\begin{array}{cc}
K^{11} & K^{12} \\
K^{21} & K^{22}
\end{array}
\right)
\, .
\end{equation}
The elements of the evolution kernels are
\begin{eqnarray}
\label{Kernel}
K^{11} ( x_1, x_2 | y_1, y_2 )
\!\!\!&=&\!\!\!
\frac{x_1}{y_1}
\left[
\frac{y_1}{x_1 - y_1} \vartheta^0_{11} (x_1, x_1 - y_1)
\right]_+
+
\frac{x_2}{y_2}
\left[
\frac{y_2}{x_2 - y_2} \vartheta^0_{11} (x_2, x_2 - y_2)
\right]_+
\, , \\
K^{12} ( x_1, x_2 | y_1, y_2 )
\!\!\!&=&\!\!\!
- \frac{4}{y_2}
\left[
x_1 \vartheta^0_{11} (x_1, x_1 - y_2)
+
x_2 \vartheta^0_{11} (x_2, x_2 - y_2)
\right]
\, , \\
K^{21} ( x_1, x_2 | y_1, y_2 )
\!\!\!&=&\!\!\!
- \frac{1}{4}
\left[
\vartheta^0_{11} (x_1, x_1 - y_1)
+
\vartheta^0_{11} (x_1, x_1 - y_2)
\right]
\, , \\
K^{22} ( x_1, x_2 | y_1, y_2 )
\!\!\!&=&\!\!\!
\left[
\frac{y_1}{x_1 - y_1} \vartheta^0_{11} (x_1, x_1 - y_1)
\right]_+
+
\left( \frac{x_2}{y_2} \right)^2
\left[
\frac{y_2}{x_2 - y_2} \vartheta^0_{11} (x_2, x_2 - y_2)
\right]_+
\nonumber\\
&&\!\!\!-
\frac{y_1}{y_2}
\vartheta^0_{111} (x_1, - x_2 , x_1 - y_2)
-
\frac{x_2}{y_2} \vartheta^0_{11} (x_1, - x_2)
\, .
\end{eqnarray}
Here the regularization of the end-point behavior $y_i \to x_i$
arises from the self-energy diagrams and results into the conventional
plus-prescription
\begin{equation}
\left[
\frac{y_1}{x_1 - y_1} \vartheta^0_{11} (x_1, x_1 - y_1)
\right]_+
\equiv
\frac{y_1}{x_1 - y_1} \vartheta^0_{11} (x_1, x_1 - y_1)
-
\delta (x_1 - y_1)
\int d x'_1
\frac{y_1}{x'_1 - y_1} \vartheta^0_{11} (x'_1, x'_1 - y_1)
\, .
\end{equation}
The kernels are diagonal in the basis of Jacobi polynomials, i.e.,
\begin{equation}
\label{PartialWave}
K^{ab} (x_1, x_2 | y_1, y_2)
=
\frac{1}{2}
\sum_{j = 0}^\infty
\frac{x_1^{\alpha_a} x_2^{\beta_a}}{n_j (\alpha_a, \beta_a)}
P^{(\alpha_a, \beta_a)}_j
\left(
\frac{
x_1 - x_2
}{
x_1 + x_2
}
\right)
\gamma_j^{ab}
\,
P^{(\alpha_b, \beta_b)}_j
\left(
\frac{
y_1 - y_2
}{
y_1 + y_2
}
\right)
\, ,
\end{equation}
where the normalization factor is
\begin{eqnarray*}
n_j (\alpha , \beta)
=
\frac{
{\mit\Gamma} (j + \alpha + 1) {\mit\Gamma} (j + \beta + 1)
}{
(2j + \alpha + \beta  + 1) {\mit\Gamma} (j + 1)
{\mit\Gamma} (j + \alpha + \beta + 1)
}
\, .
\end{eqnarray*}

The transformation of these kernels to the one acting on the light-cone
coordinates is achieved via the Fourier transformation and reads
\begin{equation}
\label{FourierTransformDO}
K (x_1, x_2 | y_1, y_2)
\equiv
-
\int_0^1 dy \int_0^1 dz \, \theta (1 - y - z) \,
\delta (x_1 - \bar y y_1 - z y_2) \,
\mathbb{K} (y, z)
\, ,
\end{equation}
under the condition of the conservation of momentum in the $t$-channel:
$x_1 + x_2 = y_1 + y_2$. Here $\bar y \equiv 1 - y$. The explicit form
of coordinate-space kernels as well as the corresponding local
anomalous dimensions is given in the main text in section
\ref{AnomalousDimension}.


\end{document}